\documentclass[11pt]{article}

\usepackage{amsmath,amssymb}
\usepackage{algorithm}
\usepackage{algpseudocode}
\usepackage{booktabs}
\usepackage{enumerate}
\usepackage{graphicx}
\usepackage{tikz}
\usepackage{hyperref}
\usepackage{microtype}
\usepackage[font=footnotesize,labelfont=bf]{caption}
\usepackage{xspace}
\newcommand{\VoroTop}{\textit{VoroTop}\xspace}
\newcommand{\VoroTypes}{\textit{VoroTypes}\xspace}

\title{Structural identification \\in general crystalline systems \\via Voronoi cell topology}

\author{Emanuel A. Lazar\thanks{Department of Mathematics, Bar Ilan University, Ramat Gan 5290002, Israel, mLazar@math.biu.ac.il}\and
Chris H. Rycroft\thanks{Department of Mathematics, University of Wisconsin--Madison, Madison, WI 53711, United States}\,\,$^,$\thanks{Mathematics Group, Lawrence Berkeley National Laboratory, Berkeley, CA 94720, United States}}

\date{\today}

\newcommand{\subfiglabel}[2]{%
  \begin{tikzpicture}[inner sep=0pt]
    \node (img) {\includegraphics[width=0.33\textwidth,
                                  trim={80 80 80 80}, clip]{#1}};
    \node[anchor=north west, fill=white, draw=black, line width=0.5pt,
          inner sep=2pt, font=\footnotesize]
         at ([xshift=0.2cm, yshift=-0.2cm] img.north west) {#2};
  \end{tikzpicture}%
}

\begin{document}
\maketitle

\begin{abstract}
Widely used methods for identifying crystal structure in atomistic
simulations are limited in practice to a handful of simple structures, leaving 
many important crystalline systems without automated classification techniques.  
Voronoi topology analysis offers a route to classifying arbitrary
crystalline and defect environments, but realizing this potential first
requires enumerating the Voronoi cell topologies that can arise under
thermal perturbation of a structure of interest.
We present an algorithm that deterministically enumerates all Voronoi cell
topologies arising from generic infinitesimal perturbations of an
arbitrary crystal structure, producing structural classifiers that
extend automated identification to general crystalline systems.  We demonstrate the
approach on several structures of general interest.
\end{abstract}

\section{Introduction}
\label{sec:intro}

Crystalline materials underpin much of modern technology, from
structural alloys and semiconductors to energy storage systems
and thermal barrier
coatings~\cite{ashcroft1976solid,kittel1976introduction,padture2002thermal}.
At the atomic scale, these materials are increasingly studied through
large-scale atomistic simulations such as molecular dynamics (MD) and Monte
Carlo methods, which track the positions and
interactions of millions of atoms over physically relevant
timescales.  A fundamental task in nearly every such study is
\emph{structural identification}: determining which atoms belong
to the bulk crystal and which are associated with defects, grain
boundaries, phase boundaries, or disordered
regions~\cite{stukowski2012structure}.  Reliable structural
identification is essential for measuring defect
densities~\cite{stukowski2010extracting}, tracking phase
transformations, characterizing grain boundary
structure~\cite{olmsted2009survey}, and quantifying the extent of
crystalline order under extreme
conditions~\cite{bringa2006shock}.

Several methods have been developed for this purpose, each typically applied to
a limited set of common crystal structures.  Common neighbor analysis (CNA)
classifies atoms based on the topology of bonds
between their shared neighbors, and in standard form distinguishes FCC, BCC,
HCP, and icosahedral environments \cite{honeycutt1987cna}.  The Ackland--Jones method
uses angular distribution functions to distinguish a similar set of structures
\cite{ackland2006jones}.  
Polyhedral template
matching (PTM) fits local atomic environments to ideal reference templates;
current implementations include FCC, BCC, HCP, simple cubic, icosahedral, and
cubic and hexagonal diamond \cite{larsen2016ptm}.  
The centrosymmetry parameter detects departures from
centrosymmetric environments but provides no classification of the structure
type itself \cite{kelchner1998dislocation}.  Energy-based approaches identify
defects through elevated potential energy~\cite{stukowski2012structure}, but
conflate structural questions with questions of interatomic physics: the
``defect'' signal depends on the potential used, and an atom can have high
energy without being structurally disordered, or vice versa.

Currently these methods share a practical limitation: they are built around a
fixed, small set of reference structures. There are many other crystal
structures encountered in materials science, e.g., Frank--Kasper
phases~\cite{frankkasper1958,frank1959complex}, Laves compounds such as
MgCu$_2$~\cite{friauf1927,stein2021laves}, multi-sublattice oxides such as
spinel~\cite{sickafus1999spinel}, and intermetallics with large unit
cells~\cite{samson1962nacd2,urban2004complex}. Lacking a matching reference, these
methods either relegate the atoms to the catch-all ``other'' category or
force-fit them to the nearest built-in type. Either way the structure goes
unidentified, and researchers resort to ad hoc indicators tuned to each
simulated system.

Voronoi tessellations have a long history in materials science as a
tool for analyzing local atomic environments. For a set of atoms in a domain,
the \emph{Voronoi cell} of a given atom is the region of the domain
closer to that atom than to any other \cite{voronoi1908nouvelles,1992okabe,lazar2022voronoi}.  
Throughout, we write
\emph{atom} when discussing atomic crystals and their
simulations, and
\emph{particle} for a point of a general configuration, which may
equally describe colloidal or other non-atomic systems. 
The volume of a particle's Voronoi cell has been
used as a measure of local density and free volume, particularly 
in glasses, granular materials, and liquids \cite{finney1970random,rycroft06a}.  The
\emph{Voronoi index}---counting, in order, the number of triangular,
quadrilateral, pentagonal, and hexagonal faces of the
cell---distinguishes broad classes of local coordination, such as
icosahedral versus crystalline order in metallic
glasses \cite{sheng2006atomic}, and has been used to track structural
order in a range of crystalline and self-assembled
systems \cite{damasceno2012predictive}.  These approaches encode 
some information about the size and shape of the Voronoi cell, but do not
determine its \emph{combinatorial type}, the precise way faces, edges,
and vertices are connected \cite{2012lazar}.  This finer combinatorial information can
distinguish crystalline structures that these coarser descriptors cannot.
FCC and HCP environments, for example, both produce Voronoi cells with
twelve four-sided faces and thus share the same Voronoi index, yet the
way these faces are connected differs between the two structures; only
the combinatorial type distinguishes
them~\cite{lazar2015topological,vorotop}.

Building on this observation, Voronoi cell topology offers a
fundamentally different approach to structural
identification~\cite{2012lazar,lazar2015topological,vorotop}.  In a perfect
crystal, atoms on equivalent crystallographic sites have identical
Voronoi cells.  Under small perturbations such as thermal vibrations, however, the
combinatorial type of each cell can change, but only within a
well-defined \emph{family} of topologies characteristic of the crystal
structure~\cite{lazar2015topological}.  An atom can therefore be
classified as locally crystalline if its Voronoi cell topology belongs
to the appropriate family, and as defective otherwise.  
This approach requires no cutoffs, is applied independently of interatomic potentials, 
and naturally accommodates thermal noise.  Implemented in the \VoroTop
framework~\cite{vorotop,lu2023extension,lazar2024characterizing}, it
has until now been limited to BCC, FCC, and HCP, because computing the
complete family for a given crystal has required structure-specific
analysis~\cite{lazar2015topological}.  Randomly
perturbing the crystal and cataloging the observed cell
types offers a partial alternative, but
provides no guarantee of completeness \cite{lazar2015topological}.

In this paper, we remove the computational barrier that has limited
Voronoi-based structural identification to simple crystals.  We present an
algorithm that deterministically enumerates all Voronoi cell topologies arising
from infinitesimal perturbations of an arbitrary crystal structure.  The
enumeration is complete in the sense that every cell topology reachable under
sufficiently small generic perturbations of the particle positions is found.  The
algorithm exploits the duality between Voronoi tessellations and Delaunay
triangulations: degenerate Voronoi vertices, where $k > 4$ particles are
equidistant from a common point, are resolved by enumerating the regular
triangulations of the equidistant point configuration~\cite{loera}.  Such
degeneracies are the norm rather than the exception in common crystals: FCC,
HCP, and simple cubic all exhibit them (with $k=6$, $6$, and $8$ respectively),
while BCC is a notable exception.  Resolutions at different vertices are then
combined to enumerate all possible global cell topologies, with site symmetry
used to identify equivalent combinations and reduce the enumeration space.
Each cell topology is identified by its Weinberg
vector~\cite{1966weinberg}, a sequence of integers that encodes the connectivity
of a polyhedron's faces, edges, and vertices.

Section~\ref{sec:2d} first introduces the central ideas in two
dimensions, where they are most easily visualized.
Section~\ref{sec:3d} develops the same ideas in three dimensions and
applies the resulting algorithm to standard reference structures,
recovering known cell-type counts.  Section~\ref{sec:practical}
examines the practical issues that arise in using it: vertex coupling,
near-equidistant neighbors, and combinatorial complexity limits.
We then demonstrate it on several crystal structures
which cannot be analyzed with current methods.  
In each case, the algorithm produces a list of 
Weinberg vectors, which we call a \emph{topological filter}, 
against which an atom's Voronoi cell topology can be matched. 
This enables subsequent automated identification of
bulk crystal, grain boundaries, and other defects in
finite-temperature MD simulations.  We provide an open-source implementation,
\VoroTypes, together with a library of pre-computed filters for a
range of crystal structures.

\providecommand{\subfigframed}[4]{%
  \begin{tikzpicture}[inner sep=0pt]
    \node (img)
      {\includegraphics[trim={#1}, clip, width=#4]{#2}};
    \draw[black, line width=0.7pt]
      (img.south west) rectangle (img.north east);
    \node[anchor=north west, fill=white, draw=black, line width=0.5pt,
          inner sep=1.5pt, font=\scriptsize]
         at ([xshift=0.1cm, yshift=-0.1cm] img.north west) {#3};
  \end{tikzpicture}%
}

\providecommand{\subfignoframe}[3]{%
  \begin{tikzpicture}[inner sep=0pt, baseline=(img.north)]
    \node (img) {\includegraphics[width=#3]{#1}};
    \node[anchor=north west, fill=white, draw=black, line width=0.5pt,
          inner sep=1.5pt, font=\scriptsize]
         at ([xshift=0.1cm, yshift=-0.1cm] img.north west) {#2};
  \end{tikzpicture}%
}

\providecommand{\edgekey}[2]{%
  \raisebox{\dimexpr(#1-#2)/2\relax}{\includegraphics[height=#2]{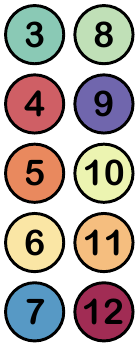}}%
}

\section{Voronoi cell types in two dimensions}
\label{sec:2d}

The framework developed in this paper is intended for
three-dimensional systems, but is most clearly explained and
most easily visualized in two dimensions.  After introducing and 
explaining the central concepts, we will return to the three-dimensional 
case in Section~\ref{sec:3d}.

\begin{figure}[b]
\centering
\subfigframed{240 195 240 195}{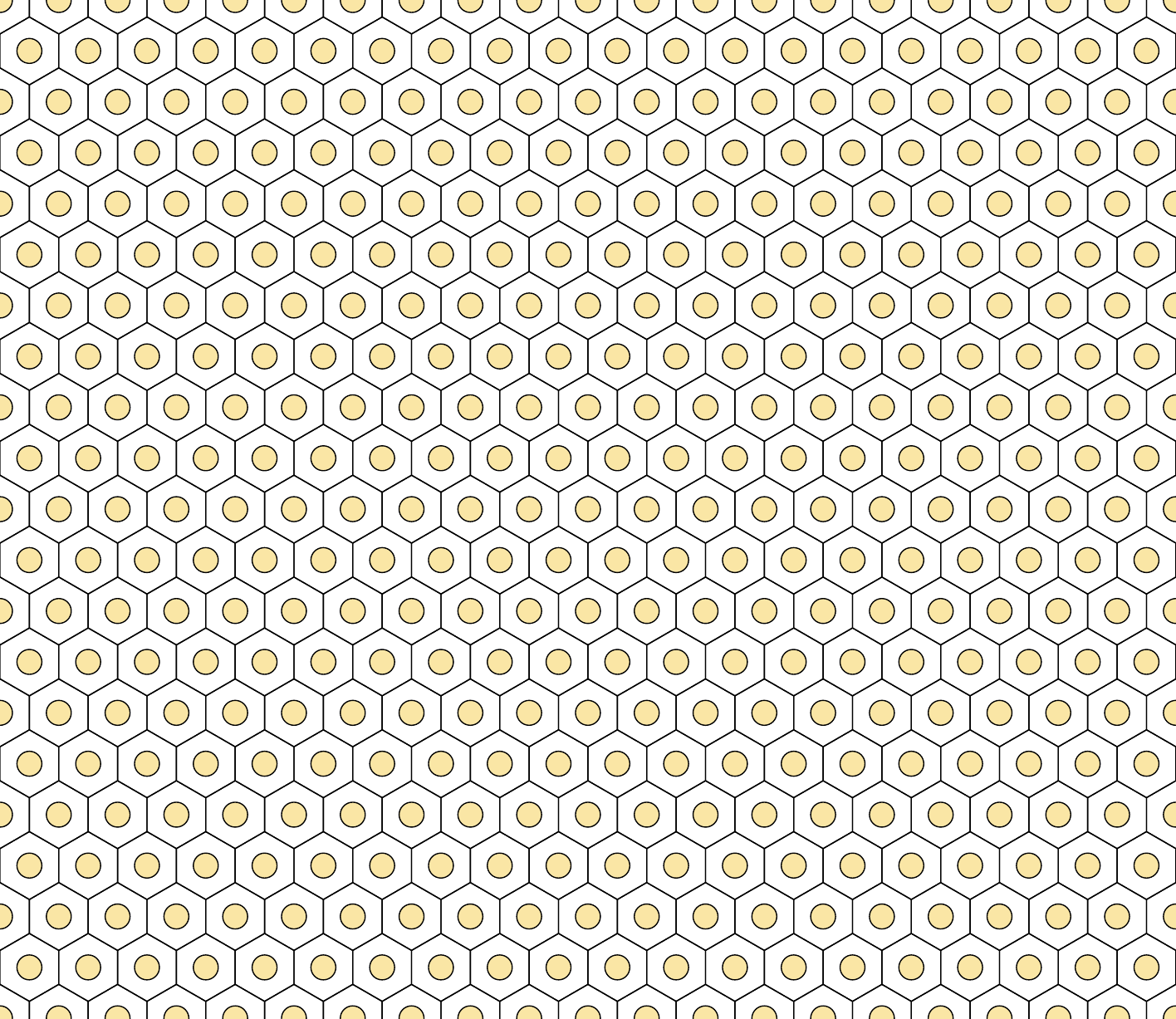}{(a)}{0.22\textwidth}\hfill
\subfigframed{58 58 58 58}{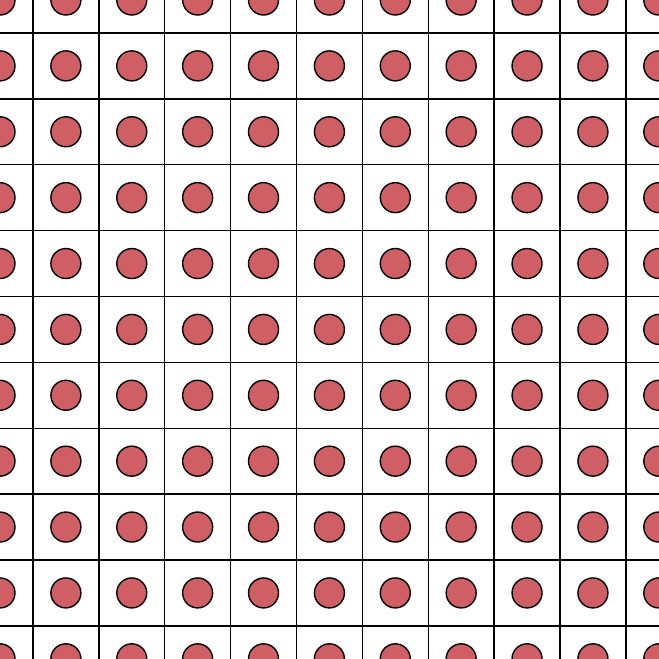}{(b)}{0.22\textwidth}\hfill
\subfigframed{195 410 195 410}{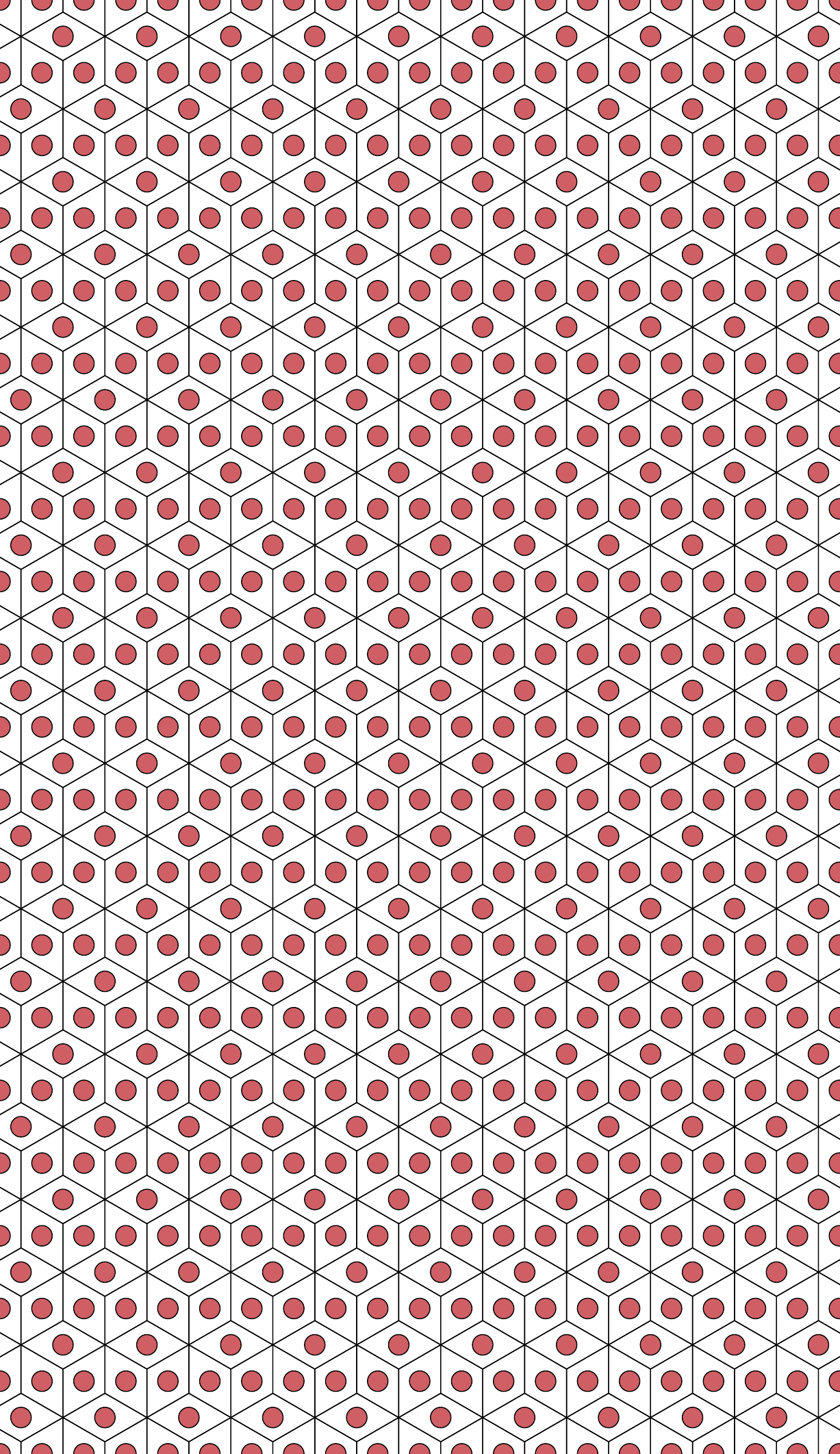}{(c)}{0.22\textwidth}\hfill
\subfigframed{312 140 320 140}{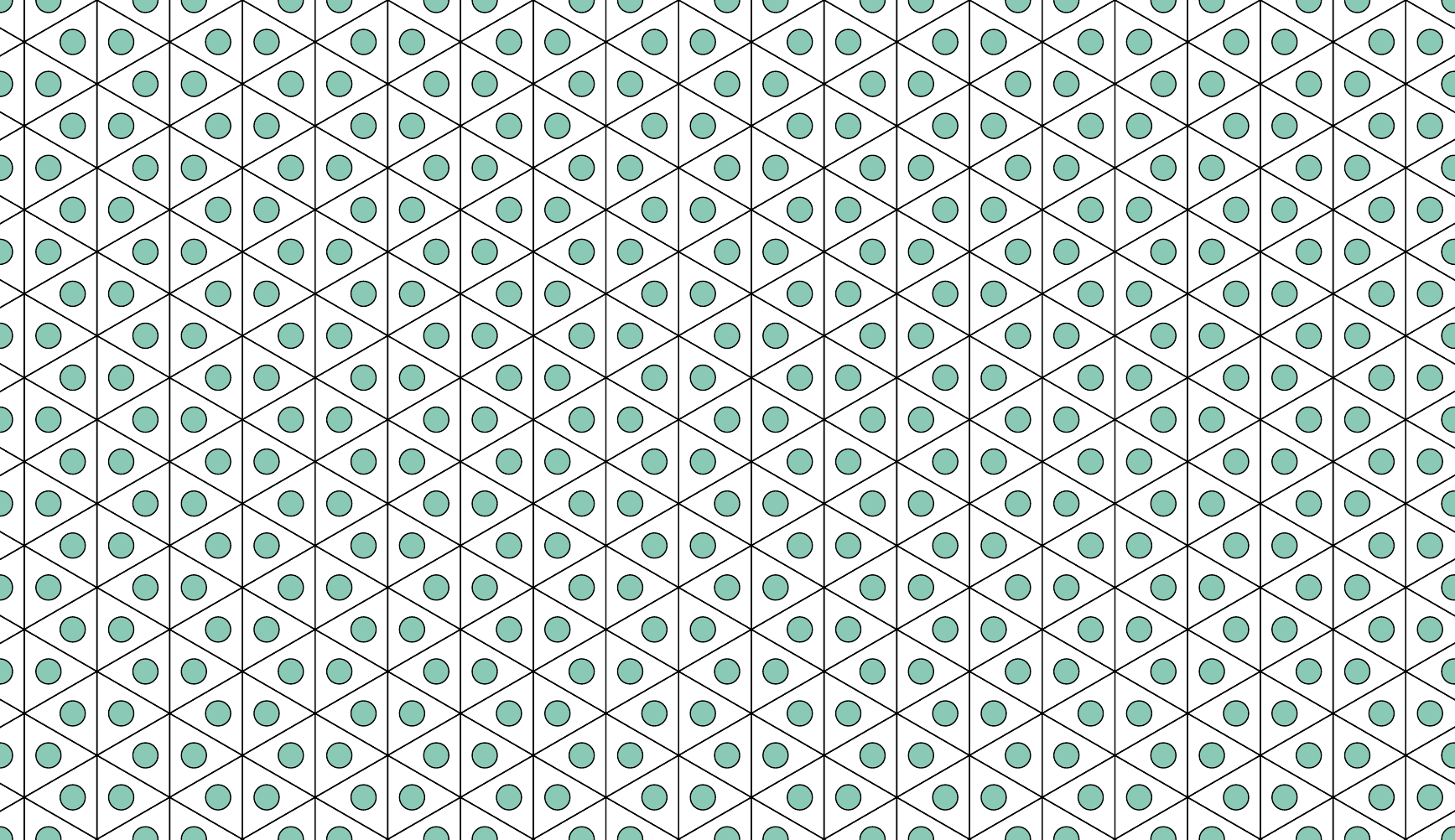}{(d)}{0.22\textwidth}\hfill
\edgekey{0.22\textwidth}{0.16\textwidth}
\caption{Two-dimensional crystals and their Voronoi
  tessellations: (a)~triangular, (b)~square, (c)~Kagome, and
  (d)~graphene lattices.  The Voronoi cell of each particle in these
  ideal crystals is an identical polygon; particles are colored by
  the number of edges of their Voronoi cell (key at right).}
\label{fig:2d_lattices}
\end{figure}

Four lattices serve as running examples
(Figure~\ref{fig:2d_lattices}): the triangular, square, Kagome,
and graphene lattices, whose Voronoi cells are respectively regular
hexagons, squares, rhombi, and equilateral triangles.  The Voronoi
tessellation of each of these ideal crystals is built from a single
repeating polygon, yet under small perturbations of the particle
positions the four crystals respond in markedly different ways.

\subsection{Voronoi cell topology and the question of stability}
\label{sec:2d_stability}

The \emph{Voronoi cell} of a particle is the region of the plane closer to it
than to any other.  Each of these regions is a convex polygon, and together
these polygons partition the plane into the \emph{Voronoi
tessellation}~\cite{voronoi1908nouvelles,1992okabe,lazar2022voronoi}. A simple
way to characterize each particle in a configuration is to record the number of
edges of its Voronoi cell. If a finer characterization is needed, then we also
record the edge counts of its immediate
neighbors~\cite{lazar2024characterizing}.

\begin{figure}[b]
\centering
\subfigframed{240 195 240 195}{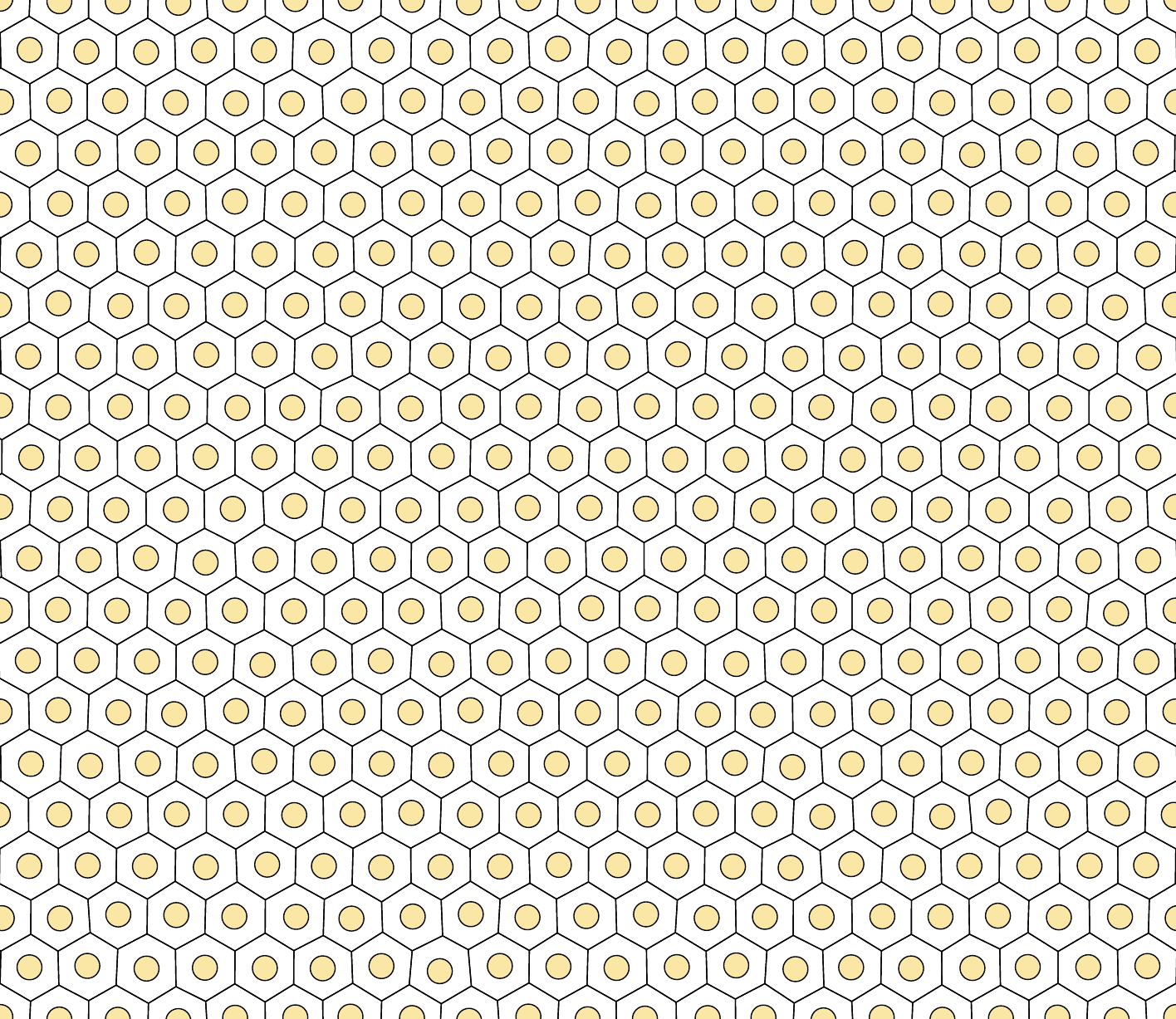}{(a)}{0.22\textwidth}\hfill
\subfigframed{58 58 58 58}{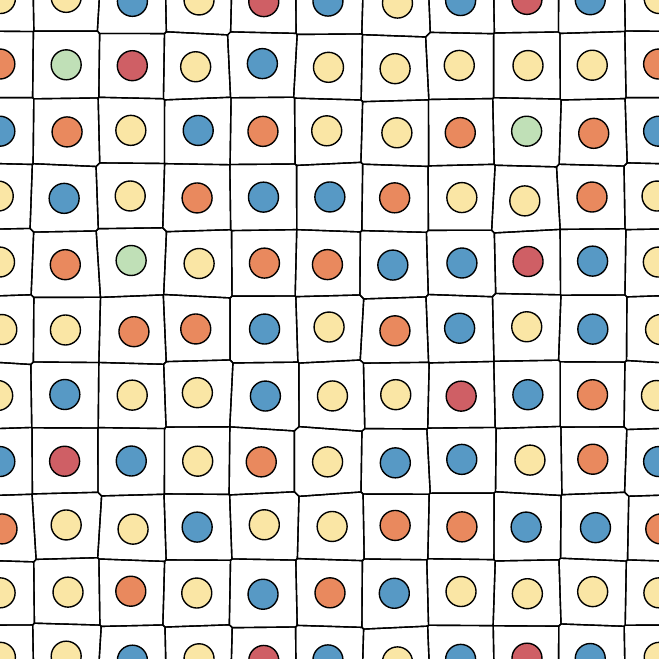}{(b)}{0.22\textwidth}\hfill
\subfigframed{195 410 195 410}{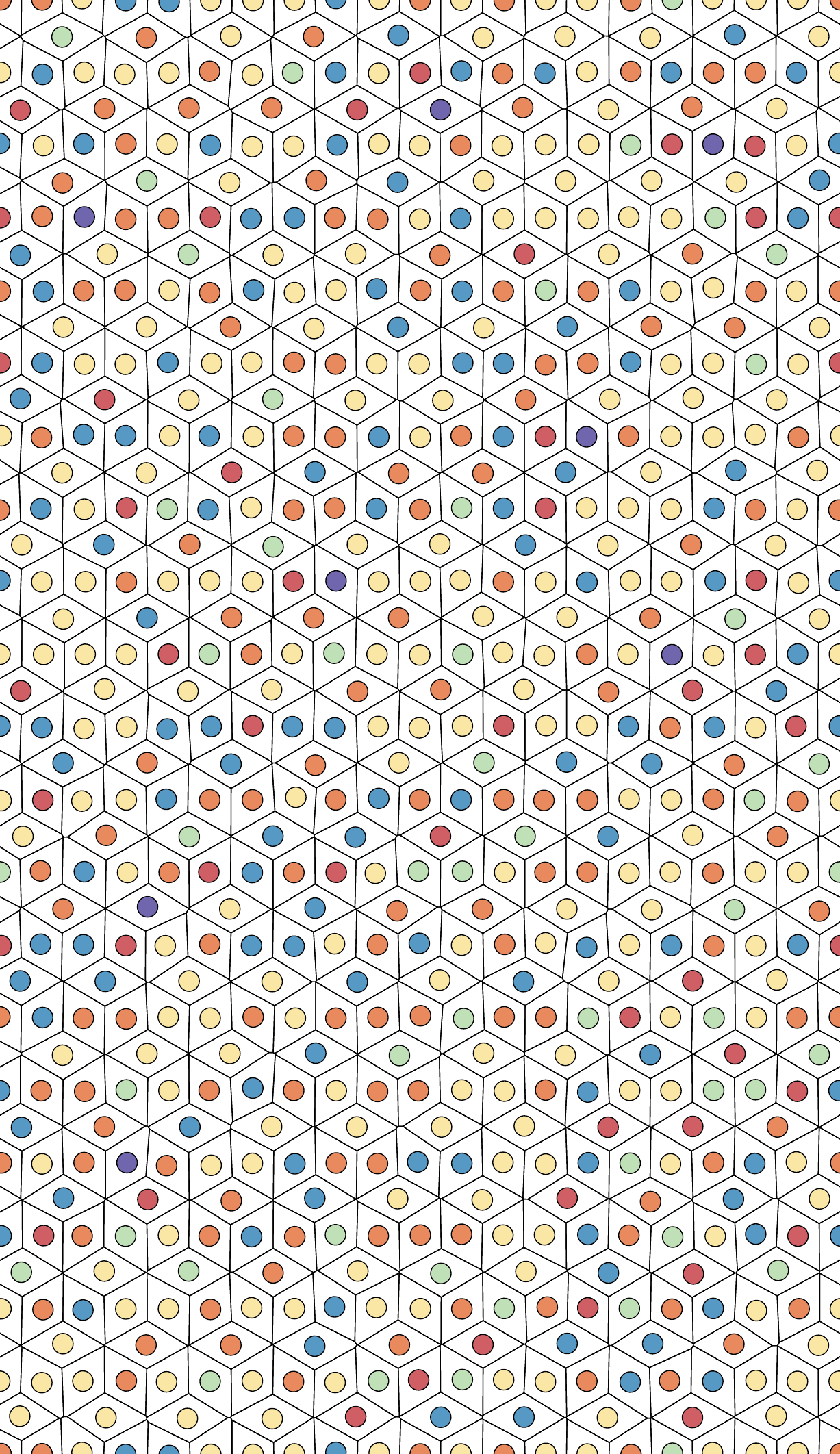}{(c)}{0.22\textwidth}\hfill
\subfigframed{312 140 320 140}{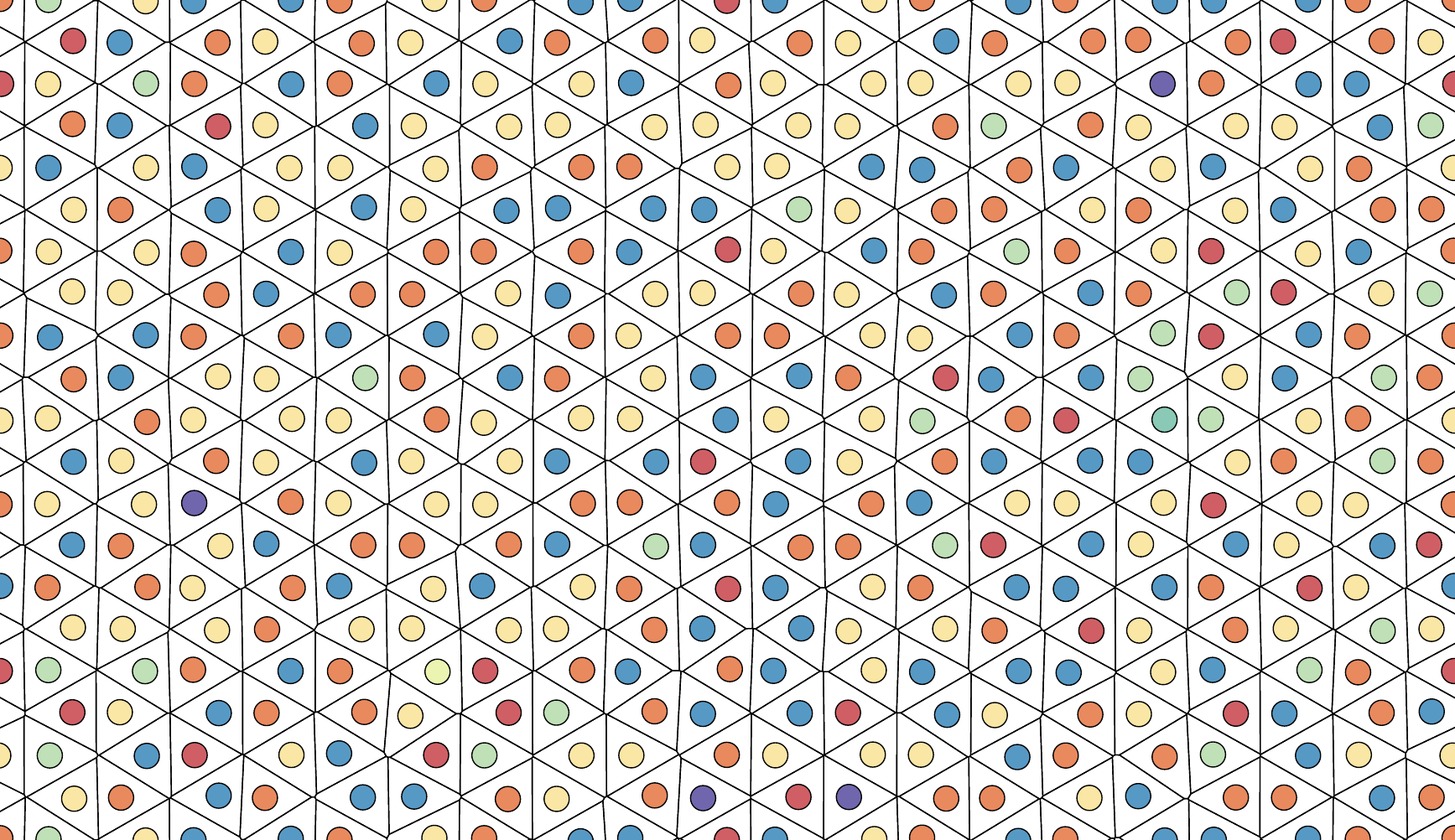}{(d)}{0.22\textwidth}\hfill
\edgekey{0.22\textwidth}{0.16\textwidth}
\caption{The same four lattices after a small generic perturbation of particle
  positions. Particles are colored by the number of edges of their Voronoi
  cell (key at right). The triangular lattice~(a) is topologically \emph{stable} in that every cell
  remains a hexagon.  The square~(b), Kagome~(c), and graphene~(d) lattices
  each develop a diversity of cell types as their degenerate Voronoi vertices
  resolve under perturbation.} \label{fig:2d_perturbed}
\end{figure}

In each of the four ideal crystals of
Figure~\ref{fig:2d_lattices} every particle has the same
characterization: six edges in the triangular lattice, four in the
square and Kagome lattices, three in graphene.  A small
perturbation of the particle positions breaks this uniformity.
Figure~\ref{fig:2d_perturbed} shows the same four crystals after
random displacements of the particle positions.  The triangular lattice is unchanged in a
combinatorial sense: even though the exact geometries of the Voronoi cells change,
they all remain hexagons.  %
The other three
crystals, however, resolve into a variety of shapes.  In particular, Voronoi
cells in the perturbed square lattice appear with between 4 and 8 edges;
those in the perturbed Kagome lattice can have between 4 and 10 edges;
and those in the perturbed graphene lattice can have between 3 and 12 edges.

This observation raises an important question: can we enumerate 
all the Voronoi cell types that can arise under
small perturbations of a given crystal?  Such perturbations
can represent thermal fluctuations and other small
displacements in physical systems.
Answering this question, in two
and then three dimensions, is the goal of this paper.  All 
diversity in Figure~\ref{fig:2d_perturbed} turns out to trace back
to a single feature of the unperturbed cells in
Figure~\ref{fig:2d_lattices}: the local geometry at their Voronoi
vertices, which we examine next.

\subsection{Generic and degenerate Voronoi vertices}
\label{sec:2d_vertices}

The behavior of a Voronoi cell under perturbation is controlled by what happens
at each of its vertices.  In two dimensions, a Voronoi vertex is \emph{generic}
if exactly three cells meet there or, equivalently, if exactly three particles
are equidistant from a common point. Generic vertices are combinatorially stable in that small
perturbations of the particle positions move them around continuously without
changing which three cells meet.  By contrast, in highly symmetric
configurations, such as the ideal crystals of
Figure~\ref{fig:2d_lattices}, more than three particles can be equidistant
from a common point, producing \emph{degenerate} Voronoi vertices.  We write
$k$ for the number of equidistant particles at a vertex, so $k = 3$ at a
generic vertex and $k > 3$ at a degenerate one.  The value of $k$ controls how
a degenerate vertex resolves under perturbation. A degenerate vertex is
combinatorially unstable: under any small perturbation the equidistant particles shift
slightly, and the single degenerate vertex \emph{resolves} into several generic
ones.  Different perturbation directions can yield different resolutions, with
different consequences for the cells nearby.

The four lattices of Figure~\ref{fig:2d_lattices} illustrate this
distinction.  The triangular lattice has only generic vertices,
and its cells are stable under perturbation, as illustrated
in Figure~\ref{fig:2d_perturbed}(a).  The square and graphene
lattices have only degenerate vertices, with $k = 4$ equidistant
particles in the square lattice and $k = 6$ in graphene.  The Kagome
lattice is intermediate: each of its rhombic Voronoi cells has two
generic vertices and two degenerate ones with $k = 6$.  Higher $k$ means more available
resolutions, and thus a richer spectrum of possible outcomes
under perturbation:
each $k = 4$ vertex of the square lattice admits just two
resolutions, while the $k = 6$ vertices of Kagome and graphene
admit many more.

The multiplicity $k$ of each Voronoi vertex therefore determines
both whether the vertex is stable and the variety of resolutions 
that small perturbations can produce.  Understanding these resolutions, 
and combining their contributions across all the vertices of a single 
two-dimensional Voronoi cell, is the subject of the rest of this section.

\subsection{How a degenerate vertex resolves}
\label{sec:2d_resolution}

A degenerate vertex with $k > 3$ equidistant particles is unstable for
a simple reason: under any generic perturbation of the
particle positions, four or more particles cannot remain exactly
equidistant from a single point.  The $k$ particles shift slightly,
no longer all equidistant from any single point, and the single
degenerate vertex splits into several three-fold generic vertices.

Different perturbation directions can produce different splittings.
Consider the simplest case, a $k = 4$ vertex of the square
lattice, where four particles sit at the corners of a unit square
(Figure~\ref{fig:2d_lattices}b).  Under a generic perturbation the
four particles no longer lie on a common circle.  
Take the circle through any three of the particles: the fourth particle now
lies either inside or outside that circle.  If it lies inside, then that particle and the one opposite it become neighbors, and their Voronoi cells meet along a new edge.  If the fourth particle lies outside that circle, then a new edge instead forms between the other pair of Voronoi cells.  Which happens depends
on the perturbation direction, so there are exactly two possibilities,
with visibly different consequences for the surrounding cells.  Under perturbation in a random direction, the two outcomes are
realized with equal probability \cite{leipold2016statistical}.

At a degenerate vertex with higher $k$ the perturbation direction
again selects the outcome, but from more possibilities.  In two dimensions the geometric
setup is the same at any $k$: of the $k$ particles equidistant at
the vertex, one is the central particle and two are existing
edge-sharing neighbors of it (those whose shared edges with the
central cell meet at the vertex); the remaining $k - 3$ touch the
central cell only at the vertex itself.  This separation underlies
the two organizing terms we use throughout the paper.  We call the
existing edge-sharing neighbors of the central particle its
\emph{full neighbors}: they share an edge with the central cell in
the unperturbed crystal and continue to share one under any
sufficiently small perturbation.  The $k - 3$ particles touching
the cell only at the vertex itself we call \emph{partial
neighbors}: each is conditional, becoming a new edge-sharing
neighbor in some resolutions of the vertex and not in others.
A resolution can promote any subset of the partial neighbors at a
vertex to new edge-sharing neighbors of the central particle,
adding between $0$ and $k - 3$ new edges to its cell.

The four lattices of Figure~\ref{fig:2d_lattices} sort cleanly along
this distinction:

\begin{center}
\begin{tabular}{lcccc}
\toprule
                  & Triangular & Square & Kagome & Graphene \\
\midrule
Full neighbors    & 6 & 4 & 4 & 3 \\
Partial neighbors & 0 & 4 & 6 & 9 \\
\bottomrule
\end{tabular}
\end{center}

\noindent
The triangular lattice has no partial neighbors and is thus topologically
stable.  The square lattice has 4 partial neighbors, so its perturbed cells 
range from 4 edges, the result of no partials being realized, to 8 edges, 
when all 4 partials are realized; this spread is visible in
Figure~\ref{fig:2d_perturbed}(b).  Kagome and graphene have $k = 6$
at each degenerate vertex, contributing 3 partial neighbors per
vertex.  With 2 degenerate vertices per Kagome cell and 3 per
graphene cell, this gives 6 and 9 partial neighbors respectively,
matching the perturbed cell-size ranges of
Section~\ref{sec:2d_stability} (4 to 10 in Kagome, 3 to 12 in
graphene).  The diversity of perturbed cell types in
Figure~\ref{fig:2d_perturbed}(b)--(d) arises from which subset of
partial neighbors each cell realizes.

The resolution at each degenerate vertex is thus a finite,
discrete choice.  The next subsection examines how these choices,
made across all the degenerate vertices of a single Voronoi cell,
combine to determine the full spectrum of cell topologies under
perturbation.

\subsection{Combining resolutions: enumerating the family}
\label{sec:2d_family}

The combinatorial type of a perturbed cell is determined by what
happens at its vertices.
Generic vertices contribute fixed local configurations regardless
of the perturbation.  In contrast, each degenerate vertex contributes 
any one of its multiple possible resolutions, selected by the perturbation direction.  The
full spectrum of cell topologies under generic perturbation is
enumerated by taking one resolution at each degenerate vertex 
of a cell, combining the choices across all the vertices,
and asking what cell topology results.

For the square lattice this counting is straightforward.  Each Voronoi cell
has four $k = 4$ corners with two resolutions apiece---each
partial neighbor at the corner is realized or not---giving
$2^4 = 16$ combinations.  The resulting cell has between 4 and 8
edges, depending on how many of the 4 partial neighbors are
realized; the 16 combinations distribute as $1, 4, 6, 4, 1$ across
these five edge counts.  Many of these combinations are related by the rotations
and reflections of the square lattice that fix the central particle,
and produce identical cell topologies.  After factoring out these
symmetry equivalences, the square lattice's family contains five
distinct cell types, one for each edge count between 4
and 8 \cite{lazar2020voronoi}.  The 6-edge count arises from two
geometrically distinct resolution patterns, with new edges at
opposite or at adjacent corners of the original square, but the
resulting cells are both hexagons, indistinguishable by edge count
alone.  This coincidence, in which symmetry-distinct resolution
patterns produce the same cell type, recurs in three dimensions
(Section~\ref{sec:symmetry}).

The graphene and Kagome lattices work similarly.  
Their degenerate vertices have higher
multiplicity ($k = 6$) and admit more resolutions apiece, so the
raw combination count is far larger than the square lattice's.
The degenerate vertices still resolve independently---the choice
made at one vertex places no constraint on its neighbors---so the raw
count is simply the product of the per-vertex possibilities.  As before, the family of
distinct cell types is obtained from this raw count by factoring
out the central particle's site symmetry and identifying
coincident cell topologies.

These effects---the raw combination count, which grows
with the multiplicity $k$, and the equivalences imposed by the
site symmetry of the central particle---together determine the size of
each lattice's family of cell types.  The result is the finite, lattice-specific
spectrum visible in Figure~\ref{fig:2d_perturbed}(b)--(d).

\subsection{Crystals with multiple inequivalent sites}
\label{sec:2d_wyckoff}

\begin{figure}[t]
\centering
\subfigframed{448 448 448 448}{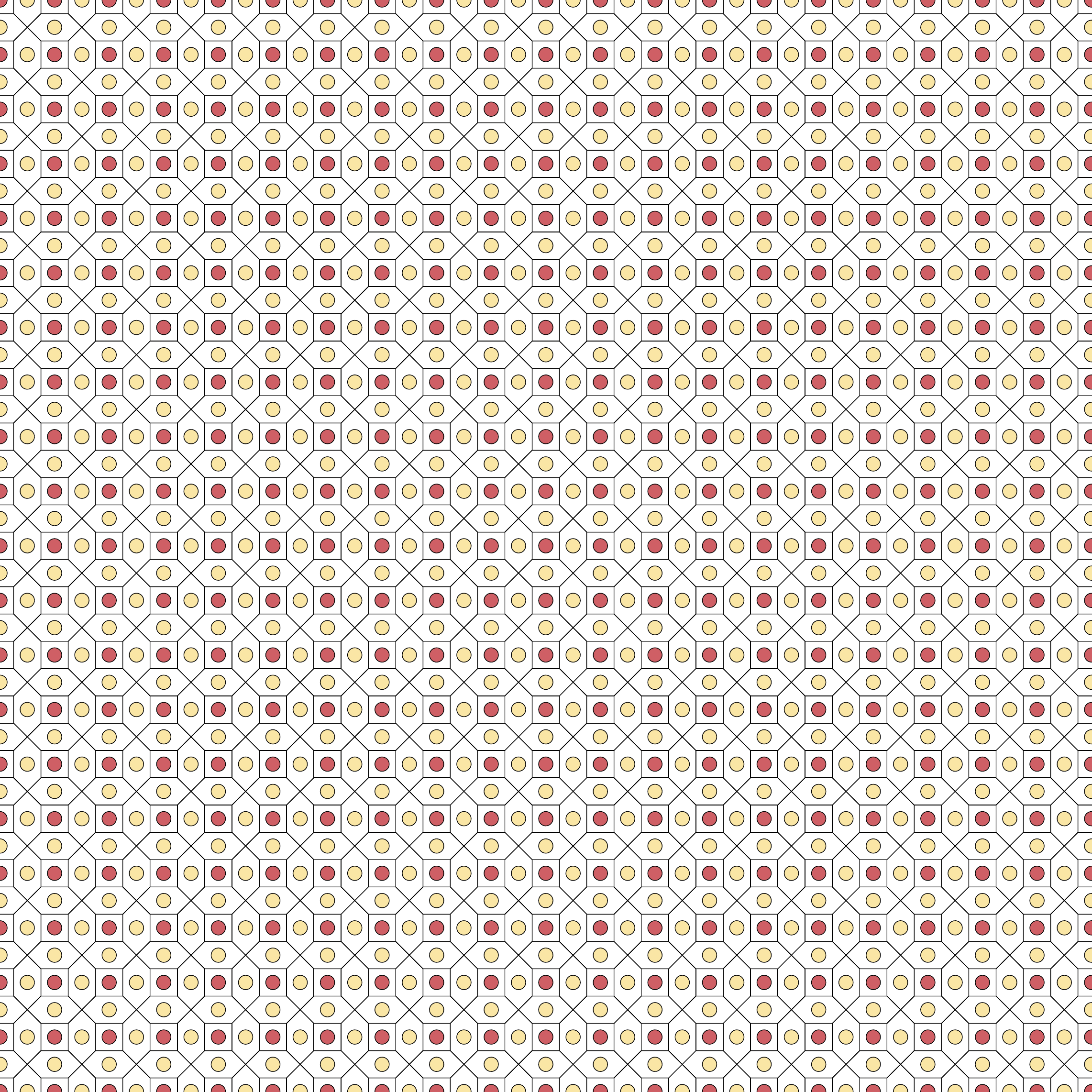}{(a)}{0.3\textwidth}\hspace{2em}\subfigframed{448 448 448 448}{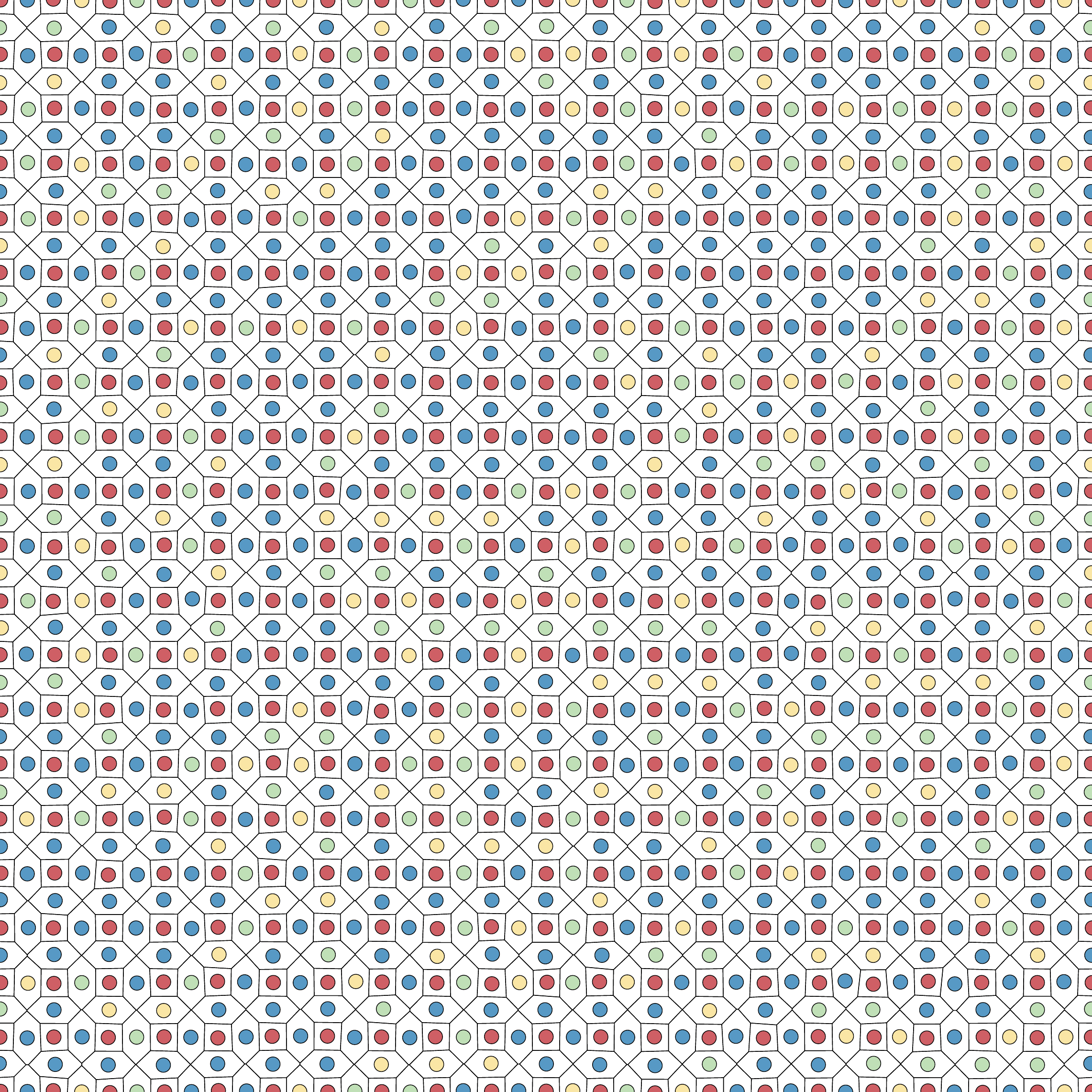}{(b)}{0.3\textwidth}%
\makebox[0pt][l]{\hspace{1.5em}\edgekey{0.3\textwidth}{0.2\textwidth}}
\caption{Lieb lattice, with two inequivalent Wyckoff sites, before
  (a) and after (b) perturbation.  Particles are colored by Voronoi
  cell edge count (key at right).  The corner sites (red squares) are topologically
  stable; the edge-midpoint sites (yellow hexagons) develop a
  spectrum of cell types.}
\label{fig:lieb}
\end{figure}

In all four examples of this section, the particles in the unit
cell are related by the crystal's symmetries. Every
particle in the crystal therefore has the same local environment, and
the perturbation-accessible family of cell topologies is the same
for every particle.  In some crystals, however, the unit cell contains
particles that are \emph{not} all related by symmetry, and different
particles occupy different \emph{Wyckoff sites} with distinct local
geometries.  The Lieb lattice \cite{lieb1989two}, illustrated in Figure~\ref{fig:lieb}, is an example.
In that lattice, each square unit cell has three particles, one at a corner and two at edge midpoints.  
The two Wyckoff sites have visibly different Voronoi cells, and they also behave differently
under perturbation: the corner-particle cells have only generic Voronoi
vertices and are topologically stable, while the edge-midpoint
cells contain degenerate vertices and develop a spectrum of cell
types.  Each Wyckoff site thus has its own
family of perturbation-accessible cell topologies, and a particle in
a simulation is classified according to which site its local cell
type belongs to.  Multiple inequivalent sites are common in three
dimensions: both structures analyzed in
Section~\ref{sec:applications}---sigma-phase CrFe and
$\omega$-Ti---contain them.  In the first, all five sites are
topologically stable; in the second, both sites carry degenerate
vertices and produce multiple cell types.

\subsection{From two dimensions to three}
\label{sec:2d_to_3d}

Most of what we developed in this section carries over directly to
three dimensions.  Voronoi cells become polyhedra rather than
polygons, but the central concepts are the same.  Each cell has generic and
degenerate \emph{vertices}, and under perturbation each degenerate
vertex resolves into several generic ones.  The family of
cell topologies a crystal can produce is then enumerated by combining
the resolution choices across all the degenerate vertices of a
single cell, up to the equivalences imposed by site symmetry.  Crystals with
multiple inequivalent Wyckoff sites carry a separate family for each site.
The one quantitative difference is that a generic vertex has $k = 3$
equidistant particles in two dimensions and $k = 4$ in three,
where four points generically determine a sphere
(Section~\ref{sec:3d_vertices}).

Two new features appear in three dimensions.
The first is the appearance of \emph{zero-volume tetrahedra}: the local
configuration of $k$ particles around a degenerate vertex can include
four-particle subsets that are exactly coplanar in the unperturbed crystal
(the FCC octahedral vertex has three such coplanar squares).  Such a subset
is flat in the ideal crystal but becomes a tetrahedron under any perturbation
that breaks the coplanarity, contributing an additional class of
resolutions, some of which persist as the perturbation vanishes and some
of which do not (Section~\ref{sec:resolution}).  Two
dimensions has no counterpart: a line meets a circle in at most two
points, so three particles equidistant from a vertex are never
collinear, whereas a plane meets a sphere in a whole circle, on which
four or more equidistant particles can lie.
The second is \emph{coupling} between degenerate vertices.  In three
dimensions a Voronoi edge can itself be degenerate, with four or more
cells meeting along it, and the degenerate vertices at its two ends then
constrain each other's resolutions: not every combination corresponds
to an actual Voronoi cell (Section~\ref{sec:coupling}).  Again, two
dimensions has no counterpart, since exactly two cells meet along every
edge, so resolutions at different vertices never constrain one another
and every combination produces a cell.  Both effects are developed in the sections that follow.

\section{Voronoi cell types in three dimensions}
\label{sec:3d}
\subsection{Voronoi cells and vertices in three dimensions}
\label{sec:3d_vertices}

In three dimensions, the \emph{Voronoi cell} of a particle is again the region
of space closer to it than to any other, and these convex polyhedra partition
space into the \emph{Voronoi tessellation}.  Most of the
conceptual structure we developed in the two-dimensional examples of
Section~\ref{sec:2d} transfers directly to three dimensions.  Each Voronoi cell
has faces, edges, and vertices, and each Voronoi vertex is a point where
multiple cells meet.  The number of equidistant particles at a vertex---the
multiplicity $k$---determines whether the vertex is generic or degenerate,
exactly as in two dimensions, but the threshold shifts.  In three dimensions,
four points generically determine a sphere, so a Voronoi vertex with $k = 4$
equidistant particles is generic and a vertex with $k > 4$ is degenerate.
Likewise, a Voronoi edge is generically shared by exactly three cells;
an edge shared by four or more cells is degenerate.  Such edges appear in
Section~\ref{sec:resolution}, in the criterion for primary resolutions,
and in Section~\ref{sec:coupling}, where they couple the
resolutions at their two ends.

The two-dimensional running examples of Section~\ref{sec:2d} have
natural three-dimensional analogues, and the distinction between full and partial neighbors
transfers with edge-sharing replaced by face-sharing.
The body-centered cubic (BCC) lattice has only generic Voronoi
vertices, and its cell is the truncated octahedron, with
14 full neighbors and no partials.  It is therefore topologically
stable, every cell remaining a truncated octahedron under any
sufficiently small perturbation.  The face-centered cubic (FCC) and
hexagonal close-packed (HCP) lattices each have a mix of generic and
degenerate Voronoi vertices, with 8 generic and 6 degenerate ($k = 6$)
per cell.  The central particle has 12 full neighbors (the 12
face-sharing neighbors of the unperturbed cell) and
6 partial neighbors, one at each degenerate vertex.  In both lattices,
full neighbors are the nearest neighbors of the central particle and
partial neighbors are the second-nearest.  Both can be seen as three-dimensional
analogues of the Kagome lattice, in that their Voronoi cells have both generic and degenerate vertices.  
The simple cubic lattice, in which every
Voronoi vertex is degenerate with $k = 8$ particles equidistant,
sits at a more unstable extreme than any of our two-dimensional
examples: under perturbation each $k = 8$ vertex can split in many
distinct ways (Section~\ref{sec:resolution}), and the resulting
family of cell topologies is enormous.
In structures with several Wyckoff sites the same distinction applies
site by site (Section~\ref{sec:2d_wyckoff}).
Figure~\ref{fig:voronoi_cells} shows two such structures: the Voronoi
cells at the two sites of A15 have only generic vertices and are
topologically stable, while those at the two sites of $\omega$-Ti have
degenerate vertices and produce multiple cell types.

\newcommand{\vorocell}[2]{%
  \begin{minipage}[t]{0.235\textwidth}\centering
    \begin{tikzpicture}[inner sep=0pt, baseline=(img.north)]
      \node (img) {\includegraphics[width=\linewidth, trim=7 7 7 7, clip]{#1}};
      \node[anchor=north west, fill=white, draw=black, line width=0.5pt,
            inner sep=1.5pt, font=\scriptsize]
           at ([xshift=0.1cm, yshift=-0.1cm] img.north west) {#2};
    \end{tikzpicture}%
  \end{minipage}}
\begin{figure}[!htb]
\centering
\begin{tabular}{@{}c@{\hspace{0.3em}}c@{\hspace{1.2em}}c@{\hspace{0.3em}}c@{}}
\multicolumn{2}{c}{\textbf{A15}} & \multicolumn{2}{c}{\textbf{$\omega$-Ti}} \\[0.2em]
\vorocell{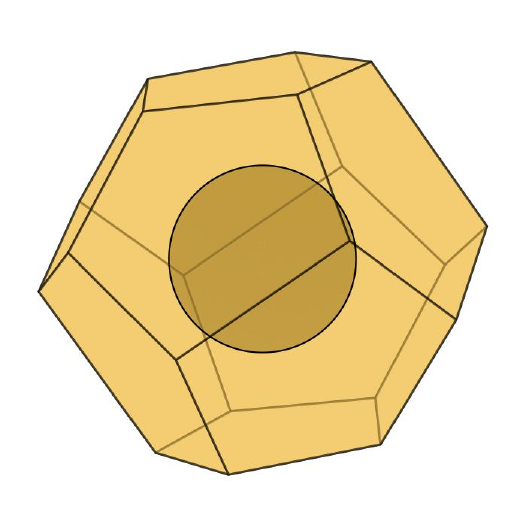}{(a)~2a} &
\vorocell{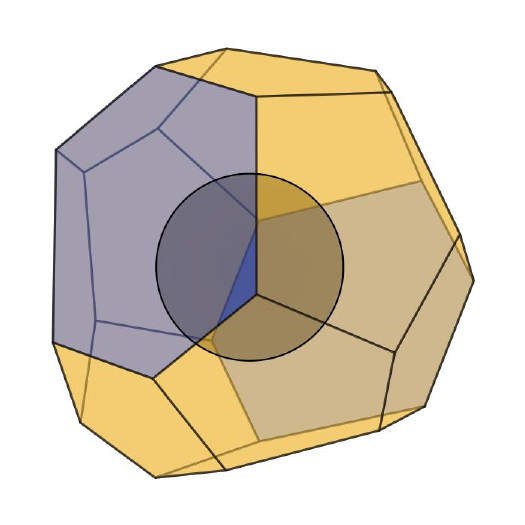}{(b)~6c} &
\vorocell{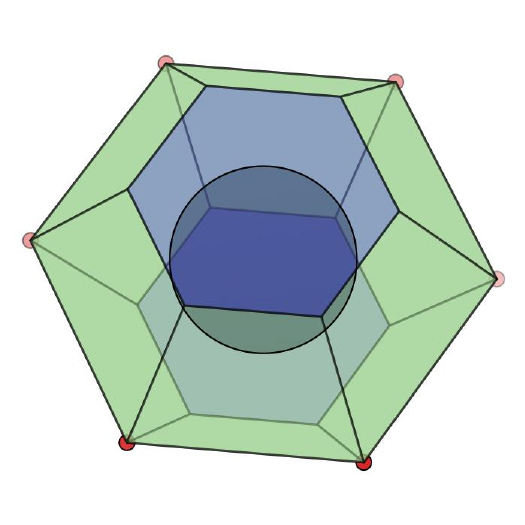}{(c)~1a} &
\vorocell{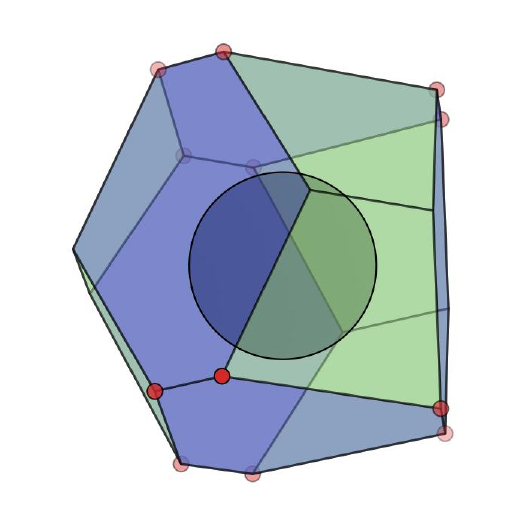}{(d)~2d} \\
\end{tabular}
\caption{Voronoi cells at the two Wyckoff sites of two
  representative multi-sublattice crystal structures:
  A15 (a,b) and $\omega$-Ti (c,d).  Faces are
  colored by number of edges: green is a quadrilateral, gold a
  pentagon, indigo a hexagon.  The dark central sphere marks the
  central particle; red dots mark degenerate Voronoi vertices.
  Topologically stable cells
  (no red dots: A15 2a/6c) admit a single Weinberg
  vector per Wyckoff site; cells with degenerate vertices
  ($\omega$-Ti 1a/2d) generate multiple cell types under
  perturbation.}
\label{fig:voronoi_cells}
\end{figure}

\subsection{The Weinberg vector}
\label{sec:weinberg}

In two dimensions we used the number of edges
of a particle's Voronoi cell to characterize its combinatorial type.
A three-dimensional Voronoi cell is a polyhedron with faces, edges,
and vertices, and a single integer no longer suffices: two polyhedra
can have the same number of faces but very different combinatorial
structures.  The natural three-dimensional label is the
\emph{Weinberg vector}~\cite{1966weinberg}, a finite sequence of
integers that uniquely identifies the combinatorial type of a
convex polyhedron.  Two polyhedra share the same Weinberg vector
if and only if their faces, edges, and vertices are connected to
one another in the same way.  Weinberg's algorithm computes the
vector by a deterministic traversal of the cell's surface, and the
vector has been used extensively in prior work to classify Voronoi
cells in particle
configurations~\cite{2012lazar,lazar2015topological,yoon2019topological}. We use it throughout the rest of this paper as the discrete output of
our enumeration: each crystal yields a finite list of Weinberg
vectors, one for each cell topology that can arise under
perturbation.

A coarser invariant that we will also use is the \emph{$p$-vector},
which records the number of triangular, quadrilateral, pentagonal,
and so on faces of the cell---the Voronoi index of
Section~\ref{sec:intro}, extended to faces of every size \cite{2012lazar}.  
The FCC Voronoi cell, for example,
has $p$-vector $(0, 12, 0, 0, \ldots)$: twelve quadrilateral faces
and no others (the rhombic dodecahedron).  The $p$-vector is
easier to compute and read than the full Weinberg vector but is
not a complete invariant: distinct Weinberg types can share the
same $p$-vector.  FCC and HCP are the canonical example
(Section~\ref{sec:intro}): their Voronoi cells share a $p$-vector
but have different Weinberg vectors.

\subsection{Resolution of a degenerate vertex in three dimensions}
\label{sec:resolution}

\begin{figure}
\centering
\begin{tikzpicture}
\node[anchor=south west,inner sep=0] (img) at (0,0){\includegraphics[width=\textwidth]{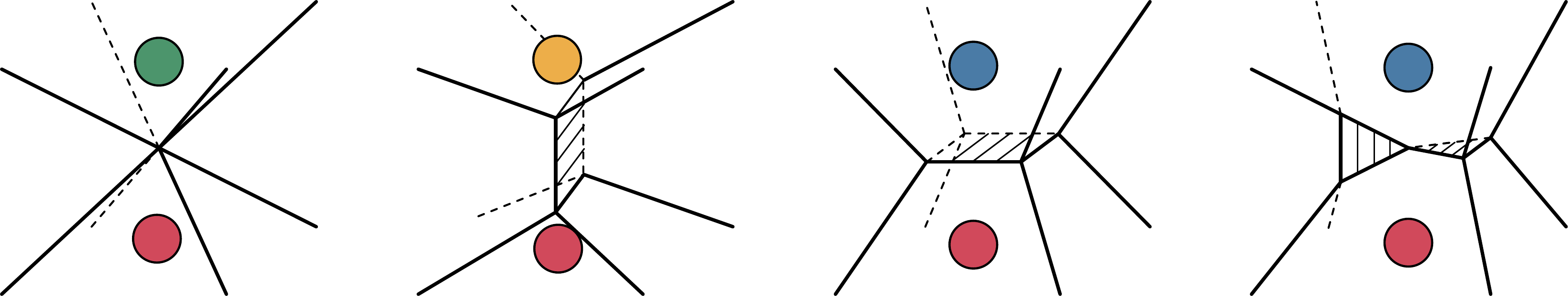}};
\begin{scope}[x={(img.south east)},y={(img.north west)}]
 \node at (0.101,-0.07){(a)};
 \node at (0.355,-0.07){(b)};
 \node at (0.621,-0.07){(c)};
 \node at (0.898,-0.07){(d)};
\end{scope}
\end{tikzpicture}
\caption{A degenerate octahedral Voronoi vertex of an FCC cell can resolve in
  several ways under a small perturbation.  The central particle (red) sits at
  the bottom of each panel, opposite the particle at top; the four equatorial
  particles, always full neighbors, are omitted.  (a)~In the unperturbed crystal
  the opposite particle is a \emph{partial neighbor} (green), touching the
  central cell only at the vertex.  The perturbation makes it (b)~a non-neighbor
  (yellow), as an edge forms and no face is shared, or a full neighbor (blue),
  sharing either (c)~a quadrilateral or (d)~a triangular face.}
\label{fig:fcc_resolution}
\end{figure}

The picture in three dimensions follows the same pattern as in
Section~\ref{sec:2d_resolution}, with the distinction between full and partial neighbors
carrying over.  At a degenerate vertex with $k > 4$
equidistant particles, the central particle is one of the $k$.
At least three of the others are \emph{full neighbors} (those
whose existing face with the central cell meets at the vertex),
and the remainder are \emph{partial neighbors} (which share no face
with the central cell, meeting it only along an edge or at the vertex).  A degenerate vertex is unstable: under any generic
perturbation of the particle positions, more than four particles
cannot remain exactly equidistant from a single point.  The $k$
particles shift slightly, the vertex splits into multiple generic
four-fold vertices, and which partial neighbors at the vertex
become new face-sharing neighbors of the central particle depends
on the perturbation direction.  A \emph{resolution type} is a combinatorially distinct way of
resolving the vertex.  Two resolutions can realize the same neighbors
yet differ in the shapes of the new faces and how they fit
together---as the quadrilateral and triangular faces of
Figure~\ref{fig:fcc_resolution}(c,d) do---and are therefore different
resolution types.

The $k$ equidistant particles at a 3D degenerate vertex lie on a
sphere centered at the vertex, and their
arrangement determines the possible resolution types.  For the
main structures we examine, the configurations are familiar: the
six particles at an FCC or HCP degenerate vertex form a regular
octahedron, and the eight particles at a simple-cubic vertex form a
cube.  The formal machinery for enumerating resolution
types---the Delaunay--Voronoi duality, the star of the central
particle, and the regular triangulations of the equidistant point
set---is collected in
Appendix~\ref{sec:algorithm}.

One feature is genuinely new in three dimensions.  Each resolution of a
degenerate vertex corresponds to a triangulation of its $k$ equidistant
particles into tetrahedra.  Some of these triangulations involve four-particle
subsets that are exactly coplanar in the unperturbed crystal---the FCC
octahedral vertex has three such squares---and a tetrahedron built on one is a
\emph{zero-volume tetrahedron}: it vanishes in the unperturbed crystal but
acquires positive volume under any perturbation that breaks the coplanarity.
We write $\sigma$ for the typical magnitude of the particle displacements
producing the perturbation---for instance, the root-mean-square thermal
displacement at finite temperature.  We call a resolution or a cell type
\emph{primary} if its frequency remains positive as $\sigma \to 0$
and \emph{secondary} if it vanishes in that limit.  Equivalently, in
the configuration space of neighbor arrangements, the region
occupied by a primary type meets the ideal configuration at a
positive solid angle, while that of a secondary type meets it at a
cusp~\cite{lazar2015topological}.

Which resolution results depends on the direction of the
perturbation, not on its magnitude $\sigma$.  A primary resolution
is produced by a positive fraction of directions at every $\sigma$,
so its frequency persists as $\sigma \to 0$.  Resolutions whose
tetrahedra all have positive volume are of this kind.  Resolutions
that contain a zero-volume tetrahedron may be primary or secondary,
depending on the position of the tetrahedron's four coplanar particles.  
When other equidistant particles lie on both sides
of their plane, as for the equatorial squares of the FCC octahedron,
the tetrahedron survives the perturbation only if the center of its
circumsphere stays close to the unperturbed vertex, an event of probability
proportional to $\sigma$. Such
a resolution is secondary, present at any finite $\sigma$ but
vanishing in the limit.  When all other equidistant particles lie on
one side, the four span a face of the equidistant set's convex hull,
and in the Voronoi tessellation they share a \emph{degenerate edge}
along which four cells meet instead of the usual three, as at the
base of the $\omega$-Ti square pyramid and on every face of the
cube of particles around a simple-cubic vertex.  After a perturbation, either one pair of
opposite cells around such an edge become neighbors along its whole
length, sharing a narrow face, or both pairs become neighbors, each
sharing a narrow face that tapers to a point, the two faces meeting tip
to tip at the Voronoi vertex of the zero-volume tetrahedron.  Each case
occurs for a positive fraction of directions, so a
resolution is primary whenever each of its zero-volume tetrahedra is
of this kind.  A cell type is primary if at least one combination of
resolutions producing it uses only primary ones.  Supplementary
Section~\ref{app:primary_validation} gives the geometric argument
and confirms the rule by sampling.

At the
FCC $k = 6$ octahedral vertex, for example, the partial neighbor is the
second-nearest neighbor directly opposite the central particle on the
octahedron, with the 4 equatorial particles being full neighbors meeting at
this vertex. Of the 7 resolutions (Figure~\ref{fig:fcc_resolution}), 3 are primary: 2 in which the partial
neighbor is not realized as a new face-sharer (an edge between opposite
equatorial particles forms instead), and 1 in which it is realized as a
quadrilateral face.  The other 4 are secondary: in each, a coplanar
quadruple---the central particle, its partial neighbor, and an opposite pair of
equatorial particles---must stay nearly on a common circle, and the partial
neighbor is realized as a triangular face.

\begin{figure}
\centering
\includegraphics[width=0.134\textwidth]{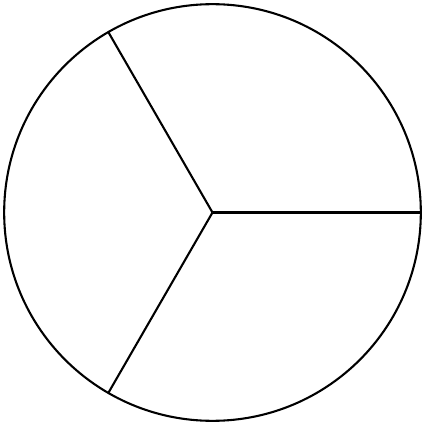}\hspace{0.01\textwidth}%
\includegraphics[width=0.134\textwidth]{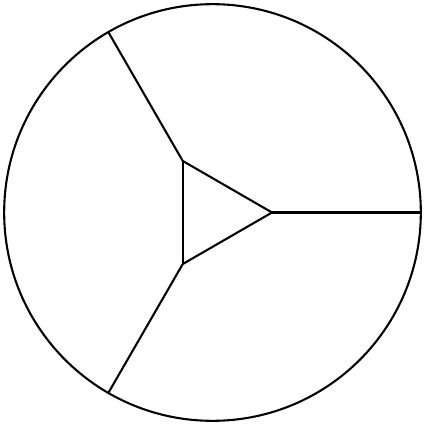}\hspace{0.01\textwidth}%
\includegraphics[width=0.134\textwidth]{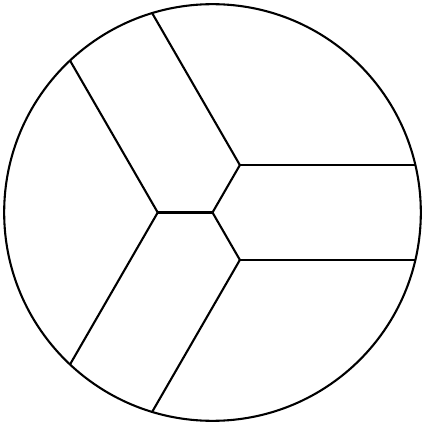}\hspace{0.01\textwidth}%
\includegraphics[width=0.134\textwidth]{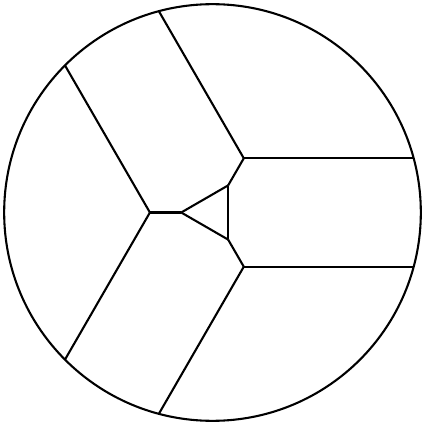}\hspace{0.01\textwidth}%
\includegraphics[width=0.134\textwidth]{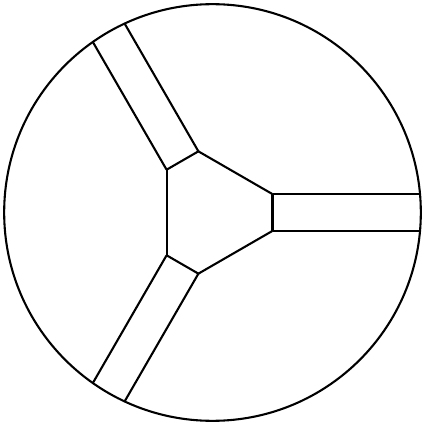}\hspace{0.01\textwidth}%
\includegraphics[width=0.134\textwidth]{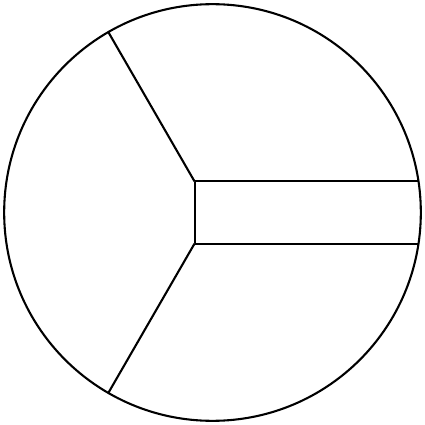}\hspace{0.01\textwidth}%
\includegraphics[width=0.134\textwidth]{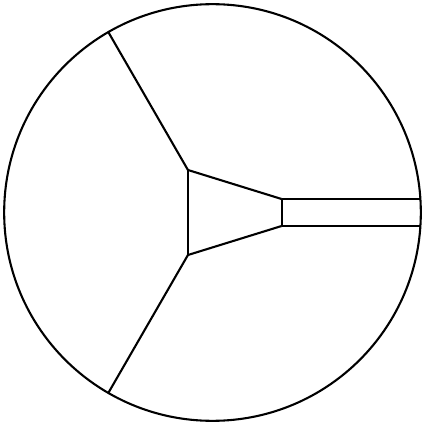}\\[0.6em]
\includegraphics[width=0.134\textwidth]{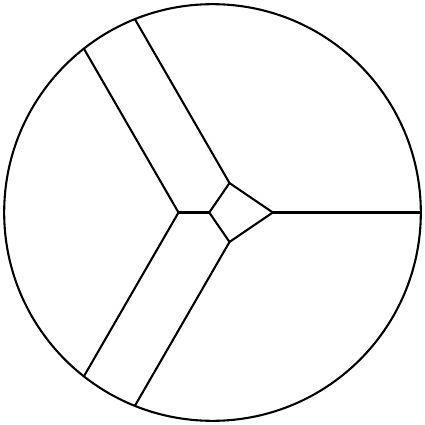}\hspace{0.01\textwidth}%
\includegraphics[width=0.134\textwidth]{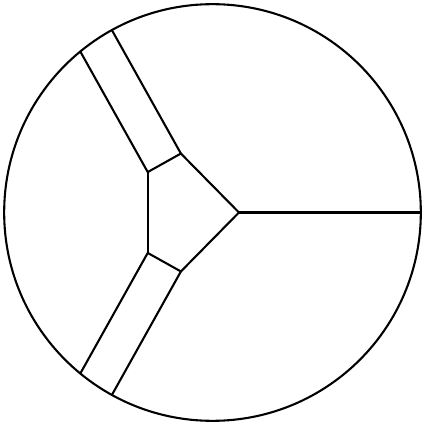}\hspace{0.01\textwidth}%
\includegraphics[width=0.134\textwidth]{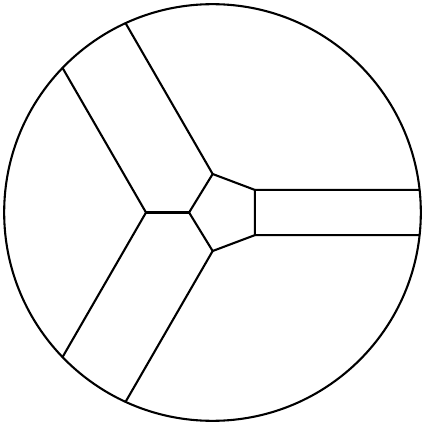}\hspace{0.01\textwidth}%
\includegraphics[width=0.134\textwidth]{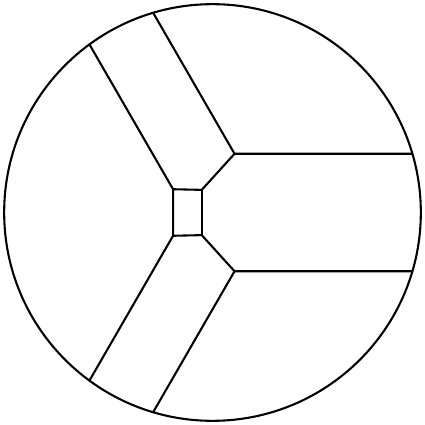}\hspace{0.01\textwidth}%
\includegraphics[width=0.134\textwidth]{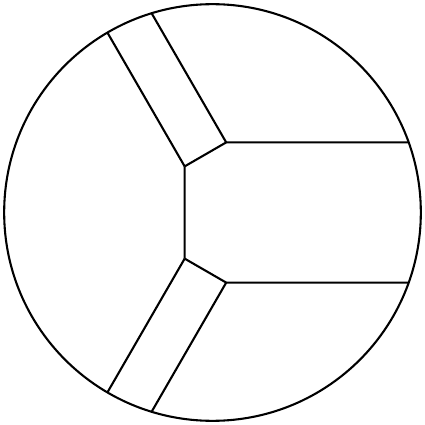}\hspace{0.01\textwidth}%
\includegraphics[width=0.134\textwidth]{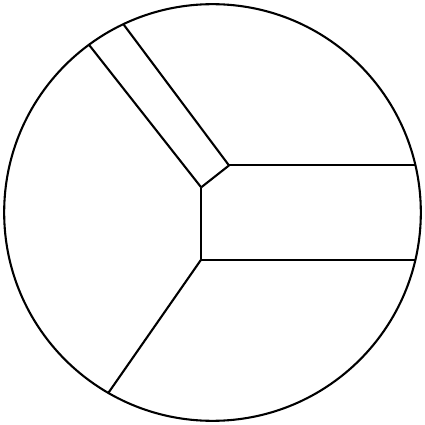}
\caption{The 13 symmetry classes of resolutions of the simple-cubic
  Voronoi vertex ($k = 8$) that involve no zero-volume tetrahedron,
  viewed from outside the central cell.  Each panel shows how the
  central particle's Voronoi cell terminates near the vertex once a
  small perturbation breaks the equidistance.  Five classes have the
  full symmetry of the vertex, with 3-fold rotations and three mirror
  planes, seven have a single mirror plane, and one is asymmetric;
  counted with multiplicity they give 32 resolutions.
  Primary resolutions that involve zero-volume tetrahedra look the same
  near the vertex and differ only farther along the edges; secondary
  resolutions are not shown.}
\label{fig:resolution}
\end{figure}

Figure~\ref{fig:resolution} illustrates the resolutions of the
simple-cubic $k = 8$ vertex.  The eight particles equidistant from
this vertex sit at the corners of a cube---a cube of particles,
distinct from the cubic Voronoi cell of the lattice.  Close to the vertex,
a perturbed cell takes one of 32 primary forms, which fall into the 13
symmetry classes shown.  Farther out, each of the three cell edges that
leave the vertex is degenerate and resolves in one of the two ways
described above, so the vertex has $32 \times 2^3 = 256$ primary
resolutions, out of 544 in all.

\subsection{Enumerating the family}
\label{sec:3d_family}

The family of perturbation-accessible cell topologies for a three-dimensional
crystal is enumerated by the same procedure as in two dimensions
(Section~\ref{sec:2d_family}), with two refinements specific to three
dimensions.  First, at each degenerate Voronoi vertex of the central cell, we
enumerate all distinct resolutions.  Next, we take all combinations of
resolutions across the cell's degenerate vertices, keeping only those whose
resolutions at adjacent vertices are geometrically consistent across their
shared Voronoi edge.  This check is applied during the enumeration itself, so
inconsistent combinations are pruned rather than generated and discarded
(Appendix~\ref{sec:algorithm}).  Then we apply the central particle's site
symmetry to collapse symmetry-equivalent combinations.  Finally, we label each
resulting cell by its Weinberg vector (Section~\ref{sec:weinberg}).  The output for each Wyckoff site of the
crystal is a finite list of Weinberg vectors---the family of cell topologies
accessible at that site under generic small perturbation.  A crystal with
multiple inequivalent Wyckoff sites (Section~\ref{sec:2d_wyckoff}) yields one
family per site.

As a concrete example, consider FCC.  The Voronoi cell of an FCC particle has
14 vertices: 8 generic and 6 degenerate, the latter all with $k = 6$ particles
equidistant in an octahedral configuration. Each degenerate vertex admits seven
distinct resolutions, so a raw enumeration would need to consider $7^6 =
117{,}649$ combinations.  The six degenerate vertices are pairwise non-adjacent
in the FCC cell, so every combination produces a valid cell.  The site symmetry
of the central particle---the rotations and reflections of the FCC lattice that
fix it---together with classification of the resulting cells by their Weinberg
vectors collapses these to $2{,}815$ distinct cell types, comprising the FCC
family in the \emph{default} enumeration mode (Section~\ref{sec:symmetry} details the
reduction; Supplementary Section~\ref{app:results} gives per-structure
breakdowns of cell counts and $p$-vectors for BCC, FCC, HCP, A15, fluorite, and
$\omega$-Ti).

The distinction between primary and secondary resolutions developed in
Section~\ref{sec:resolution} gives rise to three nested enumeration
modes.  The \emph{primary} mode includes only primary resolutions
at each vertex and captures the cell types realized in the
$\sigma \to 0$ limit.  The \emph{default} mode adds secondary
resolutions, capturing the cell types realized at any small
$\sigma > 0$.  The \emph{all} mode additionally permits any
degenerate vertex to remain unresolved, yielding so-called
\emph{partially resolved} cell types in which the central cell
retains a high-valence vertex.  The modes form a nested
inclusion \emph{primary}~$\subseteq$~\emph{default}~$\subseteq$~\emph{all}, with
substantially different counts: for FCC, for example, the three
modes give 44, 2{,}815, and 6{,}294 distinct cell types
respectively.

Which mode to use depends on the system under study.  For
finite-temperature MD simulations the \emph{default} mode is the
natural choice and is used throughout the applications of
Section~\ref{sec:applications}: thermal fluctuations of amplitude
$\sigma > 0$ resolve every degenerate Voronoi vertex into a generic
configuration, and the partially resolved cell types of the \emph{all}
mode are measure-zero events in this regime.  For systems quenched
to zero temperature, however, exact coincidences can persist in
the minimum-energy geometry---degenerate vertices that would
resolve under any nonzero perturbation but do not in the idealized structure---and the \emph{all} mode is
then the appropriate filter.  The \emph{primary} mode isolates the $\sigma \to 0$ limit and may
be useful when only leading-order topological behavior is of
interest, or when a more tractable family is needed for structures whose
\emph{default} or \emph{all}-mode families are large.

Implementation details of the algorithm---including the
backtracking enumeration of regular triangulations, the
combinatorial pruning that handles validity and adjacency
coupling, and the orderly generation that exploits site
symmetry---are collected in
Appendix~\ref{sec:algorithm}.

\section{Practical considerations}
\label{sec:practical}

This section discusses three issues that arise in practice:
coupling between adjacent degenerate vertices, near-equidistant
neighbors in noisy reference structures, and combinatorial limits
that make some structures intractable.  Before turning to
applications (Section~\ref{sec:applications}), it is useful to
understand both what the algorithm does cleanly and where its
limits lie.

\subsection{Coupling between adjacent degenerate vertices}
\label{sec:coupling}

In the two-dimensional examples of Section~\ref{sec:2d_family}, degenerate
vertices resolve independently of one another whether or not they are
adjacent.  In graphene, for instance, all three vertices of a Voronoi cell are
degenerate and mutually adjacent, yet every combination of their resolutions
yields a valid cell.  In three dimensions this independence can be lost.  Two
degenerate vertices joined by a generic Voronoi edge share only three
particles and remain independent.  When the edge joining them is itself
degenerate (Section~\ref{sec:3d_vertices}), however, the resolutions
chosen at the two must be compatible on the particles they share, and
nominal combinations can be geometrically inconsistent.  This coupling can dominate the family combinatorics, and the most
striking example in our library is the $\omega$ phase.

The $\omega$ phase (space group $P6/mmm$, three atoms per
hexagonal cell at ideal $c/a = \sqrt{3/8}$)~\cite{hickman1969omega}
has two inequivalent Wyckoff sites with strikingly different
combinatorial behavior, visible in their Voronoi cells
(Figure~\ref{fig:voronoi_cells}(c,d)).  At the \textbf{1a site}, the Voronoi cell
has 18 vertices---12 generic and 6 degenerate (with $k = 5$
particles equidistant), each degenerate vertex admitting 2
resolutions.  Some of the 6 degenerate vertices are adjacent, but the
edges joining them are generic, so every pair of their resolutions is
compatible and all $2^6 = 64$ combinations produce valid cells, giving
8 distinct cell types.

The \textbf{2d site} contains the same number of vertices (18) but with a
very different distribution: only 6 are generic and 12 are degenerate ($k =
5$), each admitting 4 resolutions.  A naive count of combinations gives $4^{12}
= 16{,}777{,}216$.  But the 12 degenerate vertices of a 2d-site Voronoi cell
are the endpoints of six degenerate Voronoi edges, and the two vertices
at the ends of each such edge share four particles.  The resolutions
chosen at the two must be compatible on these particles, or the
candidate cell is malformed, and only 8 of their 16 combinations are.
Discarding combinations that violate these consistency
constraints, which we refer to as \emph{validity-pruning}, leaves only
$8^6 = 262{,}144$ combinations ($1/64$ of the nominal product).  These
resolve to 382 distinct Weinberg types across 106 $p$-vectors, with face counts
ranging from 11 to 17 (Table~\ref{tab:omega2d}, Supplementary Section~\ref{app:results}).

The contrast between the two sites of a single crystal is substantial---8 cell
types versus 382, from degenerate vertices whose resolutions never
conflict versus six coupled pairs---a clear illustration of coupling-dominated
combinatorial structure.  Without validity-pruning, the
$\omega$-2d enumeration would consider over 16 million candidate cells, most of
them spurious.  Its savings grow with the number of coupled pairs.  Roughly 130 of the $1{,}054$ structures in the library of
Section~\ref{sec:app_similarity} have at least one such coupled pair.

\subsection{Near-equidistant neighbors and structural noise}
\label{sec:near_equidistant}

The algorithm identifies the set of particles equidistant from each
Voronoi vertex by clustering the distances from the vertex to all
nearby particles.  This clustering is unambiguous when the distances form a
sharp group separated by a clear gap from the rest.  Two issues complicate that
in practice.
First, reference structures from databases such as the Materials Project are
typically DFT-relaxed: their atomic coordinates are numerically very close to
but not exactly at the symmetric positions of the crystal's space group, with
residual deviations above machine precision and typically reflecting incomplete
relaxation rather than rounding error.  Particles that should be exactly
equidistant from a Voronoi vertex by symmetry end up at slightly different
distances, and the clustering step can split them into separate groups,
producing spurious additional Voronoi vertices.  We avoid this by symmetrizing the input coordinates---mapping each particle to
the average of its symmetry-equivalent positions under the crystal's space
group---before running the enumeration.  This averaging is a projection onto the
symmetric positions: it returns each particle precisely to a symmetric position
rather than merely closer to one, so symmetry-enforced equidistances are
restored exactly.  The projection, however, only enforces symmetries of the entire 
cell.  Distortions confined to a symmetric subregion, e.g., a cluster
of particles forming an octahedron, are not corrected this way and are instead left
to the gap-threshold mechanism described next.

Second, even after symmetrization, a particle's distance to a
Voronoi vertex may only slightly exceed that of the particles
exactly equidistant from it.  Under a small perturbation of the
particle positions, such a near-equidistant particle can become a
new Voronoi neighbor and produce a cell topology outside the strict
analytical enumeration.  Conceptually, it functions as a partial
neighbor (Section~\ref{sec:2d_resolution}) that the strict
equidistance criterion failed to identify: at a tolerance wider
than zero, the partial-neighbor set at each vertex becomes larger.  The
algorithm supports an optional gap threshold for this situation:
any particle whose distance to a vertex is within a small relative
tolerance $\tau$ of the equidistant set at that vertex is treated
as a \emph{borderline} participant---a partial-neighbor
candidate.  Such a particle may or may not
contribute under a given perturbation, so the enumeration considers
every subset of the equidistant and borderline particles at the
vertex.  The resulting filter is constructed to cover the cell
topologies reachable under perturbations up to $\tau$ times the
distance from the vertex to its defining particles
(Section~\ref{sec:step2}).  The enumerated filters are
validated against stochastic \VoroTop sampling for the library of
Section~\ref{sec:app_similarity}; the results are reported in
Supplementary Section~\ref{app:nearequidistant}.

\subsection{Complexity limits}
\label{sec:complexity}

The cost of the enumeration scales as the product of the number of
resolutions over all degenerate vertices of a Voronoi cell.  For
most crystals encountered in practice this product is moderate
(typically $10^5$--$10^7$ before site-symmetry reduction), and the
algorithm runs to completion in seconds.  Some crystals, however,
have many degenerate vertices each with high multiplicity.  For example, the simple
cubic lattice has $k = 8$ particles equidistant at each of its eight Voronoi vertices
(forming a cube); the same configuration arises at the F site of the common fluorite structure.
A single $k = 8$ vertex admits hundreds of distinct resolution
types, and combined across the eight degenerate vertices of a
simple cubic Voronoi cell the enumeration space, before
site-symmetry reduction, is of order $10^{19}$ in the \emph{primary} mode
and $10^{22}$ in the \emph{default} mode---intractable by the present
algorithm.

For such structures, two fallback options are available.  First, restricting to
primary resolutions (Section~\ref{sec:3d_family}) substantially reduces the
per-vertex count, bringing some otherwise-intractable structures within
reach---though simple cubic remains out of reach even then.  Second, stochastic
Monte Carlo sampling at small $\sigma$ can estimate the family empirically,
supplementing or replacing the analytical enumeration when even the \emph{primary}
mode is too large.  We adopt the
analytical-only approach throughout, which suffices for the structures of
interest in the applications below.

\section{Applications}
\label{sec:applications}

We now apply the enumeration algorithm to several crystal structures
of broad interest that are not currently accessible to conventional structural
identification methods.  For each system, we compute the complete
family of Voronoi cell topologies, construct a topological filter, and
use it to classify local environments in finite-temperature molecular
dynamics simulations containing defects.  The results demonstrate that
Voronoi topology provides effective structural identification in systems
that fall outside the built-in reference sets of CNA, PTM, and other 
conventional methods.

The two structures examined in detail pose contrasting
challenges.  Sigma-phase CrFe (Section~\ref{sec:app_sigma}) is
structurally complex---30 atoms on five inequivalent sites, a
Frank--Kasper phase for which template-based methods have no
reference---yet it has only generic Voronoi vertices, so its filter
follows from a trivial enumeration.  The $\omega$-Ti polycrystal
(Section~\ref{sec:app_omega}) is the reverse: a simple three-atom cell that exercises the full
machinery of
degenerate-vertex resolution and the adjacency-coupled validity
pruning of Section~\ref{sec:practical}.  Finally,
Section~\ref{sec:app_similarity} surveys a library of 1{,}054
structures.

\subsection{Sigma phase CrFe: a Frank--Kasper phase}
\label{sec:app_sigma}
We begin with the simplest case: a structure whose enumeration is
trivial, so that the workflow---from a crystallographic description
to a per-atom classification of a finite-temperature simulation---can
be followed in its clearest form.  The sigma phase of the Fe--Cr
system ($P4_2/mnm$, 30 atoms per unit cell, 5 Wyckoff sites) is a
Frank--Kasper structure~\cite{frankkasper1958} of considerable metallurgical importance:
it forms as a brittle intermetallic precipitate in stainless
steels and other high-chromium alloys~\cite{hall1966sigma}, and in
nickel-base
superalloys~\cite{rae2001precipitation}.  As with other
topologically close-packed phases, standard identification tools have no
reference for the sigma phase~\cite{stukowski2012structure, xie2021laca}:
CNA, PTM, and the Ackland--Jones method all either abstain or else assign its
sites the labels of other structures, as quantified below.

We apply the enumeration algorithm to the sigma-phase CrFe unit cell
and find that all five Wyckoff sites have generic (non-degenerate)
Voronoi vertices: no vertex has more than four equidistant atoms.
The enumeration is therefore trivial, producing exactly one Weinberg
vector per site.  Although the unit cell contains five Wyckoff sites,
some share the same Voronoi cell topology, yielding only three distinct
Weinberg vectors in total.

To test this filter under realistic conditions, we construct a
$16 \times 16 \times 32$ supercell of sigma-phase CrFe.  A
spherical grain boundary is introduced at the center of the
simulation box by rotating all atoms within a sphere of radius
35\,\AA\ by $30^\circ$ about the $z$-axis and deleting one atom of
each overlapping pair (separation $< 1.5$\,\AA), leaving 245{,}274
atoms.  The system is energy-minimized and then annealed at 300\,K for
20\,ps under NPT dynamics using the Bonny \textit{et~al.}\ Fe--Cr potential~\cite{bonny2011fecr} in
LAMMPS~\cite{thompson2022lammps}.

\begin{figure}[t]
\centering
\subfiglabel{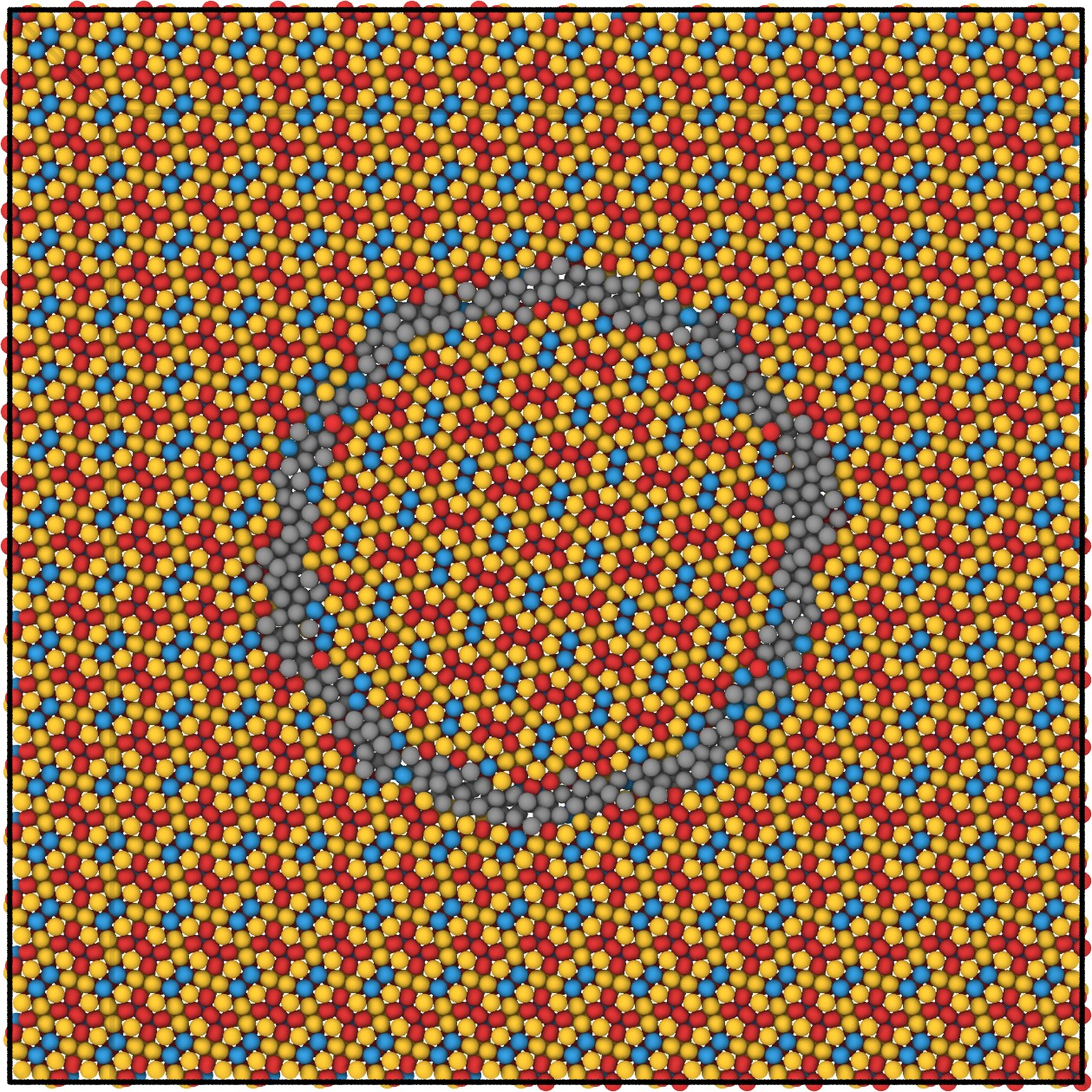}{(a)}\hfill
\subfiglabel{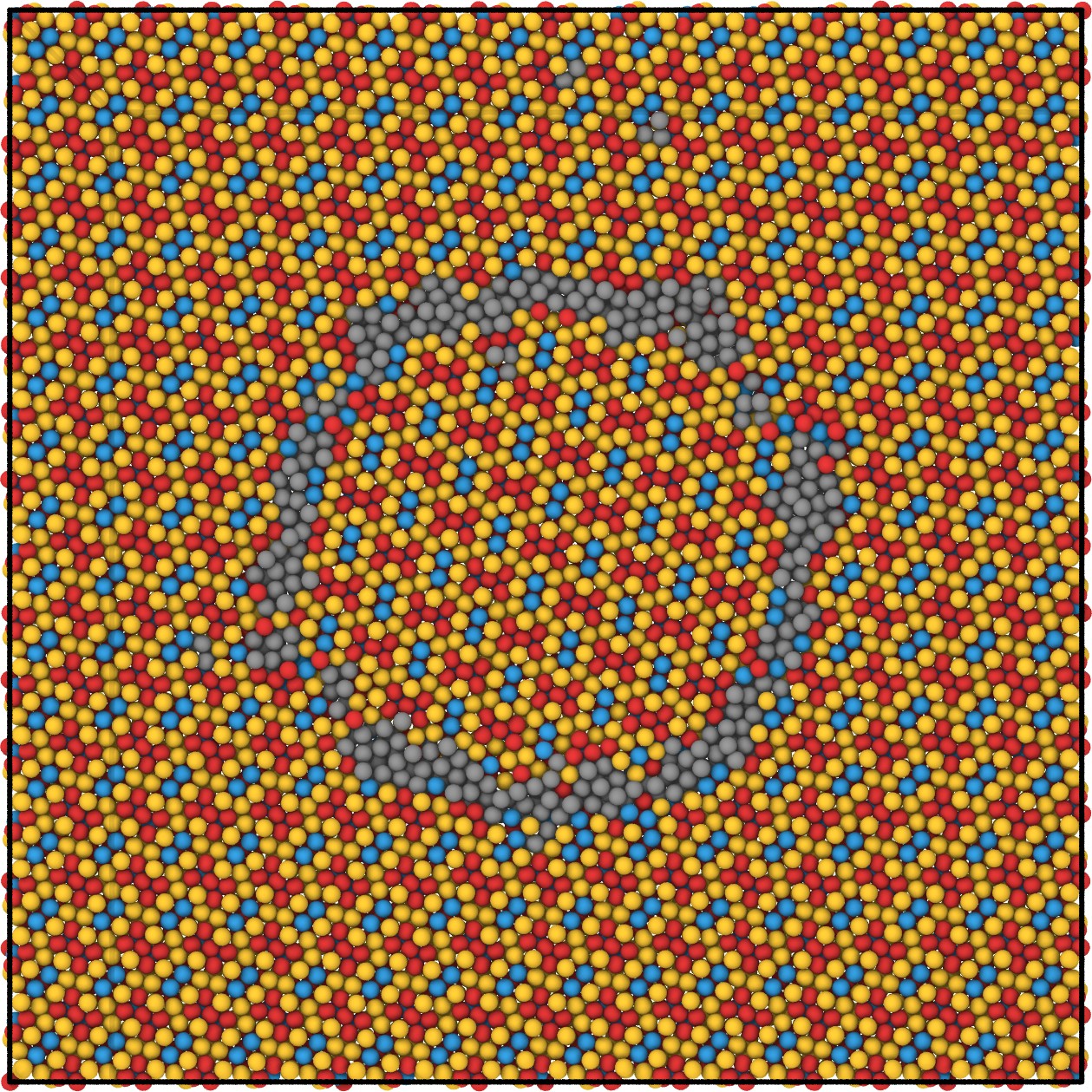}{(b)}\hfill
\subfiglabel{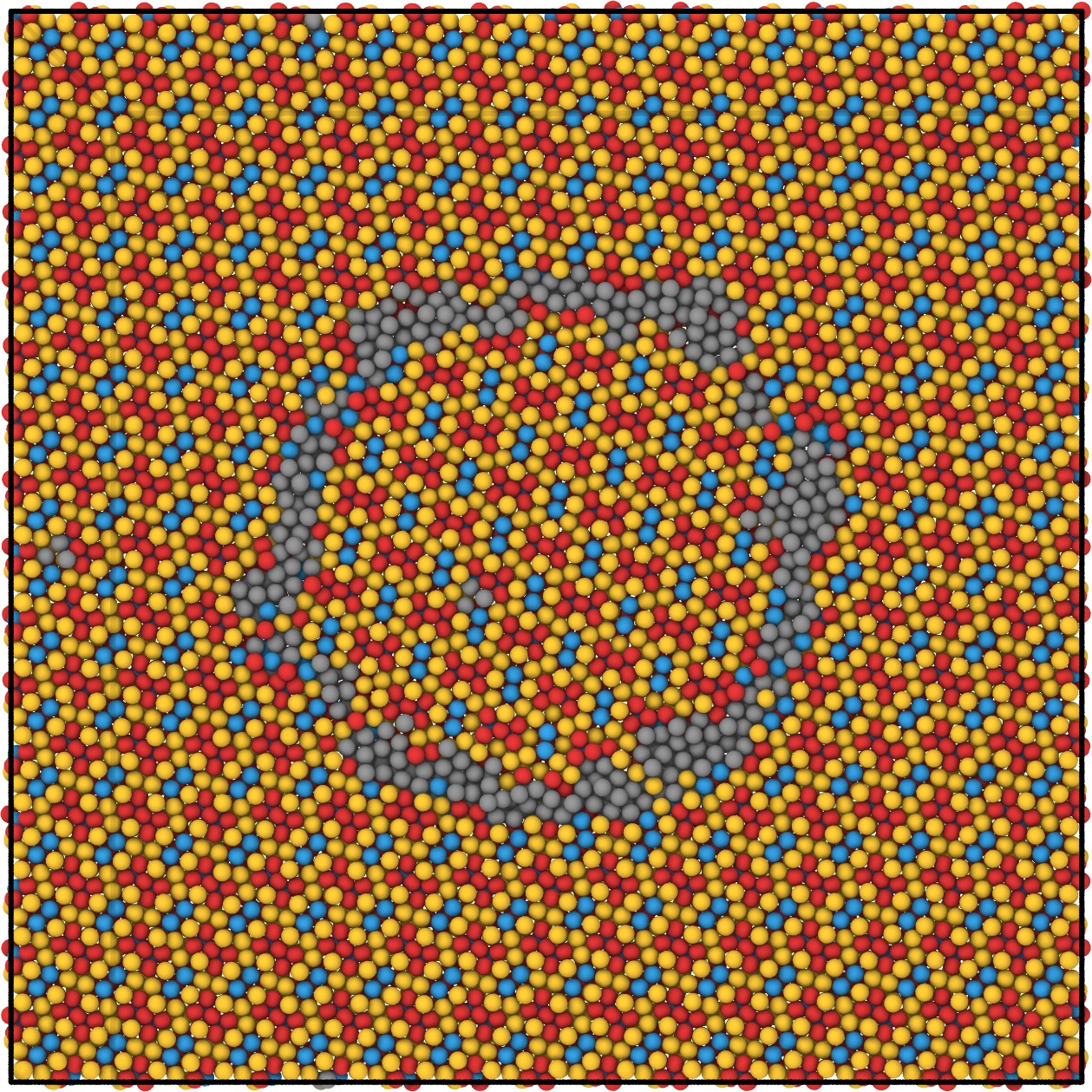}{(c)}
\caption{Cross-sections through a 245{,}274-atom sigma-phase CrFe
  system containing a spherical grain boundary.  Atoms are colored
  by \VoroTop classification: crystalline atoms, matched to one of
  the three sigma-phase Weinberg vectors, are shown in color by
  type, while defective atoms are shown in gray.
  (a)~After energy minimization, the spherical grain boundary
  appears as a sharp shell of defective atoms enclosing the
  rotated grain.
  (b,~c)~After 10 and 20\,ps of annealing at 300\,K, the boundary
  roughens under thermal disorder but remains clearly
  resolved against the crystalline bulk.}
\label{fig:bicrystal}
\end{figure}

Each snapshot is analyzed with \VoroTop using the enumerated filter.
Atoms whose Weinberg vectors match one of the three sigma-phase types
are classified as crystalline; all others are classified as defective.
Figure~\ref{fig:bicrystal} shows cross-sections through the system
at three stages.  In the minimized structure (panel~a), the
spherical grain boundary is sharply delineated as a thin shell of
defective atoms separating the rotated grain from the surrounding
matrix; \VoroTop identifies $97.6\%$ of all atoms as sigma phase,
the remaining $2.4\%$ being the grain-boundary shell itself.
During the 300\,K anneal (panels~b and~c, at 10 and 20\,ps),
thermal displacements roughen the boundary and push a small
additional fraction of atoms outside the filter, lowering the
identified fraction to $96.9\%$; the grain boundary remains a
clearly resolved shell against the crystalline bulk.

By contrast, none of the conventional methods are equipped with a sigma-phase reference,
and they fall short in two ways.  Polyhedral template matching,
run with the FCC, HCP, and BCC templates that are enabled by default,
finds no match for any sigma environment: the best fits exceed its RMSD cutoff, so it labels every atom
``other.''  This reflects coverage rather than capability: PTM admits custom
templates, and a sigma template would let it recognize the phase.
Common-neighbor analysis and the Ackland--Jones method instead
label a substantial fraction of the atoms: on the minimized
structure CNA labels $32\%$ of atoms as icosahedral---the Frank--Kasper sites
of twelvefold coordination---and the rest ``other,'' as does PTM once
its icosahedral template is enabled, while Ackland--Jones
labels $99\%$ of atoms as crystalline ($66\%$ BCC, $33\%$ icosahedral), the
BCC label falling on the sites of fourteen- and fifteenfold coordination.
These labels describe each site's coordination polyhedron under the name
of another structure; they do not identify the sigma phase, and they
persist under the $300$\,K anneal ($22\%$ icosahedral for CNA, $92\%$
crystalline for Ackland--Jones).  The centrosymmetry parameter uses no template
and provides only a continuous per-atom score, not a structural classification.
None of these methods identifies the sigma phase; \VoroTop does, distinguishing the three cell
topologies that its five Wyckoff sites adopt, because its filter is
enumerated automatically from the crystal's structure rather than
drawn from a fixed library.  Sites that share a cell topology could
be distinguished by a higher-order descriptor that
considers the cell types of an atom's neighbors alongside its
own, though we do not pursue that extension here.

\subsection{$\omega$-Ti polycrystal: degenerate-vertex resolution and
  same-species sublattice identification}
\label{sec:app_omega}

Our second application is to polycrystalline $\omega$-Ti, the
high-pressure, shock-induced polymorph of titanium and a standard
precipitate phase in metastable $\beta$-Ti alloys (Ti--Mo, Ti--V,
Ti--Nb)~\cite{hickman1969omega}, and a subject of recent atomistic
simulation studies of its formation under
shock~\cite{zong2014anisotropic} and its mechanical
response~\cite{zahiri2021transformation}.  The two Wyckoff sites of $\omega$-Ti
(1a and 2d) are both occupied by titanium, so, unlike for the
chemically distinct sublattices of a C15 Laves phase, distinguishing
the two sites cannot
exploit any chemical signal.  Both
sites also carry degenerate Voronoi vertices, and the 2d site
additionally has partial neighbors and the
adjacency-coupled vertex pairs that require validity pruning during
enumeration (Section~\ref{sec:coupling}).
$\omega$-Ti is therefore an application in which the
algorithm's degenerate-vertex resolution and validity-pruning
machinery (Appendix~\ref{sec:algorithm}) are fully
exercised, and one in which sublattice identification rests entirely on local
topology rather than chemistry.  We show below that \VoroTop
distinguishes the two sites and resolves the grain-boundary network
throughout an 8-grain polycrystalline sample, at temperatures up to
95\% of the $\omega \to \beta$ transition.

The enumeration for the two $\omega$-Ti sites was carried out in
Section~\ref{sec:coupling}: the 1a site yields 8 Weinberg vectors, and
the 2d site---where adjacency-coupled validity pruning cuts the
nominal $4^{12}$ combinations to $262{,}144$ valid
ones---yields 382.  Two Weinberg vectors are common to the two sites
(as discussed below), so the combined filter lists 390 site-labeled
entries covering 388 distinct vectors.  The
nearest atom beyond each vertex's equidistant set is, at both
sites, $34\%$ farther from the vertex than the equidistant atoms
themselves---the relative-gap quantity against which the threshold
$\tau$ of Section~\ref{sec:near_equidistant} is compared.  Closing
such a gap would require displacements of approximately
$0.7$\,\AA\ in the equilibrium $\omega$-Ti lattice, comfortably
above thermal vibration amplitudes throughout the temperature
range.

We construct a polycrystalline $\omega$-Ti sample as an orthorhombic
slab of dimensions $300\,\text{\AA} \times 300\,\text{\AA} \times 40\,\text{\AA}$, containing 8
grains with random orientations and $204{,}523$
titanium atoms.  Each grain is generated by populating its seed's
Voronoi region (under periodic boundary conditions) with a randomly oriented copy of the
$\omega$-Ti orthohexagonal cell; one atom of each pair separated
by less than
$1.5$\,\AA\ is removed. The slab geometry---$L_z = 40$~\AA, much smaller than the in-plane
grain size of roughly $100$\,\AA---forces every grain to span the full $L_z$
thickness, so every grain boundary is a vertical sheet perpendicular
to the slab face and the cross-section projection along $z$ is
unambiguous.  After conjugate-gradient minimization, the system is
annealed at seven temperatures spanning $0.1\,T_{\omega\beta}$ to
$0.95\,T_{\omega\beta}$ (117\,K to 1112\,K, where $T_{\omega\beta}
\approx 1170$\,K is the $\omega \to \beta$ transition temperature
reproduced by the potential~\cite{hennig2008meam}), with $6$\,ps of NPT equilibration at zero pressure
per stage.  Interatomic interactions are described by the
spline-based MEAM potential for titanium~\cite{hennig2008meam}, which was fit
specifically to the $\alpha/\beta/\omega$ polymorph energetics with
first-principles input.  Simulations are performed in
LAMMPS~\cite{thompson2022lammps}.

We apply the combined $\omega$-Ti filter to each equilibrated snapshot,
classifying every atom as $\omega$-1a, $\omega$-2d, or other, and group the
other population into connected components by Voronoi-cell face adjacency---a
parameter-free notion of connectivity in which two atoms are adjacent when
their Voronoi cells share a face---reclassifying clusters smaller than a size
threshold (set to $1{,}000$ atoms) as bulk thermal fluctuations.  Because the
cluster-size distribution is sharply bimodal (below), this threshold is not a
sensitive parameter: any value from a few hundred to tens of thousands of atoms
reclassifies the same thermal-noise clusters, leaving the bulk fraction
unchanged to within $0.1$ percentage points. 

\begin{figure}[t]
\centering
\subfiglabel{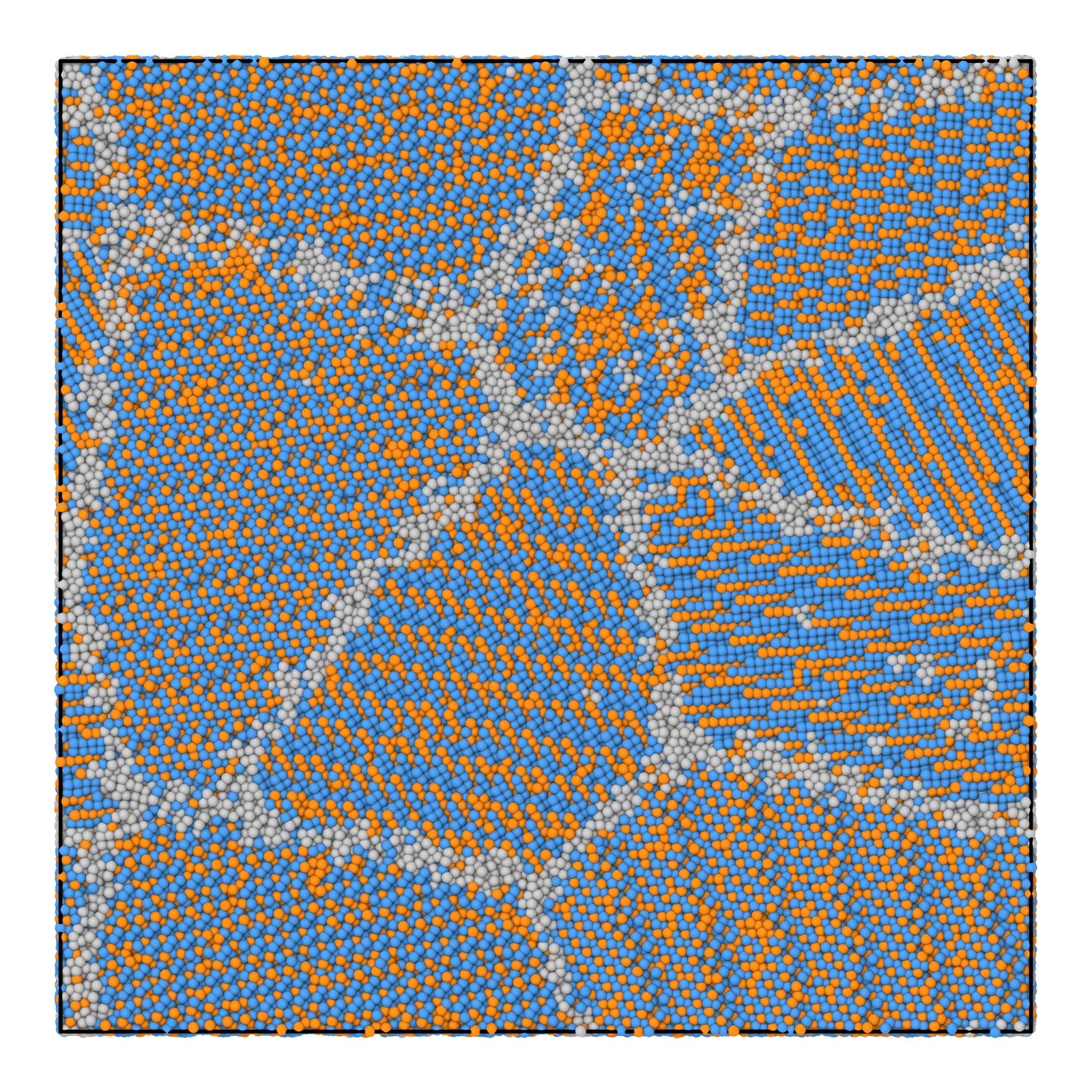}{(a)}\hfill
\subfiglabel{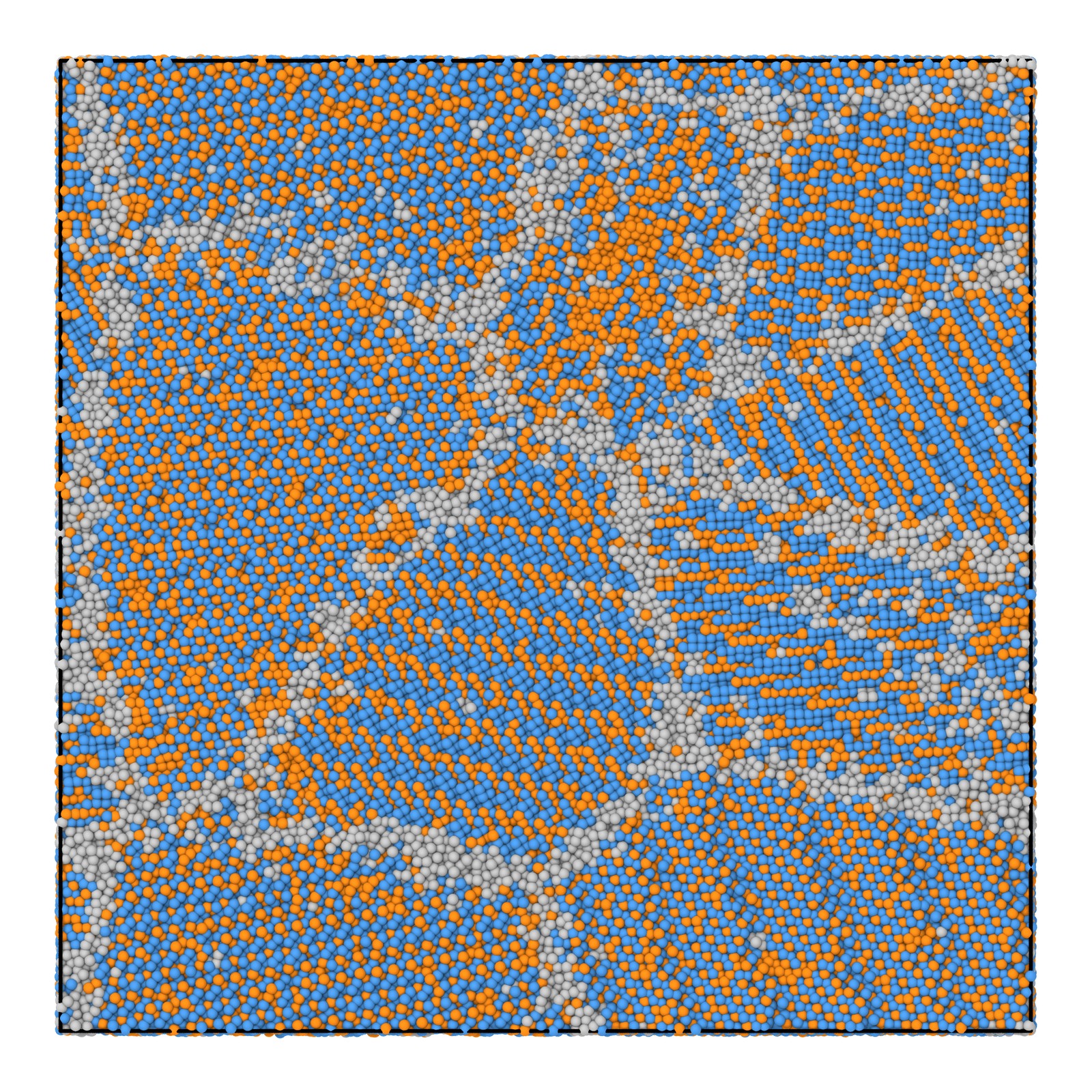}{(b)}\hfill
\subfiglabel{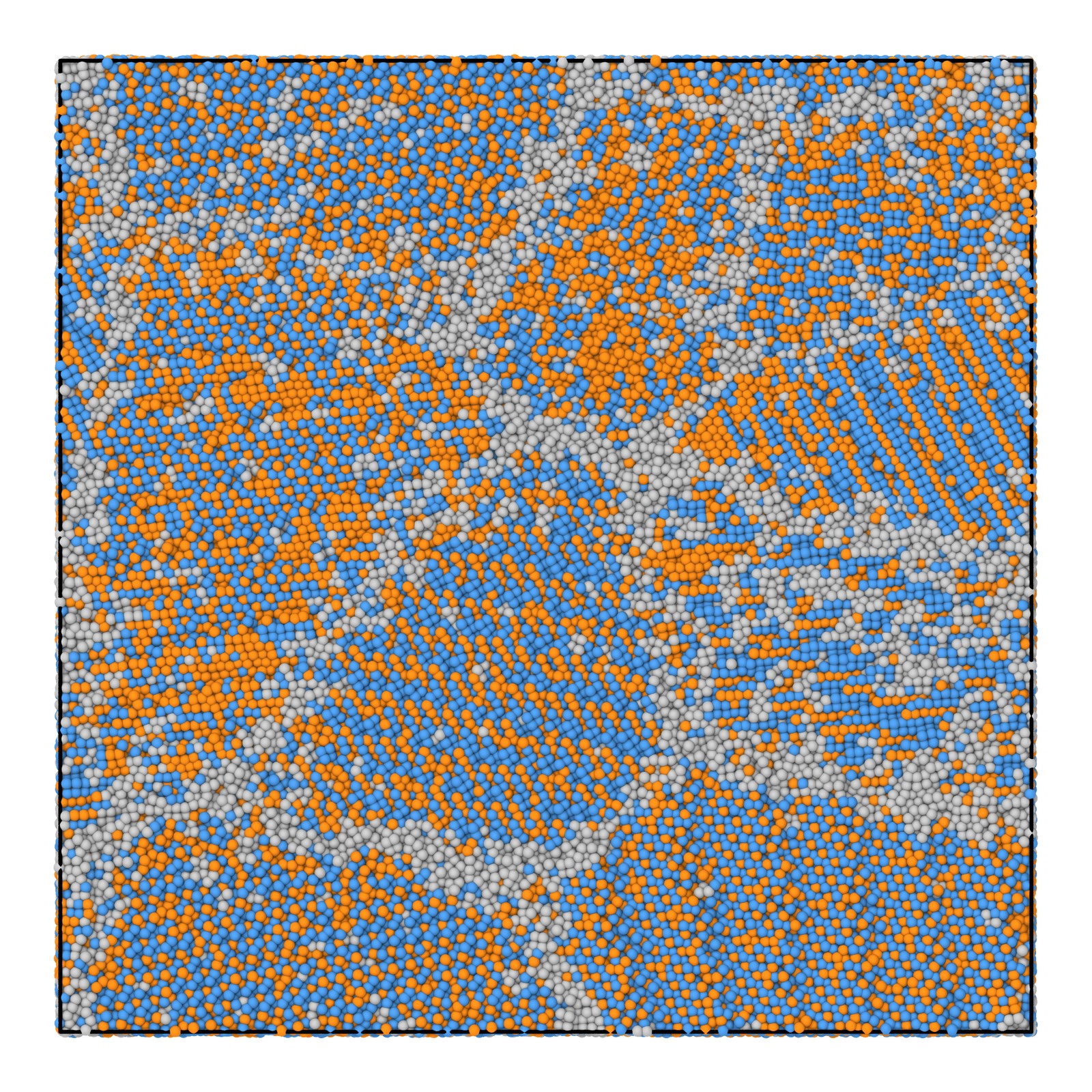}{(c)}
\caption{Polycrystalline $\omega$-Ti classified by \VoroTop with
  the combined $\omega$ filter.  Atoms are colored by Voronoi
  cell topology: $\omega$-1a in orange, $\omega$-2d in blue, and
  other atoms (mainly grain boundary) in gray; atoms exhibiting
  one of the two cell types common to both sites are colored as
  1a (Supplementary Section~\ref{app:omega_shared}).  Each panel is a
  top-down view of the same 8-grain polycrystalline slab at
  (a)~$0.5\,T_{\omega\beta}$, (b)~$0.7\,T_{\omega\beta}$, and
  (c)~$0.8\,T_{\omega\beta}$, with $T_{\omega\beta} \approx 1170$\,K
  the $\omega\to\beta$ transition.  Each grain
  interior shows a regular orange/blue mosaic---the $\omega$-Ti
  unit-cell tiling viewed along the slab's $z$-axis---whose in-plane
  orientation rotates between grains.  As temperature rises the
  grain boundaries broaden and the 1a/2d ratio within the bulk
  drifts.}
\label{fig:omega_poly}
\end{figure}

The two sublattices are
distinguished through their unshared cell types.  Two cell types,
however, are reachable from both sites; \VoroTop assigns atoms
exhibiting either of them the filter's first-listed label, here 1a,
so such atoms are still classified as crystalline $\omega$ but
carry a sublattice label that the topology alone cannot determine.
These two types are not rare: both are primary types of both sites,
and together they account for roughly $12\%$ of all atoms at
$0.1$ and $0.5\,T_{\omega\beta}$ alike.  A perturbation analysis
(Supplementary Section~\ref{app:omega_shared}) shows that the more
common of the two arises predominantly from 1a sites, so the 1a
assignment is correct for most such atoms; the net effect is that a
few percent of atoms carry an uncertain sublattice label, which
does not disturb the mosaic pattern of
Figure~\ref{fig:omega_poly}.

Every grain interior shows a regular mosaic of 1a and
2d atoms, and the mosaic's orientation rotates with each grain's lattice
orientation. In Figure~\ref{fig:omega_poly}, where 1a and 2d atoms are colored
separately and grain-boundary atoms are shown in gray, this produces visibly
distinct sublattice ``textures'' for grains of different orientation---even
though all atoms are chemically titanium and all grains share the same crystal
structure.  The viewer's eye reads the grain boundaries from the texture
changes alone.  The periodic color pattern within each grain is precisely the
$\omega$-Ti unit-cell tiling viewed along the slab's $z$-axis, and its
persistence at finite temperature confirms the crystallographic identity of the
bulk on a per-atom basis. The conventional methods examined below have no
reference for $\omega$ and do not provide this per-atom identification of
crystallographically inequivalent sites within a single chemical species.

The bulk identification rate is high throughout the temperature range, and the
cluster-size distribution is strongly bimodal at every stage: one giant cluster
(the grain-boundary network, growing from roughly $35{,}000$ atoms at
$0.1\,T_{\omega\beta}$ to roughly $59{,}000$ at $0.95\,T_{\omega\beta}$) is
separated by a clear gap from the remaining clusters (each below about
$1{,}000$ atoms). Bulk classification via atoms identified as either 1a or
2d decreases smoothly with temperature, from $82.3\%$ at
$0.1\,T_{\omega\beta}$ to $79.1\%$ at $0.5\,T_{\omega\beta}$, $73.2\%$ at
$0.8\,T_{\omega\beta}$, and $69.2\%$ at $0.95\,T_{\omega\beta} = 1112$\,K, only
$5\%$ below the $\omega \to \beta$ transition.

The cluster-size filter
adds at most about two percentage points to the bulk rate, under one at low
temperature and rising toward two near the transition.  This is because the
other population here is concentrated almost entirely in the grain-boundary
network, with little of it scattered through the grains as small
thermal-noise clusters.  At every temperature studied, the grain-boundary network
remains the dominant connected component, and the polycrystalline structure is
cleanly resolved in cross-section throughout.

We additionally observe that the
measured 1a:2d count ratio departs from the ideal stoichiometric value of $0.5$
at every temperature---it is already $0.62$ at $0.1\,T_{\omega\beta}$ and rises
to $1.31$ at $0.95\,T_{\omega\beta}$.  This traces to an overlap between the
two families: of the 8 perturbation-accessible 1a cell topologies and the 382
of the 2d site, two Weinberg vectors are common to both (14-face cells
reachable from either site).  A 2d atom whose Voronoi cell realizes one of
these shared topologies is indistinguishable from a 1a atom and is counted as
1a, biasing the ratio upward.  The shared types account for $11$--$12\%$ of
all atoms up to $0.5\,T_{\omega\beta}$, which produces the offset at low
temperature; at higher temperature their share grows, to $16\%$ at
$0.8\,T_{\omega\beta}$ and $27\%$ at $0.95\,T_{\omega\beta}$, while the
fraction of atoms with unshared 2d types falls from $51\%$ to $30\%$, and
the ratio rises accordingly (Supplementary Section~\ref{app:omega_shared}).  This affects only the within-bulk sublattice split: a
shared topology still lies within the combined filter, so the
bulk-versus-grain-boundary classification, which carries the structural
identification claim, is unaffected.  The increase in the shared types is
entirely in the fraction of atoms whose Voronoi cell is the truncated
octahedron, $4\%$ at $0.5\,T_{\omega\beta}$ and $23\%$ at
$0.95\,T_{\omega\beta}$, while the other shared type becomes rarer.  The
truncated octahedron is the Voronoi cell of BCC, so near the transition
these atoms probably sit in $\beta$-like local environments, which the
filter cannot distinguish from $\omega$.

We compare these results with the standard methods on the $0.5\,T_{\omega\beta}$ snapshot
(detailed in Supplementary Section~\ref{app:omega_methods_comparison}).  None carries a
reference for $\omega$-Ti, and as with the sigma phase they fall short in
different ways.  Common-neighbor analysis classifies essentially every atom as
``other'' ($97.85\%$ at $0.5\,T_{\omega\beta}$), an honest abstention.
Polyhedral template matching behaves likewise only at a strict RMSD cutoff; at
the more permissive cutoff used here it labels roughly $30\%$ of atoms
BCC or HCP (Table~\ref{tab:omega_ptm_sweep}).  The Ackland--Jones
method~\cite{ackland2006jones}, which has no acceptance threshold, labels
about $82\%$ of atoms FCC, HCP, BCC, or icosahedral.  These labels are
not arbitrary: most 1a atoms are labeled BCC and most 2d atoms HCP.  But
they name other structures, and most grain-boundary atoms receive them as
well.  The centrosymmetry parameter distinguishes the two sites at low
temperature (the 1a site is centrosymmetric, the 2d is not), but its
distributions broaden and overlap as temperature rises.  The Laves-phase
crystal analysis framework~\cite{xie2021laca} targets the structurally related
Laves phases and provides no $\omega$ reference.  Among the methods evaluated
here, \VoroTop with the analytical filter is the only one that identifies
the $\omega$ phase itself, with its two inequivalent titanium sites, and
resolves its grain-boundary network.

\subsection{A library of Voronoi cell families}
\label{sec:app_similarity}

To assess the algorithm's reach beyond the structures examined above,
we have applied it to a random sample of 1{,}300 stable crystal
structures drawn from the Materials
Project~\cite{horton2025accelerated}, spanning all seven crystal
systems.  Of these, 1{,}054 complete the \emph{default}-mode
enumeration at the recommended gap threshold $\tau = 0.005$ within a
per-structure compute budget of 300\,s and form the library analyzed here.  The qualitative outcome divides the
library into two regimes.  About $73\%$ of structures are
topologically stable: every Wyckoff site admits a single Voronoi cell
topology under small perturbation, so the filter for these crystals is
a short list of Weinberg vectors---often one per inequivalent site.
The remaining $27\%$ have at least one Wyckoff site with degenerate
Voronoi vertices and produce families of multiple cell types, ranging
from a handful to nearly twenty thousand.  This upper end reflects the budget:
still more degenerate structures, such as simple cubic
(Section~\ref{sec:complexity}), have far larger families and do not complete
within it.  The resulting filters form a \VoroTop{}-compatible library, so each
of these structures can be identified without re-running the enumeration.  Detailed
statistics---counts per crystal system, the distribution of family
sizes, and the sharing of Weinberg vectors across
structures---are given in Supplementary
Section~\ref{app:library}.

\section{Discussion}
\label{sec:discussion}

Computing a topological filter and applying it are different operations
with different costs.  The enumeration described in this paper is
performed once per crystal structure, independently of the data to be
analyzed, and the resulting filter is stored as a plain text file.  At
analysis time, \VoroTop
computes the Voronoi cell and Weinberg vector of each particle and matches the
vector against the filter by hash-table lookup.  The per-particle cost is
shared roughly equally between the Voronoi construction (performed by
\textsc{Voro++}~\cite{rycroft09c}) and the Weinberg-vector
computation, an all-integer canonical-labeling pass over the cell's
face network; the filter lookup adds negligible overhead.  Because
the per-particle computations are independent, they parallelize with
near-perfect speedup through the multithreaded \textsc{Voro++}
extension~\cite{lu2023extension}: classifying the $204{,}523$-atom
polycrystal snapshot of Section~\ref{sec:app_omega} takes about
$4$\,s of CPU time in total, under one second of wall-clock time on a
multicore laptop.

A filter is tailored to the exact geometry of its reference
structure, not merely to its space group or lattice type.  This is a
feature when the goal is structural identification, but it means the
reference structure must match the simulation.  Different
representations of nominally the same crystal---for example,
different DFT relaxations or database versions---can produce
different Weinberg vector sets. For AlCu ($C2/m$, mp-2500), the
Materials Project's GGA-relaxed structure (task mp-2500, cell
volume $140.31$~\AA$^3$) and its r$^2$SCAN-relaxed structure (task
mp-2050840, $136.38$~\AA$^3$) differ by $2.9\%$ in volume.  In
the strict enumeration, share only 3 of 6 Weinberg types; at a gap
threshold of $\tau = 0.01$ both yield the same 13 cell types.  A
related complication arises in alloys: a random
solid solution on a fixed lattice produces the same Voronoi families
as the pure element, but size mismatch between species introduces
local distortions that activate borderline partial neighbors
(Section~\ref{sec:near_equidistant}), generating additional cell
types beyond the ideal-lattice enumeration.  In both cases the gap-threshold
mechanism of Section~\ref{sec:near_equidistant} addresses the issue:
setting the tolerance commensurate with the expected geometric
mismatch extends the filter to cover the additional types.  For
ordered intermetallics with distinct sublattices (e.g., A15,
fluorite), each sublattice carries its own family, and the method
applies directly.

Because Voronoi cell topology captures local geometry rather than
chemical identity, isostructural compounds with the same local
geometry produce identical Weinberg vector families.  This is by
design---topology identifies \emph{structural type}, not chemical
species---and means that distinguishing such compounds requires
combining the filter with chemical descriptors.
Conversely, structures with distinct local geometries are readily
discriminated: of all pairs of library structures, $83\%$ have no cell
type in common, and only $5\%$ have identical filters (Supplementary
Section~\ref{app:library}).
Shared cell types can also arise within a single crystal: the perturbation-accessible
families of two inequivalent Wyckoff sites need not be disjoint.  The
$\omega$-Ti 1a and 2d sites share two cell topologies
(Section~\ref{sec:app_omega}), so an atom realizing one of them
cannot be assigned to a sublattice by topology alone---Voronoi
topology resolves inequivalent sites well, but not always
perfectly.

The framework is also not limited to bulk crystal identification. The same
enumeration procedure applies, without modification, to ideal defect
environments: a vacancy, interstitial, or substitutional site in any host
lattice has its own Voronoi cell and its own family of perturbation-accessible
topologies, and a filter generated for that environment can be used to identify
the defect directly in simulation.  We do not pursue this extension here, but
it broadens the scope of Voronoi-topology-based identification beyond the
bulk-versus-``other''
dichotomy adopted in the applications above.

Finally, the \emph{frequencies} with which family members appear
carry information beyond membership.  A primary type occurs with a
$\sigma$-independent leading frequency---its fixed fraction of
perturbation directions---while a secondary type requiring $m$
zero-volume tetrahedra with equidistant particles on both sides of
their plane occurs with frequency of order
$\sigma^m$ (Supplementary Section~\ref{app:primary_validation}).  Computing these leading
coefficients, and inverting measured topology statistics to
estimate the effective perturbation magnitude---and hence the
temperature---is an attractive direction for future
work.

\section{Conclusions}
\label{sec:conclusions}

Identifying local crystal structure in atomistic simulations has long
been limited to a small set of simple reference structures by the
design of existing methods.  We have shown that Voronoi cell topology
provides a systematic and general framework for structural
identification in arbitrary crystals, and have removed the main
practical barrier---the computation of complete topological
filters---by presenting an algorithm that deterministically enumerates
all Voronoi cell topologies arising from perturbations of any crystal
structure.  The algorithm exploits the Delaunay--Voronoi duality to
decompose the enumeration into local resolution problems at degenerate
vertices, with site symmetry and validity pruning keeping the computation
fast for the vast majority of structures encountered in practice.

The applications demonstrate that this approach provides effective
structural identification in systems where existing automated methods
provide little or no classification capability, including Frank--Kasper
phases and other structures with multiple inequivalent sites.  The
method is purely geometric, requires no interatomic potential or
templates, and naturally accommodates the thermal
fluctuations of finite-temperature simulations.

\section*{Acknowledgements}

EAL acknowledges support from the Israel Science Foundation under
grant 202/24. CHR was partially supported by the Applied Mathematics Program of
the U.S.~DOE Office of Science Advanced Scientific Computing Research under
Contract No.~DE-AC02-05CH11231. This research was supported in part by Grant No.~2018170 from the United States--Israel Binational Science Foundation (BSF). The authors gratefully acknowledge extensive
collaboration with Anthropic's Claude Code throughout this project, in
algorithm development, code implementation, computational verification, and
manuscript drafting.

\section*{Data availability statement}

This study concerns an algorithm and its implementation.  \VoroTypes is available at
\url{https://github.com/VoroTop/VoroTypes}, and the accompanying
library of pre-computed filter files at \url{https://www.vorotop.org}.
The molecular dynamics configurations and input scripts underlying the
applications in Section~\ref{sec:applications} are available from the
authors upon reasonable request.

\bibliographystyle{ieeetr}

\begin{thebibliography}{10}

\bibitem{ashcroft1976solid}
N.~W. Ashcroft and N.~D. Mermin, {\em Solid State Physics}.
\newblock Holt, Rinehart and Winston, New York, 1976.

\bibitem{kittel1976introduction}
C.~Kittel, {\em Introduction to Solid State Physics}.
\newblock John Wiley \& Sons, 5th~ed., 1976.

\bibitem{padture2002thermal}
N.~P. Padture, M.~Gell, and E.~H. Jordan, ``Thermal barrier coatings for
  gas-turbine engine applications,'' {\em Science}, vol.~296, no.~5566,
  pp.~280--284, 2002.

\bibitem{stukowski2012structure}
A.~Stukowski, ``Structure identification methods for atomistic simulations of
  crystalline materials,'' {\em Model. Simul. Mater. Sci. Eng.}, vol.~20,
  no.~4, p.~045021, 2012.

\bibitem{stukowski2010extracting}
A.~Stukowski and K.~Albe, ``Extracting dislocations and non-dislocation crystal
  defects from atomistic simulation data,'' {\em Modelling Simul. Mater. Sci.
  Eng.}, vol.~18, no.~8, p.~085001, 2010.

\bibitem{olmsted2009survey}
D.~L. Olmsted, S.~M. Foiles, and E.~A. Holm, ``Survey of computed grain
  boundary properties in face-centered cubic metals: {I}. {G}rain boundary
  energy,'' {\em Acta Materialia}, vol.~57, no.~13, pp.~3694--3703, 2009.

\bibitem{bringa2006shock}
E.~M. Bringa, K.~Rosolankova, R.~E. Rudd, B.~A. Remington, J.~S. Wark,
  M.~Duchaineau, D.~H. Kalantar, J.~Hawreliak, and J.~Belak, ``Shock
  deformation of face-centred-cubic metals on subnanosecond timescales,'' {\em
  Nature Materials}, vol.~5, no.~10, pp.~805--809, 2006.

\bibitem{honeycutt1987cna}
J.~D. Honeycutt and H.~C. Andersen, ``Molecular dynamics study of melting and
  freezing of small {Lennard-Jones} clusters,'' {\em J. Phys. Chem.}, vol.~91,
  pp.~4950--4963, 1987.

\bibitem{ackland2006jones}
G.~J. Ackland and A.~P. Jones, ``Applications of local crystal structure
  measures in experiment and simulation,'' {\em Phys. Rev. B}, vol.~73,
  p.~054104, 2006.

\bibitem{larsen2016ptm}
P.~M. Larsen, S.~Schmidt, and J.~Schi{\o}tz, ``Robust structural identification
  via polyhedral template matching,'' {\em Modelling Simul. Mater. Sci. Eng.},
  vol.~24, p.~055007, 2016.

\bibitem{kelchner1998dislocation}
C.~L. Kelchner, S.~Plimpton, and J.~Hamilton, ``Dislocation nucleation and
  defect structure during surface indentation,'' {\em Phys.~Rev.~B}, vol.~58,
  no.~17, p.~11085, 1998.

\bibitem{frankkasper1958}
F.~C. Frank and J.~S. Kasper, ``Complex alloy structures regarded as sphere
  packings. {I}. {D}efinitions and basic principles,'' {\em Acta
  Crystallographica}, vol.~11, no.~3, pp.~184--190, 1958.

\bibitem{frank1959complex}
F.~C. Frank and J.~S. Kasper, ``Complex alloy structures regarded as sphere
  packings. {II}. {A}nalysis and classification of representative structures,''
  {\em Acta Crystallographica}, vol.~12, no.~7, pp.~483--499, 1959.

\bibitem{friauf1927}
J.~B. Friauf, ``The crystal structures of two intermetallic compounds,'' {\em
  Journal of the American Chemical Society}, vol.~49, no.~12, pp.~3107--3114,
  1927.

\bibitem{stein2021laves}
F.~Stein and A.~Leineweber, ``{L}aves phases: a review of their functional and
  structural applications and an improved fundamental understanding of
  stability and properties,'' {\em Journal of Materials Science}, vol.~56,
  no.~9, pp.~5321--5427, 2021.

\bibitem{sickafus1999spinel}
K.~E. Sickafus, J.~M. Wills, and N.~W. Grimes, ``Structure of spinel,'' {\em
  Journal of the American Ceramic Society}, vol.~82, no.~12, pp.~3279--3292,
  1999.

\bibitem{samson1962nacd2}
S.~Samson, ``Crystal structure of {NaCd$_2$},'' {\em Nature}, vol.~195,
  no.~4838, pp.~259--262, 1962.

\bibitem{urban2004complex}
K.~Urban and M.~Feuerbacher, ``Structurally complex alloy phases,'' {\em
  Journal of Non-Crystalline Solids}, vol.~334--335, pp.~143--150, 2004.

\bibitem{voronoi1908nouvelles}
G.~Vorono{\"\i}, ``{Nouvelles applications des param{\`e}tres continus {\`a} la
  th{\'e}orie des formes quadratiques. Deuxi{\`e}me m{\'e}moire. Recherches sur
  les parall{\'e}llo{\`e}dres primitifs.},'' {\em J. Reine Angew. Math.},
  vol.~134, pp.~198--287, 1908.

\bibitem{1992okabe}
A.~Okabe, B.~N. Boots, K.~Sugihara, and S.~N. Chiu, {\em Spatial Tessellations:
  Concepts and Applications of Voronoi Diagrams}.
\newblock John Wiley \& Sons, 2nd~ed., 2000.

\bibitem{lazar2022voronoi}
E.~A. Lazar, J.~Lu, and C.~H. Rycroft, ``Voronoi cell analysis: The shapes of
  particle systems,'' {\em Am. J. Phys.}, vol.~90, no.~6, pp.~469--480, 2022.

\bibitem{finney1970random}
J.~L. Finney, ``Random packings and the structure of simple liquids. {I.} the
  geometry of random close packing,'' {\em Proceedings of the Royal Society of
  London A}, vol.~319, pp.~479--493, 1970.

\bibitem{rycroft06a}
C.~H. Rycroft, G.~S. Grest, J.~W. Landry, and M.~Z. Bazant, ``Analysis of
  granular flow in a pebble-bed nuclear reactor,'' {\em Phys. Rev. E}, vol.~74,
  p.~021306, Aug 2006.

\bibitem{sheng2006atomic}
H.~W. Sheng, W.~K. Luo, F.~M. Alamgir, J.~M. Bai, and E.~Ma, ``Atomic packing
  and short-to-medium-range order in metallic glasses,'' {\em Nature},
  vol.~439, pp.~419--425, 2006.

\bibitem{damasceno2012predictive}
P.~F. Damasceno, M.~Engel, and S.~C. Glotzer, ``Predictive self-assembly of
  polyhedra into complex structures,'' {\em Science}, vol.~337, pp.~453--457,
  2012.

\bibitem{2012lazar}
E.~A. Lazar, J.~K. Mason, R.~D. MacPherson, and D.~J. Srolovitz, ``Complete
  topology of cells, grains, and bubbles in three-dimensional
  microstructures,'' {\em Phys. Rev. Lett.}, vol.~109, no.~9, p.~095505, 2012.

\bibitem{lazar2015topological}
E.~A. Lazar, J.~Han, and D.~J. Srolovitz, ``Topological framework for local
  structure analysis in condensed matter,'' {\em Proc. Natl. Acad. Sci.
  U.S.A.}, vol.~112, no.~43, pp.~E5769--E5776, 2015.

\bibitem{vorotop}
E.~A. Lazar, ``{VoroTop}: Voronoi cell topology visualization and analysis
  toolkit,'' {\em Model. Simul. Mater. Sci. Eng.}, vol.~26, no.~1, p.~015011,
  2018.

\bibitem{lu2023extension}
J.~Lu, E.~A. Lazar, and C.~H. Rycroft, ``{An extension to VORO++ for
  multithreaded computation of Voronoi cells},'' {\em Computer Physics
  Communications}, vol.~291, p.~108832, 2023.

\bibitem{lazar2024characterizing}
E.~A. Lazar, J.~Lu, C.~H. Rycroft, and D.~Schwarcz, ``Characterizing structural
  features of two-dimensional particle systems through {V}oronoi topology,''
  {\em Modelling and Simulation in Materials Science and Engineering}, vol.~32,
  no.~8, p.~085022, 2024.

\bibitem{loera}
J.~A.~D. Loera, J.~Rambau, and F.~Santos, {\em Triangulations: Structures for
  Algorithms and Applications}.
\newblock Springer-Verlag, 2010.

\bibitem{1966weinberg}
L.~Weinberg, ``A simple and efficient algorithm for determining isomorphism of
  planar triply connected graphs,'' {\em IEEE Trans. Circuit Theory},
  vol.~{CT13}, no.~{2}, pp.~142--148, {1966}.

\bibitem{leipold2016statistical}
H.~Leipold, E.~A. Lazar, K.~A. Brakke, and D.~J. Srolovitz, ``Statistical
  topology of perturbed two-dimensional lattices,'' {\em J. Stat. Mech. Theory
  Exp.}, vol.~2016, no.~4, p.~043103, 2016.

\bibitem{lazar2020voronoi}
E.~A. Lazar and A.~Shoan, ``Voronoi chains, blocks, and clusters in perturbed
  square lattices,'' {\em J. Stat. Mech. Theory Exp.}, vol.~2020, no.~10,
  p.~103204, 2020.

\bibitem{lieb1989two}
E.~H. Lieb, ``Two theorems on the {Hubbard} model,'' {\em Phys. Rev. Lett.},
  vol.~62, no.~10, pp.~1201--1204, 1989.

\bibitem{yoon2019topological}
T.~J. Yoon, M.~Y. Ha, W.~B. Lee, Y.-W. Lee, and E.~A. Lazar, ``Topological
  generalization of the rigid--nonrigid transition in soft-sphere and
  hard-sphere fluids,'' {\em Phys.~Rev.~E}, vol.~99, no.~5, p.~052603, 2019.

\bibitem{hickman1969omega}
B.~S. Hickman, ``The formation of omega phase in titanium and zirconium alloys:
  A review,'' {\em J. Mater. Sci.}, vol.~4, pp.~554--563, 1969.

\bibitem{hall1966sigma}
E.~O. Hall and S.~H. Algie, ``The sigma phase,'' {\em Metallurgical Reviews},
  vol.~11, pp.~61--88, 1966.

\bibitem{rae2001precipitation}
C.~M.~F. Rae and R.~C. Reed, ``The precipitation of topologically close-packed
  phases in rhenium-containing superalloys,'' {\em Acta Materialia}, vol.~49,
  pp.~4113--4125, 2001.

\bibitem{xie2021laca}
Z.~Xie, D.~Chauraud, E.~Bitzek, S.~Korte-Kerzel, and J.~Gu{\'e}nol{\'e},
  ``Laves phase crystal analysis ({LaCA}): Atomistic identification of lattice
  defects in {C14} and {C15} topologically close-packed phases,'' {\em Journal
  of Materials Research}, vol.~36, pp.~2010--2024, 2021.

\bibitem{bonny2011fecr}
G.~Bonny, R.~C. Pasianot, D.~Terentyev, and L.~Malerba, ``Iron chromium
  potential to model high-chromium ferritic alloys,'' {\em Phil. Mag.},
  vol.~91, no.~12, pp.~1724--1746, 2011.

\bibitem{thompson2022lammps}
A.~P. Thompson, H.~M. Aktulga, R.~Berger, D.~S. Bolintineanu, W.~M. Brown,
  P.~S. Crozier, P.~J. in~'t Veld, A.~Kohlmeyer, S.~G. Moore, T.~D. Nguyen,
  {\em et~al.}, ``{LAMMPS}---a flexible simulation tool for particle-based
  materials modeling at the atomic, meso, and continuum scales,'' {\em Comp.
  Phys. Comm.}, vol.~271, p.~108171, 2022.

\bibitem{zong2014anisotropic}
H.~Zong, T.~Lookman, X.~Ding, S.-N. Luo, and J.~Sun, ``Anisotropic shock
  response of titanium: {R}eorientation and transformation mechanisms,'' {\em
  Acta Materialia}, vol.~65, pp.~10--18, 2014.

\bibitem{zahiri2021transformation}
A.~H. Zahiri, J.~Ombogo, T.~Ma, P.~Chakraborty, and L.~Cao,
  ``Transformation-induced plasticity in omega titanium,'' {\em Journal of
  Applied Physics}, vol.~129, p.~015105, 2021.

\bibitem{hennig2008meam}
R.~G. Hennig, T.~J. Lenosky, D.~R. Trinkle, S.~P. Rudin, and J.~W. Wilkins,
  ``Classical potential describes martensitic phase transformations between the
  {$\alpha$}, {$\beta$}, and {$\omega$} titanium phases,'' {\em Physical Review
  B}, vol.~78, p.~054121, 2008.

\bibitem{horton2025accelerated}
M.~K. Horton, P.~Huck, R.~X. Yang, J.~M. Munro, S.~Dwaraknath, A.~M. Ganose,
  R.~S. Kingsbury, M.~Wen, J.~X. Shen, T.~S. Mathis, {\em et~al.},
  ``Accelerated data-driven materials science with the {M}aterials {P}roject,''
  {\em Nature Materials}, vol.~24, no.~10, pp.~1522--1532, 2025.

\bibitem{rycroft09c}
C.~H. Rycroft, ``{Voro++}: A three-dimensional {V}oronoi cell library in
  {C}++,'' {\em Chaos: An Interdisciplinary Journal of Nonlinear Science},
  vol.~19, no.~4, p.~041111, 2009.

\bibitem{scipy2020}
P.~Virtanen, R.~Gommers, T.~E. Oliphant, {\em et~al.}, ``{SciPy} 1.0:
  Fundamental algorithms for scientific computing in {Python},'' {\em Nature
  Methods}, vol.~17, pp.~261--272, 2020.

\bibitem{barber96}
C.~B. Barber, D.~P. Dobkin, and H.~Huhdanpaa, ``The quickhull algorithm for
  convex hulls,'' {\em ACM Trans. Math. Softw.}, vol.~22, no.~4, pp.~469--483,
  1996.

\bibitem{pymatgen}
S.~P. Ong, W.~D. Richards, A.~Jain, G.~Hautier, M.~Kocher, S.~Cholia,
  D.~Gunter, V.~L. Chevrier, K.~A. Persson, and G.~Ceder, ``{P}ython
  {M}aterials {G}enomics (pymatgen): A robust, open-source {P}ython library for
  materials analysis,'' {\em Comput. Mater. Sci.}, vol.~68, pp.~314--319, 2013.

\bibitem{read1978}
R.~C. Read, ``Every one a winner or how to avoid isomorphism search when
  cataloguing combinatorial configurations,'' {\em Annals of Discrete
  Mathematics}, vol.~2, pp.~107--120, 1978.

\bibitem{caroli2009periodic}
M.~Caroli and M.~Teillaud, ``Computing 3{D} periodic triangulations,'' in {\em
  Algorithms -- {ESA} 2009}, vol.~5757 of {\em Lecture Notes in Computer
  Science}, pp.~59--70, Springer, 2009.

\bibitem{ong2015materials}
S.~P. Ong, S.~Cholia, A.~Jain, M.~Brafman, D.~Gunter, G.~Ceder, and K.~A.
  Persson, ``The {M}aterials {A}pplication {P}rogramming {I}nterface ({API}): A
  simple, flexible and efficient {API} for materials data based on
  {RE}presentational {S}tate {T}ransfer ({REST}) principles,'' {\em
  Computational Materials Science}, vol.~97, pp.~209--215, 2015.

\bibitem{wigner1933constitution}
E.~Wigner and F.~Seitz, ``On the constitution of metallic sodium,'' {\em Phys.
  Rev.}, vol.~43, no.~10, p.~804, 1933.

\end{thebibliography}
\clearpage
\appendix
\renewcommand{\thesection}{\Alph{section}}
\setcounter{section}{0}

\section{Algorithm}
\label{sec:algorithm}

We now describe our algorithm for enumerating all Voronoi cell topologies
arising from infinitesimal perturbations of a crystal structure.  For a
unit cell with $N$ particles, the algorithm is applied independently
to each particle in the basis, which then serves as the central
particle $p_0$.  Particles related by
symmetry belong to the same \emph{Wyckoff site},
and thus have identical local environments and produce identical
families of cell types, which are automatically merged in the output; 
particles at inequivalent Wyckoff sites
(e.g., the 2a and 6c sites in A15) generally produce distinct
families.

The algorithm for a single particle operates in four stages.
\textbf{Stage~1} constructs a supercell,
computes its Delaunay triangulation, identifies the Voronoi vertices
of $p_0$, and classifies each as generic or degenerate
(Section~\ref{sec:step1}).
\textbf{Stage~2} enumerates, at each
degenerate vertex, the \emph{resolution types}: all ways the vertex
can split under perturbation into generic vertices, each realizing
some subset of the partial neighbors at that vertex
(Section~\ref{sec:step2}).
\textbf{Stage~3} combines resolution types
across vertices, discarding incompatible combinations and using
crystallographic site symmetry to reduce the
search to one representative per equivalence class
(Section~\ref{sec:step3}).
\textbf{Stage~4} reconstructs the Voronoi
cell from each combination and identifies its combinatorial type via
its Weinberg vector (Section~\ref{sec:step4}).
The procedure is summarized in Algorithm~\ref{alg:main}.
Two further sections elaborate on the enumeration:
Section~\ref{sec:extended} on the enumeration modes, and
Section~\ref{sec:symmetry} on the symmetry reduction applied in
Stage~3.

\begin{algorithm}[htbp]
\caption{Enumerate Voronoi cell types under perturbation}
\label{alg:main}
\begin{algorithmic}[1]
\Require Crystal $(\mathbf{a}_1,\mathbf{a}_2,\mathbf{a}_3;\{p_i\})$,
         basis particle $p_0$
\Ensure Set $\mathcal{W}$ of Weinberg vectors of cell types accessible
        to $p_0$
\Statex The Voronoi analysis of Section~\ref{sec:step1} produces:
        $\mathcal{V}$, the vertices of the Voronoi cell of $p_0$,
        partitioned into a set $\mathcal{V}_g$ of generic vertices ($k=4$), 
        and a set $\mathcal{V}_d$ of $n$ degenerate vertices ($k>4$).
\Statex Each generic vertex is dual to a single Delaunay tetrahedron
        incident to $p_0$; write $\mathcal{T}_{\mathrm{gen}}$ for the
        set of these tetrahedra.
\Statex
\For{$i = 1, \ldots, n$}
  \State $\mathcal{R}_i \gets \Call{EnumerateResolutions}{v_i}$
         \Comment{Section~\ref{sec:step2}}
\EndFor
\State $\mathcal{C} \gets
        \Call{OrderlyEnumerate}{\mathcal{R}_1, \ldots, \mathcal{R}_n}$
\Statex \Comment{pairwise-compatible, symmetry-canonical resolution combinations (Algorithm~\ref{alg:orderly}, Supplementary Section~\ref{app:impl})}
\State $\mathcal{W} \gets \emptyset$
\For{each $c = (r_1, \ldots, r_n) \in \mathcal{C}$}
  \State $S \gets \mathcal{T}_{\mathrm{gen}} \cup
          \bigcup_{i=1}^n r_i$
         \Comment{$r_i$ identified with its set of tetrahedra}
  \State construct the Voronoi cell of $p_0$ from $S$
         \Comment{Section~\ref{sec:step4}}
  \If{some face's adjacency graph is not a single cycle}
    \State \textbf{continue}
           \Comment{face-construction failure; \S\ref{sec:step4}}
  \EndIf
  \State compute the Weinberg vector $w$ of the cell
  \If{the Weinberg trace fails}
    \State \textbf{continue}
           \Comment{global-assembly failure; \S\ref{sec:step4}}
  \EndIf
  \State $\mathcal{W} \gets \mathcal{W} \cup \{w\}$
\EndFor
\State \Return $\mathcal{W}$
\end{algorithmic}
\end{algorithm}

\subsection{Voronoi analysis}
\label{sec:step1}

Given a crystal with lattice vectors $\mathbf{a}_1, \mathbf{a}_2, \mathbf{a}_3$
and a set of basis particles, we construct a supercell containing, for each
basis particle $p_i$, all periodic images $p_i +
n_1\mathbf{a}_1 + n_2\mathbf{a}_2 + n_3\mathbf{a}_3$ for $n_j \in
\{-N_j, \ldots, N_j\}$.  The image count $N_j$ must be large enough that the
Voronoi cell of $p_0$ is unaffected by the supercell boundary; Supplementary
Section~\ref{app:impl} gives a sufficient bound in terms of the maximal
circumradius of the Delaunay tetrahedra incident to $p_0$.  We use $N_j = 3$
in each direction, which satisfies the bound for every crystal structure
considered in this paper.

We compute a Delaunay triangulation of the supercell using
SciPy's~\cite{scipy2020} Qhull~\cite{barber96} interface, extract
the star of the central particle $p_0$ (the tetrahedra incident to
$p_0$), and compute the circumcenter of each star tetrahedron.  At
a degenerate vertex the $k \geq 5$ equidistant particles can be
triangulated in many ways, and Qhull, like any standard Delaunay
code, returns just one of them, chosen by floating-point rounding.
We assume nothing about which one.  Instead, degeneracy detection
rests on a geometric fact: all tetrahedra inscribed in a common
sphere have that sphere's center as their circumcenter.  Whichever
triangulation Qhull returns, the tetrahedra covering the degenerate
region therefore have circumcenters that coincide, up to
floating-point error, at the true Voronoi vertex, just as the two
triangles obtained by drawing either diagonal across a square share
the square's center as their circumcenter.  We identify the vertex
by clustering these near-coincident circumcenters.  Because the
scatter is set by floating-point error rather than by any fixed
scale, we cluster by \emph{adaptive gap detection}: the pairwise
distances between circumcenters are sorted, and a large gap in the
sorted distances separates coincident centers from distinct ones.
The numerical tolerances are given in Supplementary
Section~\ref{app:impl}.

For each cluster, we sort the distances of all supercell particles to
the cluster center and again use gap detection to identify the
equidistant set: particles are included until a gap much larger than 
the within-cluster spread is found, by the same criterion as above.  
This yields the full set of
equidistant particles and hence the multiplicity $k$ and point
configuration $Q$ of the corresponding Voronoi vertex.  All
subsequent per-vertex analysis (Section~\ref{sec:step2}) operates
on $Q$ alone; the particular tetrahedra Qhull returned play no
further role there.  Vertices
with $k = 4$ are generic; those with $k > 4$ are
degenerate.

The tetrahedra from generic vertices form the \emph{fixed} part of
the Delaunay star: they are unchanged under any sufficiently small
perturbation, since the corresponding circumspheres have exactly
four particles on their boundary.

\subsection{Per-vertex resolution enumeration}
\label{sec:step2}

Stage~2 implements the \textsc{EnumerateResolutions} procedure of
Algorithm~\ref{alg:main}.  At each degenerate vertex with point
configuration $Q = \{q_0, q_1, \ldots, q_{k-1}\}$ (where $q_0 = p_0$ is the
central particle), we enumerate all resolution types by a deterministic
combinatorial search over the triangulations of the convex hull
of $Q$.  Each distinct star of $q_0$ across these triangulations
defines a resolution type, characterized by its tetrahedra,
neighbor set, and face valences contributed to the cell.

The $k$ equidistant particles are co-spherical, and for the small
co-spherical configurations enumerated here ($k \leq 8$) every face-matching
triangulation of the convex hull is realizable as the Delaunay
triangulation of some small perturbation~\cite{loera}, as the sampling
in Supplementary Section~\ref{app:primary_validation} confirms for the
vertices tested there.  The enumeration is therefore purely
combinatorial: we seek all subsets of the $\binom{k}{4}$ candidate
tetrahedra (the 4-element subsets of $Q$) that satisfy the
face-matching constraints of a valid triangulation, namely that
each boundary face of the convex hull appears in exactly one
tetrahedron and each internal triangular face appears in exactly
zero or exactly two tetrahedra.

The search is a backtracking algorithm:
\begin{enumerate}
  \item Compute the convex hull of $Q$ to obtain the boundary face
    set $\mathcal{B}$.
  \item Generate all $\binom{k}{4}$ candidate tetrahedra.
  \item Search for subsets satisfying the face-matching
    constraints.
\end{enumerate}
The backtracking maintains a set of \emph{open faces}: uncovered boundary
faces and exposed internal faces (those appearing in exactly one tetrahedron so
far).  At each step it selects an open face and tries every candidate
tetrahedron containing it.  A branch is pruned if any face count would exceed
its maximum (1 for boundary, 2 for internal) or if the accumulated tetrahedron
volume would exceed the convex hull volume.  When no open faces remain, a valid
triangulation has been found.  

The volume condition is essential
rather than an optimization: with zero-volume tetrahedra among the
candidates, the face constraints alone also admit complexes that
over-cover the hull.  A flat tetrahedron has all four faces in one
plane, so it can be attached to a face without adding volume, and
complexes built from such tetrahedra can fold back and cover parts
of the hull more than once while satisfying every face count.  At
the simple-cubic $k = 8$ vertex, face matching alone accepts $3{,}002{,}678$
closed complexes, of which only $11{,}648$ respect the volume
bound; at the smaller configurations in this paper the two
conditions agree.

A key subtlety is that the candidate set must include
\emph{degenerate} (zero-volume) tetrahedra formed by four coplanar
particles.  Such tetrahedra are flat in the ideal configuration but
become non-degenerate under any perturbation that breaks the
coplanarity, so they represent physically valid resolutions.  For
example, the octahedral configuration at each FCC degenerate
vertex has three squares of four coplanar particles, and 6 of its 9
valid triangulations use exactly one such degenerate tetrahedron each.

The algorithm is exact and deterministic: it finds every valid
triangulation.  Every structure whose enumeration is reported here has $k \leq 8$ at
all of its vertices, where the search at a vertex takes seconds at
most.  The implementation applies the same search up to $k = 10$ and,
for larger $k$, falls back to a breadth-first search of the flip graph
of regular triangulations of a slightly perturbed configuration; no
reported result depends on that fallback.

When the near-equidistant gap threshold (Section~\ref{sec:near_equidistant})
is active, the equidistant set at a vertex is partitioned into a
\emph{core} $C$ (the particles identified by adaptive gap detection;
always at least four) and a possibly-empty \emph{borderline} set
$B$ (particles within $\tau \cdot r$ of the core distance, where $r$
is the vertex-to-defining-particle distance).  Writing
$E = C \cup B$ for the full near-equidistant set and
$E^\ast = E \setminus \{p_0\}$ for the near-equidistant neighbors
of $p_0$, the per-vertex resolution set at a vertex with at least
one borderline particle is the union of the resolutions produced by
every subset $S \subseteq E^\ast$ of three or more neighbors:
\begin{equation*}
  \mathcal{R}_v \;=\; \bigcup_{\substack{S \subseteq E^\ast \\ |S| \geq 3}}
  \bigl\{\text{stars of $p_0$ in regular triangulations of
  $\operatorname{conv}(\{p_0\} \cup S)$}\bigr\};
\end{equation*}
a vertex with no borderline particle uses the ordinary single
enumeration described above.  Ranging over all subsets of
$E^\ast$, rather than varying only the borderline particles, serves two
purposes.  First, it makes the
construction monotone in $\tau$: a larger threshold can only
enlarge $E^\ast$, and hence the family of subsets, so the filter
at threshold $\tau$ contains the filter at every smaller threshold
$\tau' < \tau$.  Second, it makes the result insensitive to
exactly where adaptive gap detection draws the core/borderline
boundary; that partition determines only
\emph{whether} subset enumeration is triggered at a vertex, not
which subsets it then ranges over. In practice the implementation
caps $|E^\ast|$ at seven per vertex, a guard against combinatorial
explosion in pathological configurations.  Each triangulation
therefore runs on at most eight co-spherical points ($p_0$ plus up
to seven neighbors), and within this cap the union has fewer
than $2^{|E^\ast|} \leq 128$ terms.  No reported
result is affected by this cap: every structure whose enumeration
completes has at most eight near-equidistant particles at any vertex,
the maximum the cap permits.

\subsection{Combining resolutions across vertices}
\label{sec:step3}

Given $n$ degenerate vertices with $m_1, m_2, \ldots, m_n$
resolution types respectively, we enumerate all
$\prod_{i=1}^n m_i$ combinations.  For each combination, we
construct the global Delaunay star by combining the fixed generic
tetrahedra with the local star from each chosen resolution, with
local indices mapped to global supercell indices.

Whether resolutions at different vertices are independent depends
on the geometry of the central cell.  Two degenerate vertices are
\emph{adjacent} if they share at least one neighbor besides the
central particle $p_0$, a broader notion than sharing a Voronoi edge.
This is a conservative superset of the
pairs that can actually conflict.  Two vertices joined by a generic
Voronoi edge share two neighbors $a, b$ besides $p_0$, and the face
$\{p_0, a, b\}$ lies in exactly one tetrahedron of every resolution of
each, so their resolutions are always compatible.  A conflict requires
a degenerate Voronoi edge, across which the two vertices share three or
more neighbors (Section~\ref{sec:coupling}).  We write
$\mathrm{adj}(i)$ for the set of indices $j$ such that $v_j$ is
adjacent to $v_i$.  The compatibility test below rejects no pair of resolutions in FCC,
HCP, the fluorite Ca site, or the $\omega$-1a site
(Figure~\ref{fig:voronoi_cells}(c)), so all combinations are valid
there; in FCC no two degenerate vertices even share a Voronoi edge,
while at the other three some do, but none of those edges is degenerate.  At the $\omega$-2d site (Figure~\ref{fig:voronoi_cells}(d)),
in contrast, six pairs of degenerate vertices are joined by degenerate
edges and their resolutions are \emph{coupled}: the choices at the two
ends of such an edge may be incompatible, and most combinations produce
malformed cells.

For coupled crystals, validity pruning is essential.  For each pair of adjacent
vertices $(v_i, v_j)$ and each pair of resolutions $(r_i, r_j)$, we precompute
a compatibility indicator $\mathrm{compat}_{ij}(r_i, r_j) \in \{0,1\}$
recording whether the combined tetrahedra on their shared faces produce a valid
partial structure: for each shared neighbor $a$, we collect all tetrahedra
containing the triangle $\{p_0, a, \cdot\}$ contributed by the generic vertices
and by resolutions $r_i$ and $r_j$, and check that no triangle $\{p_0, a, b\}$
in this partial complex is contained in more than two tetrahedra.  A face
appearing in three or more tetrahedra is a permanent failure---additional
resolutions can only add more tetrahedra to its count, so any extension is also
invalid. This pairwise compatibility test is integrated into the orderly
depth-first search (DFS)
(Section~\ref{sec:symmetry}) as a table lookup: after assigning a resolution,
the algorithm tests it against all previously assigned adjacent vertices via
the precomputed table, pruning incompatible subtrees.

The pairwise test detects only one of the two ways in which a
combination can fail.  In a valid cell every face $\{p_0, a, b\}$ lies
in exactly two tetrahedra.  The test rejects a face lying in three or
more, but not a face lying in only one, since tetrahedra contributed by
vertices not yet assigned could still complete it.  Combinations with
such a face are caught when the cell is constructed
(Section~\ref{sec:step4}): the Voronoi face dual to $(p_0, a)$ fails to
close and the combination is discarded, so no
special-case logic is
required for coupled vertices.  Such late rejections are not rare: at
the $\omega$-2d site, $61{,}693$ of the $83{,}785$ canonical
combinations that pass pairwise pruning are rejected in this way
(Section~\ref{sec:symmetry}), while at sites with no coupled vertices
(FCC, HCP, the fluorite Ca site, and $\omega$-1a) no combination is
rejected late.

The combinations are enumerated by a DFS over the
combination space that generates only canonical representatives
under the site symmetry (Section~\ref{sec:symmetry}), using
constant memory per combination; when no symmetry is available it
reduces to a straightforward DFS with the validity pruning above.

\subsection{Cell construction and identification}
\label{sec:step4}

From a global Delaunay star $S$, the face structure of the Voronoi
cell is recovered via Delaunay--Voronoi duality.  For each
neighbor $a$ of $p_0$, the Voronoi face $F_a$ is dual to the
Delaunay edge $(p_0, a)$, and we construct it as follows:
\begin{enumerate}
  \item The Voronoi vertices of $F_a$ are the circumcenters of
    tetrahedra in $S$ containing both $p_0$ and $a$.
  \item Two such circumcenters are adjacent on $F_a$ if their
    corresponding tetrahedra share a face
    $\{p_0, a, b\}$ for some third particle $b$.
  \item The adjacency graph of each face is a cycle (since the
    face is convex); tracing this cycle yields the
    ordered vertex sequence of the face.
\end{enumerate}
If the adjacency graph of a face is not a single cycle, the cell
is malformed and the combination is discarded.  This is what happens
to every invalid combination that survives pairwise pruning
(Section~\ref{sec:step3}).

When an unresolved degenerate vertex is present
(Section~\ref{sec:extended}), all tetrahedra from that vertex are
assigned the same group identifier.  After the face cycle is
traced at the tetrahedron level, consecutive entries with the same
group identifier are collapsed into a single Voronoi vertex,
implementing the merging of multiple generic vertices back into a
single higher-valence vertex.  Faces that reduce to fewer than
three vertices after collapsing are discarded.

The faces are oriented consistently by a breadth-first search over
the face adjacency graph: an initial face is oriented arbitrarily,
and each subsequent face adjacent across a shared edge is oriented
so that the shared edge is traversed in opposite directions by the
two faces, as required for a consistently oriented closed surface.
The Weinberg vector is then computed from the oriented face
structure, providing a unique identifier for the combinatorial
type of the cell.  Should a malformed cell pass the face check above,
the Weinberg traversal would encounter duplicate or missing directed
edges, and the combination would be discarded.

\subsection{Enumeration modes}
\label{sec:extended}

The algorithm supports three nested enumeration modes:

\paragraph{Primary.}
At each degenerate vertex, only primary resolutions
(Section~\ref{sec:resolution}) are admitted.  This recovers the
family of cell types reached with positive frequency under
arbitrarily small generic perturbations~\cite{lazar2015topological}.
When the gap-threshold subset enumeration is active, primary
resolutions are filtered separately for each subset of the
near-equidistant set.

\paragraph{Default.}
All fully resolved types: each degenerate vertex is resolved into generic
vertices, with resolutions admitted whether or not they contain
zero-volume tetrahedra.
Includes the union over subsets of the near-equidistant set
when the gap threshold is active.

\paragraph{All.}
Default plus \emph{partially resolved} types, in which a degenerate vertex may
be left unresolved (preserving its high valence in the Voronoi cell). At each
degenerate vertex the resolution set is augmented with a single
\emph{unresolved option}, giving $\prod_{i=1}^{n}(m_i + 1)$ total combinations.
For FCC this increases from $7^6 = 117{,}649$ default combinations to $8^6 =
262{,}144$.  The Weinberg vector remains a valid identifier for these partially
resolved cells, because the resulting polyhedra are still convex, and the
surface graph of any convex polyhedron meets the conditions Weinberg's
algorithm requires. When the gap threshold is active, the unresolved option is
constructed from the resolution in which every near-equidistant particle
participates, preserving the inclusion \emph{default}~$\subseteq$~\emph{all}.

The three modes form a nested chain: \emph{primary}~$\subseteq$~\emph{default}~$\subseteq$~\emph{all}.
Counts for representative structures are listed in
Table~\ref{tab:modes}.

\paragraph{Choosing a mode.}
\emph{Default} is appropriate for analyzing finite-temperature
simulations, since both primary and secondary resolutions are realized
with nonzero frequency at any temperature 
$T>0$~\cite{lazar2015topological}.  \emph{All} is a more robust choice whenever an analyzed
cell may contain an unresolved degenerate vertex, either because
the structure carries an exact degeneracy (idealized crystals, or structures quenched to zero temperature), or because
thermal displacements are small enough that finite numerical precision
leaves a near-degenerate vertex unresolved, which can occur in
low-temperature data.  Because the modes are nested
(\emph{primary}~$\subseteq$~\emph{default}~$\subseteq$~\emph{all}), selecting \emph{all}
can only enlarge the set of matched cells. Its cost
is a slightly more permissive filter and a more expensive, sometimes
intractable enumeration. \emph{Primary} is appropriate for theoretical analyses of
the $T\to 0$ limit, or when \emph{default} enumeration is computationally
intractable.  The simple cubic lattice illustrates the latter: each
of its 8 degenerate vertices has $k=8$ equidistant particles and 544
resolutions in the \emph{default} mode, putting \emph{default} enumeration far beyond
reach; the \emph{primary} mode admits 256 resolutions per vertex
(Section~\ref{sec:resolution}) and is still intractable, yielding
$256^8 \approx 10^{19}$ combinations per cell.

\begin{table}[h]
\centering
\caption{Distinct Voronoi cell types in the three enumeration modes for
  representative structures.  \emph{Primary} and \emph{default} coincide for
  sites whose degenerate vertices admit no secondary resolution
  (Section~\ref{sec:resolution}).  Entries marked --- denote enumerations too large to complete in the allotted time budget;
  the simple cubic lattice is intractable in every mode
  (Section~\ref{sec:complexity}).
  The listed structures have no
  borderline atoms at the recommended gap threshold $\tau = 0.005$, so
  the subset-enumeration extension (Section~\ref{sec:step2}) does
  not change their type counts; the counts for structures with
  borderline atoms in the threshold-$\tau = 0.005$ library appear
  in Supplementary Section~\ref{app:library}.}
\label{tab:modes}
\begin{tabular}{lrrr}
\toprule
Structure / site & Primary & Default & All \\
\midrule
BCC                  & 1   & 1       & 1       \\
FCC                  & 44  & 2{,}815 & 6{,}294 \\
HCP                  & 66  & 9{,}556 & 21{,}611 \\
A15 (2a + 6c)        & 2   & 2       & 2       \\
Fluorite Ca          & 6   & 6       & 37      \\
$\omega$-1a          & 8   & 8       & 52      \\
$\omega$-2d          & 382 & 382     & 5{,}460 \\
SC                   &---    &---    &---\\
\bottomrule
\end{tabular}
\end{table}

\paragraph{Intermediate vertex subdivisions.}
The \emph{all} mode admits each degenerate vertex in only two limiting
states: fully resolved (any regular triangulation) or fully unresolved
(the vertex kept \emph{whole}).  It
does not enumerate \emph{intermediate} regular subdivisions, in which a
vertex resolves only partly, leaving a surviving lower-valence
degenerate vertex.  For the central particle, such an intermediate
subdivision enlarges the type list only at vertices whose surviving
cell rests on a \emph{boundary} face of the local point
configuration; where the relevant coplanar face is interior, keeping
it whole merely reproduces the fully unresolved cell already
enumerated.  The common vertices add no such type: the $k=5$
square-pyramidal sites admit no proper intermediate subdivision at
all, and at the $k=6$ octahedral sites the relevant square---formed
by the central particle's full neighbors---is interior.
Only boundary-face vertices, the $k=8$ cubic configuration and the
$k=6$ triangular prism, admit type-adding intermediate subdivisions.
Of the $1{,}025$ library structures whose enumeration completes, $1{,}018$
contain only interior-face vertices, at which no intermediate
subdivision adds a type.  The remaining seven contain a prismatic or cubic vertex and
are correspondingly incomplete in the \emph{all} mode: their filters
omit the intermediate-subdivision topologies at that vertex, which
would require enumerating the full family of regular subdivisions
there rather than only the triangulations and single unresolved state.
This shortfall is confined to the \emph{all} mode, however: for
finite-temperature identification one uses the \emph{default} filter,
which, for all $1{,}025$ structures alike (the seven included), enumerates
every type reachable under sufficiently small perturbations.  These filters
therefore remain fully usable at finite temperature.  Sampled data at
very small perturbation amplitudes can also contain cells in which two
nearly coincident vertices are reported as one; these too can lie
outside the \emph{all} filter (Supplementary
Section~\ref{app:nearequidistant}).

\subsection{Symmetry reduction}
\label{sec:symmetry}

Many of the $\prod m_i$ resolution combinations enumerated in
Section~\ref{sec:step3} are related by the symmetry of the central
particle.  This symmetry, the rotations and reflections of the
crystal's space group that fix~$p_0$, is its \emph{site symmetry group}.
Combinations related by such a symmetry necessarily produce identical
Voronoi cells, and hence identical Weinberg vectors, so it suffices
to enumerate one representative from each set of symmetry-related
combinations.

In practice, the site symmetry group is obtained automatically from the detected
space group of the structure~\cite{pymatgen}, as the subgroup of
its operations that fix~$p_0$; if no space group can be determined,
the enumeration proceeds without symmetry reduction, changing the
runtime but not the resulting family of cell types.

Rather than testing every combination for canonicality, the
implementation uses a depth-first \emph{orderly generation}
search~\cite{read1978}:
combinations are built one vertex at a time, any partial assignment
that can already be shown non-canonical is pruned, and each canonical
combination is yielded together with its multiplicity.  How the site
symmetry acts on combinations, the canonical-form definition, and the
full pseudocode (Algorithm~\ref{alg:orderly}) are given in
Supplementary Section~\ref{app:impl}.

\paragraph{Validity pruning.}
When degenerate vertices are joined by degenerate Voronoi edges,
their resolutions are coupled and many combinations are invalid;
the DFS prunes them using the precomputed pairwise compatibility
test of Section~\ref{sec:step3}.  For the $\omega$-2d site
(Section~\ref{sec:coupling}), pairwise pruning combined with
symmetry reduces the $4^{12} = 16{,}777{,}216$ nominal combinations
to $83{,}785$ canonical combinations, a $200\times$ reduction.
Because the pairwise test detects only one of the two ways a
combination can fail (Section~\ref{sec:step3}), some of these still
correspond to invalid cells and are rejected at cell construction:
$61{,}693$ of the $83{,}785$ are discarded there, and the
remaining $22{,}092$ valid canonical combinations expand under
symmetry to the $262{,}144$ valid combinations and resolve to the 382
distinct $\omega$-2d cell types (Table~\ref{tab:omega2d}, Supplementary
Section~\ref{app:results}).

The reductions are substantial.  For FCC, the site symmetry
group~$O_h$ (order~48) reduces the enumeration from 117{,}649
combinations to 2{,}821 canonical representatives, a 42-fold
reduction; among these, 2{,}815 produce distinct Weinberg types, with the discrepancy
of~6 reflecting pairs of symmetry classes whose Voronoi cells are
combinatorially equivalent despite not being related by the site
symmetry.  The reduction falls short of the group order of 48
because combinations left fixed by some symmetry operation have
orbits smaller than the full group.  For HCP, the site symmetry group~$D_{3h}$ (order~12) yields
a more modest reduction: the same $117{,}649$
nominal combinations reduce to $9{,}940$ canonical representatives,
of which $9{,}556$ produce distinct Weinberg types---a nearly
12-fold reduction.  The symmetry reduction is also compatible with the \emph{all}
enumeration mode of Section~\ref{sec:extended}.

\clearpage
\renewcommand{\thesection}{S\arabic{section}}
\setcounter{section}{0}
\renewcommand{\theHsection}{S\arabic{section}}
\renewcommand{\theHsubsection}{S\arabic{section}.\arabic{subsection}}

\section*{Supplementary Material}

\section{Implementation and numerical details}
\label{app:impl}

This section collects implementation-level material supporting the
algorithm of Appendix~\ref{sec:algorithm}: the supercell-sufficiency
bound mentioned in Section~\ref{sec:step1}, the numerical tolerances of the
circumcenter clustering, and the canonical-form machinery of the
symmetry reduction (Section~\ref{sec:symmetry}).

\subsection{Supercell sufficiency}
\label{app:impl_supercell}

The Voronoi cell of $p_0$ depends only on particles $q$ whose
perpendicular bisector with $p_0$ intersects the
cell---which requires $\|q - p_0\| \le 2R_{\max}$, where $R_{\max}$ is
the largest circumradius over Delaunay tetrahedra incident to
$p_0$.  It therefore suffices to choose
\[
  N_j \;=\; \left\lceil \frac{2 R_{\max}}{h_j} \right\rceil,
  \qquad
  h_j \;=\; \frac{V}{\|\mathbf{a}_k \times \mathbf{a}_\ell\|},
\]
where $h_j$ is the perpendicular distance between the two faces
of the unit cell spanned by $\mathbf{a}_k$ and $\mathbf{a}_\ell$,
$V$ is the unit-cell volume, and $(j,k,\ell)$ is a cyclic
permutation of $(1,2,3)$.
Note that $h_j \le \|\mathbf{a}_j\|$, with equality only when
$\mathbf{a}_j$ is perpendicular to the other two lattice vectors,
so using $\|\mathbf{a}_j\|$ in place of $h_j$ gives too few images for
oblique cells.  $R_{\max}$ can be bounded above by the covering
radius of the crystal, or estimated from an initial Delaunay pass.
Related conditions for finite periodic Delaunay computations are
discussed by Caroli and Teillaud~\cite{caroli2009periodic}.  For highly oblique
or sparsely populated cells, image counts larger than the $N_j = 3$ used
here may be required.

\subsection{Circumcenter-clustering tolerances}
\label{app:impl_clustering}

In the adaptive gap detection of Section~\ref{sec:step1}, the pairwise
distances between circumcenters are sorted, and the first jump that
exceeds 100 times the spread of the distances below it separates
coincident centers from distinct ones.  The spread is taken to be at
least $10^{-8}$, and the distances below the jump must be smaller than
$10^{-3}$, both in the length units of the input.  For clean,
symmetrized coordinates the spread itself is close to machine epsilon
(about $10^{-16}$) and this floor sets the threshold; for noisy inputs
the spread is orders of magnitude larger and the threshold grows with
it.  When no such jump exists, all centers are treated as coincident if
every distance is below $10^{-4}$, and otherwise only those within
$10^{-8}$ of one another.  The equidistant set of each vertex is
identified by the same jump criterion: the particle-to-vertex distances
are sorted, the four nearest are always included, and particles are
added until the next jump exceeds 100 times the spread so far, again
taken to be at least $10^{-8}$.

\subsection{Orderly generation of canonical combinations}
\label{app:impl_orderly}

The site symmetry of the central particle relates many of the
$\prod m_i$ resolution combinations of Section~\ref{sec:step3} to one
another: each symmetry
permutes the degenerate vertices among themselves and, because it
also carries the particles around each vertex onto those around its
image, correspondingly relabels the resolution chosen at each vertex.
Two combinations related in this way produce combinatorially
identical Voronoi cells, so only one representative of each symmetry
class need be classified.  To choose representatives consistently, we
fix an arbitrary order on the resolutions at each vertex (for
instance, the lexicographic order on their sorted tetrahedra); this
orders the combinations lexicographically, and the \emph{canonical}
combination is the smallest in its class.  The choice of order changes
which member is canonical but not the grouping into classes, and hence
not the final list of cell types.

Rather than testing all $\prod m_i$ combinations for canonicality, the
depth-first orderly generation builds combinations one vertex at a
time, pruning any partial assignment that can already be shown
non-canonical.  At depth~$k$, the partial combination
$(r_1, \ldots, r_k)$ has been assigned.  We maintain the set of
symmetries that could still map it to a lexicographically smaller
combination---the ones capable of showing it is non-canonical.  After
assigning $r_k$, the image of the partial combination under each such
symmetry~$g$ is compared with the combination itself, position by
position from the first: the comparison stops at the first position
where the two differ, at the first position whose image entry is not
yet determined (its value comes from a vertex not yet assigned), or
after position~$k$, whichever comes first.  Three outcomes are
possible:
\begin{itemize}
  \item \textbf{Smaller:} the comparison stops at a difference, with
    the image smaller there.  All positions up to the difference are
    already fixed, so every completion inherits it:
    none can be canonical, and the subtree is pruned.
  \item \textbf{Larger:} the comparison stops at a difference, with
    the image larger there.  Every completion inherits this difference
    as well, so $g$ can never produce a smaller image and is dropped
    for the rest of this subtree.
  \item \textbf{Undecided:} no difference is found.  The comparison
    still depends on positions beyond~$k$ or on image entries not yet
    determined, so $g$ is retained and re-checked at a later depth.
\end{itemize}
Once all $n$ vertices have been assigned, every comparison resolves,
and the surviving symmetries are those that fix the
combination; the group order divided by their number gives its
multiplicity (the size of its symmetry class). This procedure, together with the
validity pruning of Section~\ref{sec:step3}, is summarized in
Algorithm~\ref{alg:orderly}.

\begin{algorithm}[H]
\caption{Orderly DFS over combinations with validity pruning}
\label{alg:orderly}
\begin{algorithmic}[1]
\Require Totally ordered resolution sets $\mathcal{R}_1,\ldots,\mathcal{R}_n$;
  site symmetry group $G_{p_0}$, acting on combinations by permuting
  the vertices and relabeling their resolutions; adjacency $\mathrm{adj}$ and
  compatibility $\mathrm{compat}_{ij}$ (Section~\ref{sec:step3}).
\Ensure Each pairwise-compatible canonical combination is yielded with its multiplicity.
\Statex
\Procedure{OrderlyEnumerate}{$\mathcal{R}_1, \ldots, \mathcal{R}_n$}
  \State $c \gets$ scratch tuple of length $n$
  \State \Call{DFS}{$1,\; G_{p_0}$}
\EndProcedure
\Statex
\Procedure{DFS}{$k,\; A$}
  \Comment{$A$: elements of $G_{p_0}$ not yet shown to enlarge $c$}
  \If{$k > n$}
    \State \textbf{yield} $\bigl(c,\; |G_{p_0}|/|A|\bigr)$; \Return
       \Comment{$A$ is the set of symmetries fixing $c$}
  \EndIf
  \For{$r_k \in \mathcal{R}_k$ in increasing order}
    \State $c[k] \gets r_k$ \Comment{place the candidate in position $k$}
    \State for each $g \in A$, compare $g \cdot c$ with $c$ position
           by position, stopping at the first difference or the first
           entry of $g \cdot c$ not yet determined (rules above)
    \If{$g \cdot c$ is lex-smaller for some $g \in A$}
       \State \textbf{continue} \Comment{prefix non-canonical}
    \EndIf
    \If{$\mathrm{compat}_{kj}(c[k], c[j]) = 0$ for some
        $j \in \mathrm{adj}(k),\; j < k$}
       \State \textbf{continue} \Comment{validity pruning}
    \EndIf
    \State $A' \gets \{\, g \in A : g \cdot c \text{ is not lex-larger} \,\}$
    \State \Call{DFS}{$k+1,\; A'$}
  \EndFor
\EndProcedure
\end{algorithmic}
\end{algorithm}

\section{Library of Voronoi cell families}
\label{app:library}

We apply the enumeration algorithm to a broad sample of crystal
structures from the Materials Project~\cite{horton2025accelerated}.
A random sample of $1{,}300$ stable structures with 1--16 atoms per
primitive cell is downloaded via the
Materials Project API~\cite{ong2015materials}.  Each structure is
enumerated with a timeout of 300\,s at gap threshold $\tau = 0.005$
and a space-group detection tolerance (\texttt{symprec}) of
$10^{-4}$, in three nested enumeration modes
(Section~\ref{sec:extended}): $1{,}088$ complete in the \emph{primary}
mode, $1{,}054$ in the \emph{default} mode, and $1{,}025$ in the
\emph{all} mode; every structure that completes in a harder mode also
completes in the easier ones, as expected from their per-vertex cost
ordering.  The structures that do not complete include all ten
candidates with a degenerate vertex of more than 8 equidistant atoms.  The $1{,}054$ structures
completing the \emph{default} mode form the library analyzed below;
statistics are reported for that mode.  The structures that do not
complete are not concentrated in any crystal system: in each of the
seven systems, between 12\% and 23\% of the candidates fail to complete.

\subsection{Summary statistics}

The 1{,}054 structures span all seven crystal systems and $91$ distinct space
groups, producing 77{,}644 unique Weinberg vectors across 207{,}750 total
vector--structure associations (the same Weinberg vector may appear in
multiple structures).  Table~\ref{tab:library_summary} summarizes the
distribution by crystal system.

\begin{table}[h]
\centering
\caption{Library summary by crystal system.  For each crystal system,
  the number of structures enumerated, the mean and median number of
  distinct Voronoi cell types per structure (counted per Wyckoff
  site), and the fraction of
  structures with multiple types at some Wyckoff site
  are shown. Counts are for the \emph{default} enumeration mode at
  threshold $\tau = 0.005$.  The hexagonal, cubic and trigonal means are
  driven by a small number of outliers; their medians are $4$, $1$
  and $10$.}
\label{tab:library_summary}
\begin{tabular}{lrrrr}
\toprule
Crystal system & Count & Mean types & Median types & \% multi-type \\
\midrule
Triclinic    & 31   & 7.5    & 5  & 29.0 \\
Monoclinic   & 117  & 20.8   & 5  & 39.3 \\
Orthorhombic & 233  & 26.7   & 3  & 28.8 \\
Tetragonal   & 188  & 13.6   & 3  & 26.6 \\
Trigonal     & 96   & 232.9  & 10 & 54.2 \\
Hexagonal    & 115  & 1067.6 & 4  & 34.8 \\
Cubic        & 274  & 191.0  & 1  & 8.0  \\
\midrule
All          & 1{,}054 & 198.2  & 3  & 27.1 \\
\bottomrule
\end{tabular}
\end{table}

Several features of the data are noteworthy.
\begin{itemize}
  \item \textbf{Most structures are topologically stable.}  In $72.9\%$ of structures
    ($768$ of $1{,}054$), every Wyckoff site has a single Voronoi cell
    topology.  At the site level, $83.2\%$ of the $4{,}119$ Wyckoff
    sites in the library have exactly one type.
  \item \textbf{The distribution is highly skewed.}  The median
    structure has $3$ cell types, but a
    long tail of high-symmetry hexagonal and trigonal structures
    pushes the mean far above the median; the maximum is $19{,}775$
    (NaSnN), whose degenerate vertices admit many secondary
    resolutions.
  \item \textbf{Higher-symmetry systems
    have a longer tail.}  Trigonal and hexagonal mean type counts
    are inflated by a small number of structures whose degenerate
    vertices admit many secondary resolutions, as in HCP; their
    medians ($10$ and $4$) remain far below their means.  Cubic structures are the least likely in this library
    to have a multi-type Wyckoff site.  This is partly a selection
    effect: cubic structures with degenerate vertices, such as rock
    salt and simple cubic, tend to have high multiplicities and mostly
    do not complete, so those that remain are predominantly
    topologically stable ones such as BCC and CsCl.
  \item \textbf{Triclinic structures
    pick up degeneracies via borderline atoms.}  Triclinic crystals
    have no symmetry-forced degenerate vertices, yet $29\%$ of
    triclinic structures show multi-type Wyckoff sites at
    $\tau = 0.005$, produced not by symmetry but by near-equidistant
    atoms, in most cases caught by the gap-threshold extension.
\end{itemize}

\subsection{Weinberg vector sharing}

Of the 77{,}644 unique Weinberg vectors in the library, $54{,}725$ ($70.5\%$) appear in exactly one structure, while $22{,}919$ ($29.5\%$)
are shared between two or more structures.  Most of the
structure-specific vectors belong to three structures with very large
families.  The most widely shared
Weinberg vector, the truncated octahedron of BCC, appears in $338$ of
the $1{,}054$ structures ($32.1\%$).  Most of the sharing is between
structures whose filters are identical, as with isostructural
compounds: the $1{,}054$ structures have $380$ distinct filters.  Of
the $554{,}931$ pairs of library structures, $83\%$ have no Weinberg
vector in common, $5\%$ have identical filters, and $12\%$ share some
vectors but not all.

\section{Near-equidistant neighbor validation}
\label{app:nearequidistant}

We validate the analytical enumeration against stochastic \VoroTop
sampling on the library enumerated at gap threshold $\tau = 0.005$
(Section~\ref{app:library}).  For each structure that completed
enumeration, \VoroTop is invoked in perturbation mode, with each atom
displaced by independent Gaussian noise of standard deviation $\sigma$
in each coordinate, at relative
magnitude $\sigma_{\rm rel} = \sigma / d_{\min}$, where $d_{\min}$
is the nearest-neighbor distance, producing at least 125{,}000
sampled Voronoi cells per structure, about 160{,}000 on average.  We then check, for every
sampled cell, whether its Weinberg vector appears in the analytical
filter of each of the three enumeration modes
(Section~\ref{sec:extended}).

\paragraph{Coverage at the reference amplitude
$\sigma_{\rm rel} = 5 \times 10^{-4}$.}
Validation is performed on the $1{,}025$ structures that complete
  enumeration in all three modes (Section~\ref{app:library}), so
  that the three filters can be compared on equal footing: for each
  structure, a single \VoroTop perturbation run (the sampled cells
  described above) is checked
  against the \emph{primary}, \emph{default}, and \emph{all} filters at once.  Because the filters are nested
  (\emph{primary}~$\subseteq$~\emph{default}~$\subseteq$~\emph{all})
  and the comparison uses one common sample per structure, the
  fraction of sampled cells found in the filter can only increase
  from \emph{primary} to \emph{default} to \emph{all}.

For each structure, we define its \emph{containment} against a
filter as the fraction of the sampled cells whose Weinberg vector
appears in that filter; containment is \emph{perfect} when every
sampled cell's vector is in the filter.  At $\sigma_{\rm rel} = 5
\times 10^{-4}$, containment is perfect against the \emph{primary}
filter for $928$ of the $1{,}025$ structures ($90.5\%$), against
\emph{default} for $1{,}003$ ($97.9\%$), and against \emph{all} for
$1{,}004$ ($98.0\%$).  The $75$ structures perfect against default but
not primary have secondary cell types, which occur with small
but positive frequency at any $\sigma > 0$; their primary containment
is nonetheless above $99.7\%$ in every case.  One further structure
is perfect only against the \emph{all} filter: three of its
$234{,}375$ sampled cells have partially resolved types.  The remaining $21$ structures
have sampled cells outside all three filters: fewer than $1\%$ of the
cells for 15 of them, between $1\%$ and $2\%$ for two, and between
$9\%$ and $30\%$ for four; the next paragraph examines them.

\paragraph{Finite-$\sigma$ convergence.}
The analytical enumeration is exact in the limit of small
perturbation---it captures every fully resolved cell topology
reachable as $\sigma \to 0$.  The validation amplitude
$\sigma_{\rm rel} = 5 \times 10^{-4}$ is small but finite, and for
a structure in which a further particle lies only slightly beyond an
equidistant set, a finite perturbation occasionally carries a cell
beyond the enumerated family.  This is the expected finite-$\sigma$
behavior of a small-$\sigma$ enumeration, not an omission from it:
repeating the full validation at $\sigma_{\rm rel} = 5 \times
10^{-5}$ brings 20 of the 22 structures that are imperfect against the
\emph{default} filter to perfect containment, including the worst, which
rises from $69.6\%$ to $100\%$.  Of the two that remain, one rises
from $73.6\%$ to $99.98\%$.  Reducing $\sigma$ does not make every structure
perfect, however: at $5 \times 10^{-5}$, 37 structures that were
perfect at $5 \times 10^{-4}$ lose a handful of their roughly
$160{,}000$ sampled cells (containment remains above $99.99\%$).
About three quarters of these cells have partially resolved types that
the \emph{all} filter contains; the rest, one to six cells in each of
15 structures, are in none of the filters.  Both are small-$\sigma$
effects, discussed next.  The \emph{default} filter is perfect for $1{,}003$
structures at $5 \times 10^{-4}$ and for $986$ at $5 \times 10^{-5}$,
so we take the larger amplitude as the reference.  The threshold $\tau$
controls how far in $\sigma$ the family stays exact
(Section~\ref{sec:near_equidistant}); an analysis at higher
temperature, where thermal displacements are larger, is
accommodated by enumerating the filter at a correspondingly larger
$\tau$.

\paragraph{The small-$\sigma$ limit.}
As $\sigma_{\rm rel}$ decreases, two of the vertices into which a
degenerate vertex resolves increasingly often lie closer together than
the numerical tolerance of the Voronoi computation, which then reports
them as a single vertex.  When the degenerate vertex has resolved into
just those two, the result is the cell with that vertex still
degenerate, and its Weinberg vector is a partially resolved type: it
lies in the \emph{all} filter but not the \emph{default} one.  When it
has resolved into more than two, the resulting cell is usually in none
of the filters.  At $\sigma_{\rm rel} = 5 \times 10^{-5}$ each kind accounts
for about one sampled cell in a hundred thousand in the structures
affected.  For data analyzed at very small effective perturbation,
such as structures quenched toward zero temperature, the \emph{all}
filter is the appropriate choice: as $\sigma \to 0$ the sampled cells
condense onto cell types with unresolved degenerate vertices, all of
which the \emph{all} filter contains.  Cells in which a vertex has
resolved only partly, whether in the way just described or as one of
the intermediate subdivisions of Section~\ref{sec:extended}, are
confined to this nearly degenerate regime and are irrelevant at any
positive temperature.

\section{Primary resolutions: criterion and validation}
\label{app:primary_validation}

Section~\ref{sec:resolution} defines a resolution as primary or
secondary by whether its frequency persists or vanishes as
$\sigma \to 0$.  A resolution whose tetrahedra all have positive
volume is primary.  For a resolution containing a zero-volume
tetrahedron, Section~\ref{sec:resolution} states a geometric
criterion: whether the other equidistant particles lie on one side
or on both sides of the plane of the tetrahedron's four coplanar
particles.  The reason for the criterion is as follows.  In
the ideal crystal the four particles are coplanar and equidistant
from the vertex, so they lie on a single circle, and a sphere
through all four can have its center anywhere on the line through
the circle's center perpendicular to its plane.  After a random
perturbation of size $O(\sigma)$, the four particles determine a
unique sphere.  To leading order that sphere still passes through
the original circle, so the perturbation decides only where on that
line the center lies.  That position depends on the ratio of the in-plane to
the out-of-plane components of the displacements, both of order
$\sigma$, and is therefore spread over a range that does not shrink
as $\sigma \to 0$.  The perturbed tetrahedron
belongs to the Delaunay triangulation, and hence to the resolution,
only if the selected sphere contains no other particle.

Among the spheres through the circle, the one centered at the
Voronoi vertex passes through all $k$ equidistant particles.  Suppose first that the other equidistant
particles lie on both sides of the plane of the four.  Any other
sphere through the
circle has its center displaced from the vertex along the line, and
it encloses the equidistant particles on the side toward which the
center moved, each by a margin proportional to the displacement of
the center.  The perturbation moves each particle by only
$O(\sigma)$, so after the perturbation such a sphere can be empty
only if its center lies within $O(\sigma)$ of the vertex.  Since the center's
position is spread over a range that does not shrink with
$\sigma$, it falls in that window with probability proportional to
$\sigma$.  Otherwise the sphere is not empty, the
tetrahedron is absent, and the four particles belong to ordinary
tetrahedra.  The probability of a resolution containing
this tetrahedron is therefore proportional to $\sigma$ and vanishes
as $\sigma \to 0$; when a resolution requires $m$ such tetrahedra,
its probability is of order $\sigma^m$.  When the vertex lies
in the plane of the four, as for the equatorial squares of the FCC
octahedron, the center must stay within $O(\sigma)$ of that plane,
which requires the four particles to stay nearly on a common circle.

Suppose instead that all other equidistant particles lie on one
side of the plane.  A sphere whose center is displaced from the
vertex away from them encloses none of them, so as the center moves
away the sphere stays empty until it reaches a particle outside the
equidistant set.  The centers of these empty spheres fill a finite
interval of the line.  The points of that interval
are equidistant from the four particles and closer to them than to
any other, so the interval is an edge of the Voronoi tessellation
along which four cells meet, a degenerate edge; along a generic edge
only three cells meet.  The center of the perturbed sphere falls in
this interval for a fixed fraction of perturbation directions.  For those
directions the other particles lie outside the sphere by margins of
order one, which the $O(\sigma)$ displacements cannot close, so the
sphere stays empty and the tetrahedron belongs to the Delaunay
triangulation.  This establishes the criterion: a resolution is
secondary if it contains a zero-volume tetrahedron with equidistant
particles on both sides of its plane, and primary otherwise.

We verified the criterion by sampling.  For the FCC octahedral
vertex, the $\omega$-Ti 2d vertex, and the simple-cubic vertex, the
ideal lattice was perturbed by independent Gaussian displacements
with standard deviation $\sigma$ in each coordinate, for $\sigma$
from $3\%$ down to $0.03\%$ of the
nearest-neighbor distance, and the Delaunay star of one particle
was recorded ($20{,}000$ to $60{,}000$ samples per $\sigma$ unless stated otherwise).  Every
star observed was one of the enumerated resolutions.  At the FCC
vertex the three primary stars appear with $\sigma$-independent
frequencies of about $33\%$ each, while the four secondary stars
together fall from $1.8\%$ of samples at $\sigma = 3\%$ to
$0.06\%$ at $\sigma = 0.075\%$, with a fitted exponent of
$0.92 \pm 0.03$ over six values of
$\sigma$ ($40{,}000$ samples each), close to the exponent $1$ of a
probability proportional to $\sigma$.  At the $\omega$-Ti 2d vertex
all four resolutions are primary and all four appear with stable
frequencies.  At the simple-cubic vertex all 256 primary stars
appear at $\sigma = 0.3\%$, where the secondary ones together account
for under $0.1\%$ of samples; at $\sigma = 3\%$, ten million samples
produce all 544 enumerated stars and no others.

Primary resolutions that contain zero-volume tetrahedra are common.  Of the $1{,}300$ library
structures, 267 have a degenerate vertex at which some zero-volume
tetrahedron lies on a degenerate edge, so that the primary family
includes resolutions containing zero-volume tetrahedra (structures with a vertex of multiplicity $k > 8$, which complete
no enumeration mode, were not examined).  The effect on the family can be
large: at the $\omega$-Ti 2d site, only 13 of the 382 primary cell
types arise from resolutions free of zero-volume tetrahedra.

\section{Detailed computational results}
\label{app:results}

This section collects per-structure enumeration results---vertex
classifications, resolution counts, and tables of cell types---for the
crystal structures analyzed in the main text.  Two are not listed.
Sigma-phase CrFe has only generic vertices at all five of its Wyckoff
sites, so its enumeration is trivial and is fully described in
Section~\ref{sec:app_sigma}.  Simple cubic is absent because its
enumeration does not complete (Section~\ref{sec:complexity}).

\subsection{BCC}

The BCC lattice (one-atom primitive cell, conventional lattice
parameter $a = 1$) has 24 Voronoi vertices per particle, all generic ($k = 4$),
and 14 face-sharing neighbors.  Since no degenerate vertices exist,
the Voronoi cell is uniquely the truncated octahedron under any
sufficiently small perturbation.  Its $p$-vector is
$(0, 6, 0, 8, 0, \ldots)$.

The truncated octahedron is the Wigner--Seitz cell of the BCC lattice~\cite{ashcroft1976solid,wigner1933constitution}
and tiles space by translation.  Its 14 faces correspond to the 8
nearest neighbors (hexagonal faces) and 6 second-nearest neighbors
(square faces) in the BCC structure.

\subsection{FCC}

The FCC lattice (conventional lattice parameter $a = 1$) has 14 Voronoi
vertices per particle: 8 generic and 6 degenerate.  Each degenerate vertex
lies at the midpoint of a second-neighbor bond and has an octahedral point
configuration of 6 equidistant particles ($k = 6$).  The central particle
has 12 full neighbors (the 12 first-nearest neighbors, face-sharing in
the unperturbed rhombic dodecahedron) and 6 partial neighbors (the 6
second-nearest neighbors, each at the opposite pole of a degenerate
vertex's octahedron).

At each degenerate vertex the enumeration finds 9 triangulations of
the octahedron, 6 of which contain a zero-volume tetrahedron.  These fall
into 7 distinct resolution types per vertex, since two pairs of
triangulations differ only in tetrahedra not containing the central
particle.  Of these types, 2 do \emph{not} realize the partial
neighbor as a face-sharer, 4 realize it as a \emph{triangular} face,
and 1 realizes it as a \emph{quadrilateral} face.  The 2 non-realization types and the
1 quadrilateral-face type are \emph{primary}; the 4 triangular-face
types are \emph{secondary} (Section~\ref{sec:resolution}).

No two of the 6 degenerate vertices share a Voronoi edge (each is
separated from the others by generic vertices), and every pair of
resolutions is compatible.  All
$7^6 = 117{,}649$ combinations produce valid cells.
These combinations yield $2{,}815$ distinct Weinberg types, whose
distribution by face count is given in Table~\ref{tab:fcc}.

\begin{table}[h]
\centering
\caption{FCC: distribution of Voronoi cell types by face count.}
\label{tab:fcc}
\begin{tabular}{crr}
\toprule
Faces & Weinberg types & Combinations \\
\midrule
12 &     7 &      64 \\
13 &    33 &     960 \\
14 &   170 &   6{,}000 \\
15 &   479 &  20{,}000 \\
16 &   891 &  37{,}500 \\
17 &   855 &  37{,}500 \\
18 &   380 &  15{,}625 \\
\midrule
Total & 2{,}815 & 117{,}649 \\
\bottomrule
\end{tabular}
\end{table}

\noindent The combination counts
$64$, $960$, $6{,}000$, $20{,}000$, $37{,}500$, $37{,}500$, and
$15{,}625$ follow the pattern
$\binom{6}{j} \cdot 2^{6-j} \cdot 5^j$ for $j = 0, \ldots, 6$, where
$j$ is the number of partial neighbors realized: $\binom{6}{j}$ chooses
which $j$ of the 6 to realize, with 2 resolutions at each of the
$6-j$ unrealized vertices and 5 (one quadrilateral plus four
triangular) at each of the $j$ realized vertices.  The cell's face
count is then $12 + j$, ranging from 12 (no partials realized) to
18 (all 6 realized).

With the \emph{all} enumeration mode (Section~\ref{sec:extended}), each vertex
has 8 options ($7 + 1$ unresolved), giving $8^6 = 262{,}144$ combinations
and $6{,}294$ distinct Weinberg types in total.

\subsection{HCP}

Ideal HCP ($c/a = \sqrt{8/3}$) has two atoms per unit cell; both
produce identical families.  The counts are the same as for FCC
(14 vertices, 8 generic, 6 degenerate with octahedral
configurations, 7 resolution types each; 12 full and 6 partial neighbors),
but the degenerate vertices are arranged
differently on the cell: three pairs of them share a Voronoi edge in
HCP, none in FCC.  These edges are generic, so the resolutions at the
vertices they join are compatible in every pairing.

The 9,556 Weinberg types span 513 distinct $p$-vectors; their
distribution by face count is given in Table~\ref{tab:hcp}.

\begin{table}[h]
\centering
\caption{HCP: distribution of Voronoi cell types by face count.}
\label{tab:hcp}
\begin{tabular}{crr}
\toprule
Faces & Weinberg types & Combinations \\
\midrule
12 &       9 &      64 \\
13 &      69 &     960 \\
14 &     440 &   6{,}000 \\
15 &   1{,}514 &  20{,}000 \\
16 &   3{,}066 &  37{,}500 \\
17 &   3{,}103 &  37{,}500 \\
18 &   1{,}355 &  15{,}625 \\
\midrule
Total & 9{,}556 & 117{,}649 \\
\bottomrule
\end{tabular}
\end{table}

The combination counts are identical to FCC (same binomial pattern),
confirming that every combination is valid.  The $3.4\times$ increase in Weinberg
types reflects the lower site symmetry ($D_{3h}$ vs.\ $O_h$).
With the \emph{all} enumeration mode (Section~\ref{sec:extended}), each vertex
has 8 options, giving $8^6 = 262{,}144$ combinations and $21{,}611$
distinct Weinberg types in total.

\subsection{A15}

A15 (Cr$_3$Si-type, $Pm\bar{3}n$, $a = 1$) has 8 atoms per conventional
cell: 2 on the 2a Wyckoff site and 6 on the 6c site.

\textbf{2a site:} 20 generic vertices, 0 degenerate.  Unique cell type:
a pentagonal dodecahedron (dual to the Z12 icosahedral
coordination) with 12 pentagonal faces and $p$-vector
$(0, 0, 12, 0, \ldots)$.

\textbf{6c site:} 24 generic vertices, 0 degenerate.  Unique cell type:
the 14-faced cell dual to the Z14 Frank--Kasper coordination
polyhedron, with 12 pentagonal and 2 hexagonal
faces, $p$-vector $(0, 0, 12, 2, 0, \ldots)$.

Both cell types are stable under any sufficiently small perturbation.

\subsection{Fluorite}

Fluorite (CaF$_2$, $Fm\bar{3}m$) in its primitive FCC cell has Ca at
$(0,0,0)$ and F at $(\frac{1}{4},\frac{1}{4},\frac{1}{4})$ and
$(\frac{3}{4},\frac{3}{4},\frac{3}{4})$.

\textbf{Ca site:} 6 degenerate vertices (all with $k = 5$), no
generic vertices, 2 resolution types per vertex.  All $2^6 = 64$
combinations are valid.  The 6 distinct cell types all have 8 faces:
\begin{center}
\begin{tabular}{lr}
\toprule
$p$-vector & Combinations \\
\midrule
$(4, 0, 0, 4, 0, \ldots)$ & 2 \\
$(2, 0, 6, 0, \ldots)$ & 4 \\
$(2, 2, 2, 2, 0, \ldots)$ & 12 \\
$(1, 3, 3, 1, 0, \ldots)$ & 24 \\
$(0, 6, 0, 2, 0, \ldots)$ & 4 \\
$(0, 4, 4, 0, \ldots)$ & 18 \\
\midrule
Total (6 types) & 64 \\
\bottomrule
\end{tabular}
\end{center}

\textbf{F site:} 16 Voronoi vertices: 4 degenerate with $k = 8$
(the same cubic configuration as the simple-cubic lattice,
Section~\ref{sec:complexity}) and 12 degenerate with $k = 5$.  Each
$k = 5$ vertex admits 4 resolutions and each $k = 8$ vertex 544.  The
resulting space of about $10^{18}$ combinations is beyond the reach of
the present algorithm (Section~\ref{sec:complexity}).

\subsection{$\omega$ phase}

The $\omega$ phase ($P6/mmm$, ideal $c/a = \sqrt{3/8}$) has 3 atoms per
hexagonal cell: 1 on the 1a Wyckoff site and 2 on the 2d site.

\textbf{1a site:} 18 Voronoi vertices (12 generic, 6 degenerate with
$k = 5$), 2 resolution types per vertex.  The edges joining adjacent degenerate
vertices are generic, so every pair of their resolutions is
compatible and all $2^6 = 64$ combinations are valid.  The 8 distinct types all have 14 faces:
\begin{center}
\begin{tabular}{lr}
\toprule
$p$-vector & Types / Combinations \\
\midrule
$(0, 6, 0, 8, 0, \ldots)$ & 1 / 2 \\
$(0, 4, 4, 6, 0, \ldots)$ & 3 / 30 \\
$(0, 2, 8, 4, 0, \ldots)$ & 3 / 30 \\
$(0, 0, 12, 2, 0, \ldots)$ & 1 / 2 \\
\midrule
Total (8 types) & 8 / 64 \\
\bottomrule
\end{tabular}
\end{center}

\textbf{2d site:} 18 vertices (6 generic, 12 degenerate with
$k = 5$), 4 resolution types per vertex, all primary.  The five
equidistant particles at each vertex form a square pyramid with the
central particle in its base, and the zero-volume tetrahedron lies on
that base, whose four particles share a degenerate edge
(Section~\ref{sec:resolution}).
The 12 degenerate vertices are the endpoints of six such edges, and
the resolutions at the two ends of each are coupled through their four
shared atoms.  This coupling and the resulting validity pruning
are discussed in Section~\ref{sec:coupling}; the distribution of the
382 distinct Weinberg types by face count is given in
Table~\ref{tab:omega2d}.

\begin{table}[h]
\centering
\caption{Distribution of cell types at the $\omega$-2d site by face
  count, after validity-pruning of the $4^{12}$ nominal resolution
  combinations.}
\label{tab:omega2d}
\begin{tabular}{crr}
\toprule
Faces & Weinberg types & Valid combinations \\
\midrule
11 &   1 &         1 \\
12 &   2 &        42 \\
13 &  16 &       735 \\
14 &  49 &     6{,}860 \\
15 & 114 &    36{,}015 \\
16 & 126 &   100{,}842 \\
17 &  74 &   117{,}649 \\
\midrule
Total & 382 & 262{,}144 \\
\bottomrule
\end{tabular}
\end{table}

\section{Cell types common to the two $\omega$-Ti sites}
\label{app:omega_shared}

Of the 8 cell types of the $\omega$-Ti 1a site and the 382 of the
2d site (Section~\ref{sec:coupling}), two are reachable from both.
Both are primary types of both sites, so their frequencies do not
vanish as thermal displacements shrink.  Indeed every cell type of
both sites is primary, since the only zero-volume tetrahedra at
these sites lie on degenerate edges
(Section~\ref{sec:resolution}).  Types are identified by their full Weinberg vectors; the
$p$-vectors below are descriptive only.  One is one of the three 1a types with $p$-vector
$(0, 4, 4, 6, 0, \ldots)$; the other, with $p$-vector
$(0, 6, 0, 8, 0, \ldots)$, is the
truncated octahedron---the Voronoi cell of the
BCC lattice---which is natural given that the $\omega$ structure is
derived from BCC by the collapse of alternating pairs of $(111)$
planes~\cite{hickman1969omega}.  The combined filter lists each of
the two vectors under both site labels; the OVITO implementation of
\VoroTop used for Figure~\ref{fig:omega_poly} assigns the
first-listed label, here 1a.

These two types are frequent in the polycrystalline sample.  At
$0.1\,T_{\omega\beta}$ the shared $(0,4,4,6)$ type is exhibited by
$8.0\%$ of all atoms and the truncated octahedron by $3.2\%$; at
$0.5\,T_{\omega\beta}$ the figures are $7.6\%$ and $4.1\%$.
Together they account for $11.2\%$ and $11.7\%$ of all atoms at
the two temperatures, essentially independent of temperature in
this range, as expected for primary types.  At higher temperatures
their combined share grows, to $12.7\%$, $15.9\%$, $24.0\%$ and
$27.1\%$ at $0.7$, $0.8$, $0.9$ and $0.95\,T_{\omega\beta}$.  The
growth is entirely in the truncated octahedron, whose share of all
atoms rises from $4.1\%$ at $0.5\,T_{\omega\beta}$ to $23.1\%$ at
$0.95\,T_{\omega\beta}$, while that of the other shared type falls from
$7.6\%$ to $4.0\%$.

To estimate how often the 1a assignment is correct, we perturb the
ideal $\omega$-Ti lattice with independent Gaussian displacements
of standard deviation $\sigma$ equal to $1\%$, $2\%$ and $3\%$ of the
nearest-neighbor distance and classify the cell of a single 1a or
2d atom ($12{,}000$ samples per site and $\sigma$).  The shared
$(0,4,4,6)$ type arises in $25\%$ of the 1a samples but only $3\%$ of
the 2d samples; the truncated octahedron arises in $7\%$ and $5\%$,
respectively.  Weighting by the $1{:}2$ multiplicity of the two
sites, an atom exhibiting the shared $(0,4,4,6)$ type occupies a 1a
site with probability $0.82$, and an atom exhibiting the truncated
octahedron with probability $0.39$; both values are insensitive to
$\sigma$ over the range considered.  The first-listed rule
therefore labels most atoms exhibiting the shared $(0,4,4,6)$ type
correctly and most atoms exhibiting the truncated octahedron
incorrectly.  At $0.1$ and $0.5\,T_{\omega\beta}$ this amounts to a net
mislabeling of roughly $3.5$--$4\%$ of all atoms.  The
mosaic pattern of Figure~\ref{fig:omega_poly} remains clearly visible,
since the shared-type atoms are scattered through the grain
interiors.

The estimate does not extend to higher temperatures.  In the same
model the truncated octahedron never accounts for more than about
$6\%$ of the atoms, at any displacement amplitude up to $12\%$ of the
nearest-neighbor distance, so the growth of its share in the simulation
more likely reflects BCC-like local environments appearing as the
$\omega \to \beta$ transition is approached.  The filter cannot
distinguish these from $\omega$, because the truncated octahedron
belongs to both $\omega$ families.

\begin{figure}[t]
\centering
\includegraphics[width=0.55\textwidth]{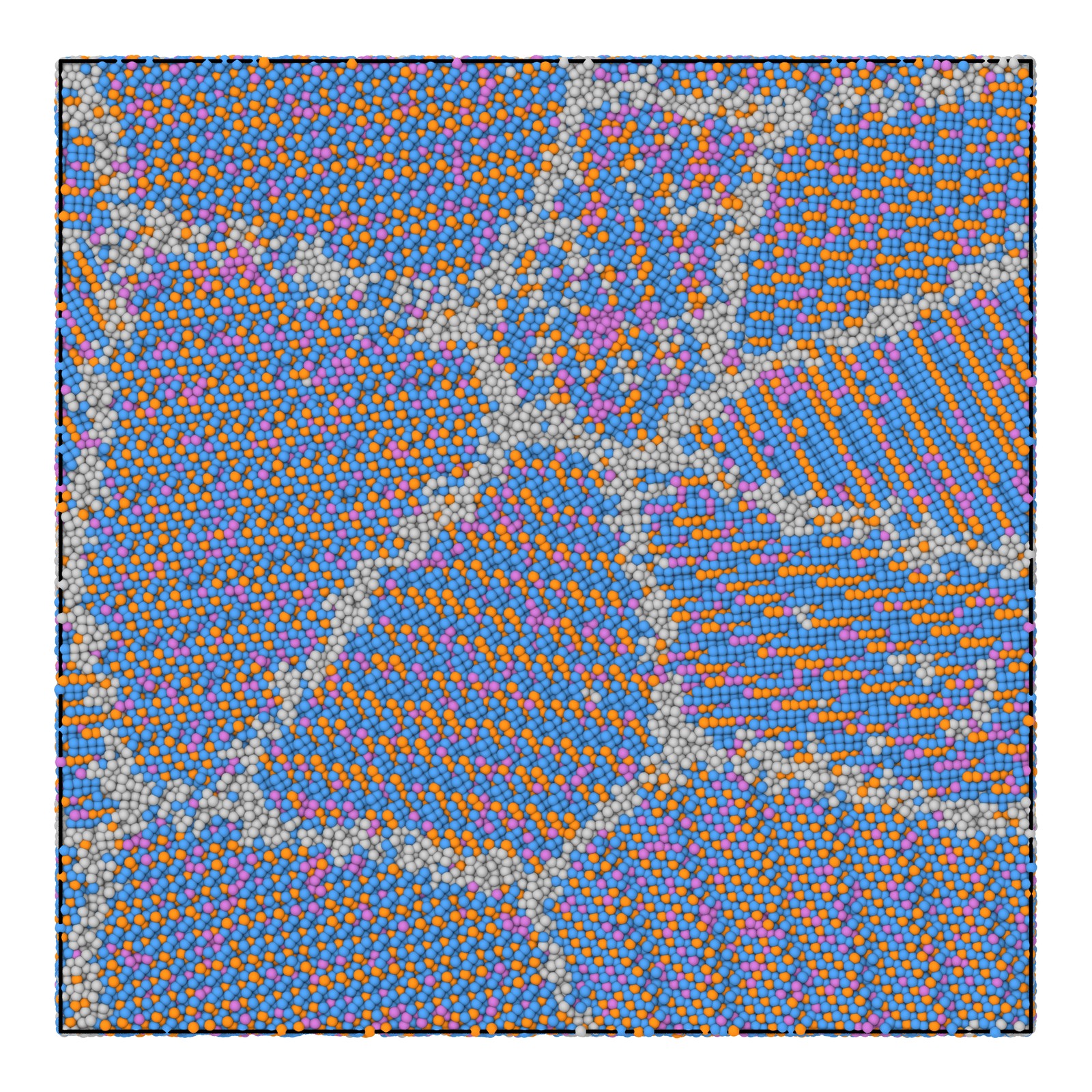}
\caption{The $0.5\,T_{\omega\beta}$ snapshot of
  Figure~\ref{fig:omega_poly}(a), with atoms exhibiting one of the
  two cell types common to both sites shown in violet rather than
  as 1a (orange).  All remaining atoms are colored as before:
  unshared 1a types in orange, unshared 2d types in blue, and other
  atoms in gray.
  The violet atoms are $11.7\%$ of the total.}
\label{fig:omega_shared}
\end{figure}

Figure~\ref{fig:omega_shared} shows the $0.5\,T_{\omega\beta}$
snapshot with the shared-type atoms rendered in a third color.

\section{Conventional methods on the $\omega$-Ti polycrystal}
\label{app:omega_methods_comparison}

Section~\ref{sec:app_omega} shows that \VoroTop with the analytical
$\omega$ filter classifies polycrystalline $\omega$-Ti
across the $0.1$--$0.95\,T_{\omega\beta}$ temperature
range, distinguishing the 1a and 2d sublattices within each grain
through their unshared cell types.
Here we apply three standard discrete structural-identification
methods to the $0.5\,T_{\omega\beta}$ snapshot.  None has a reference for
$\omega$-Ti, and their outcomes differ qualitatively.  The centrosymmetry parameter, a continuous score rather than a
classifier, is not included; its behavior is described in
Section~\ref{sec:app_omega}.

\begin{figure}[t]
\centering
\subfiglabel{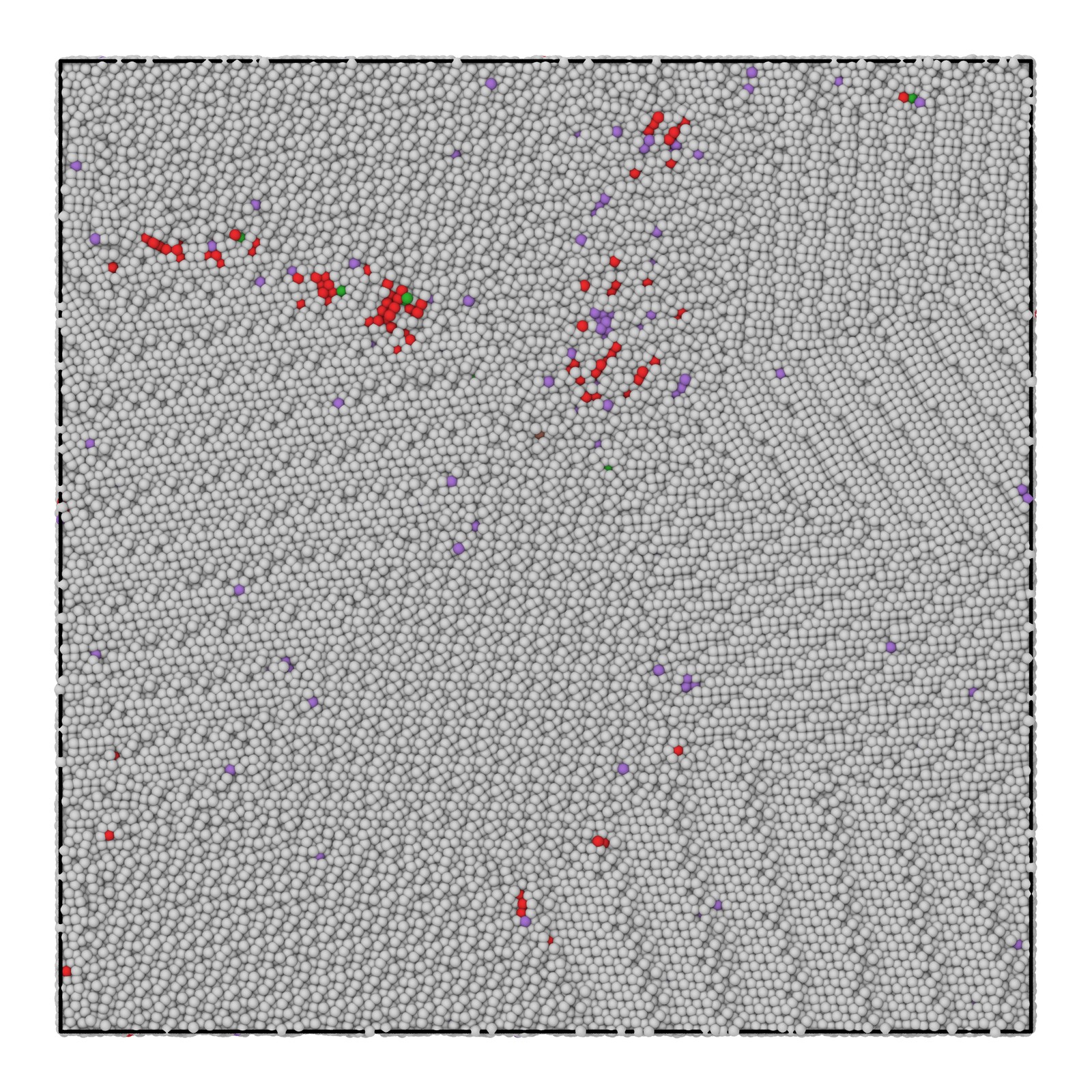}{(a)}\hfill
\subfiglabel{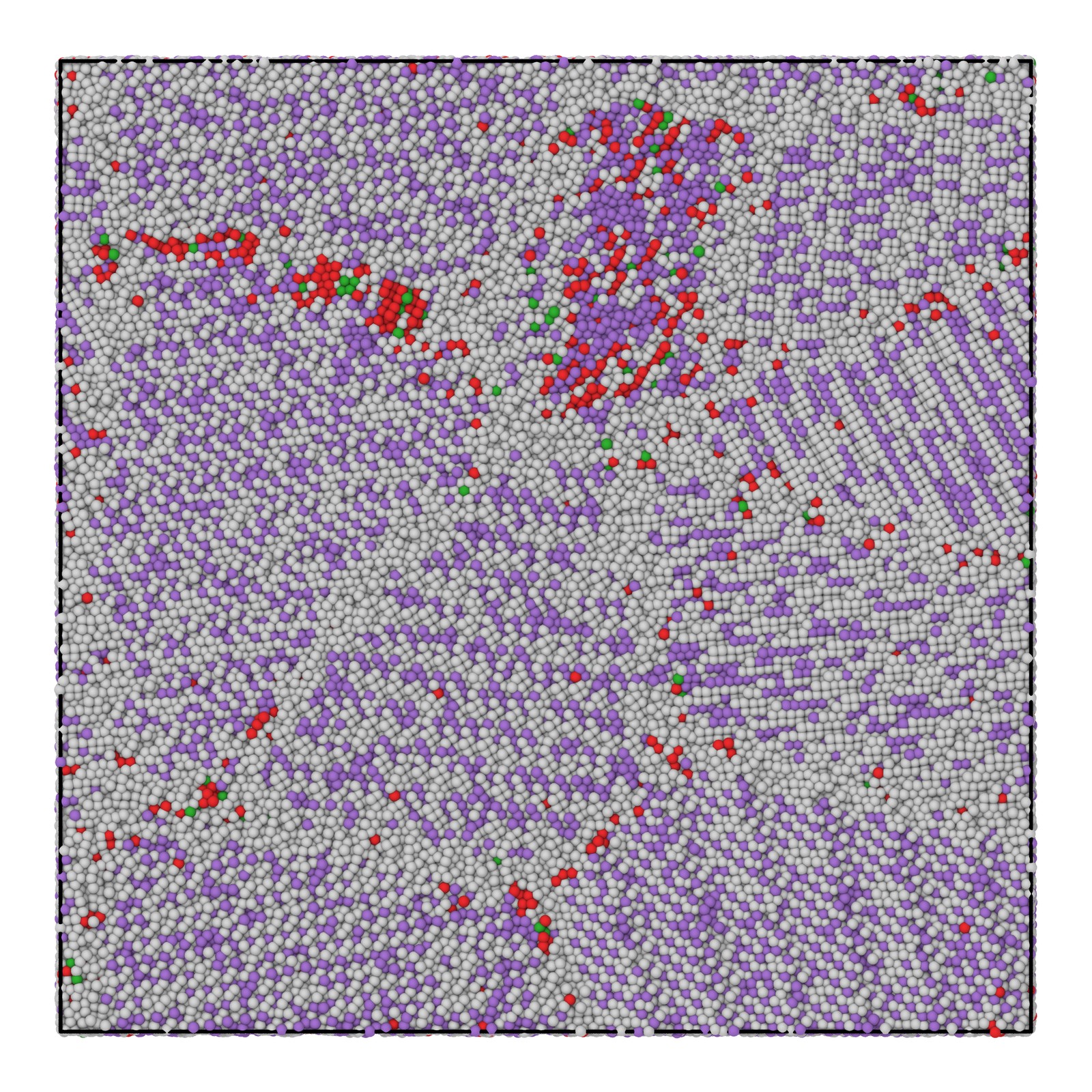}{(b)}\hfill
\subfiglabel{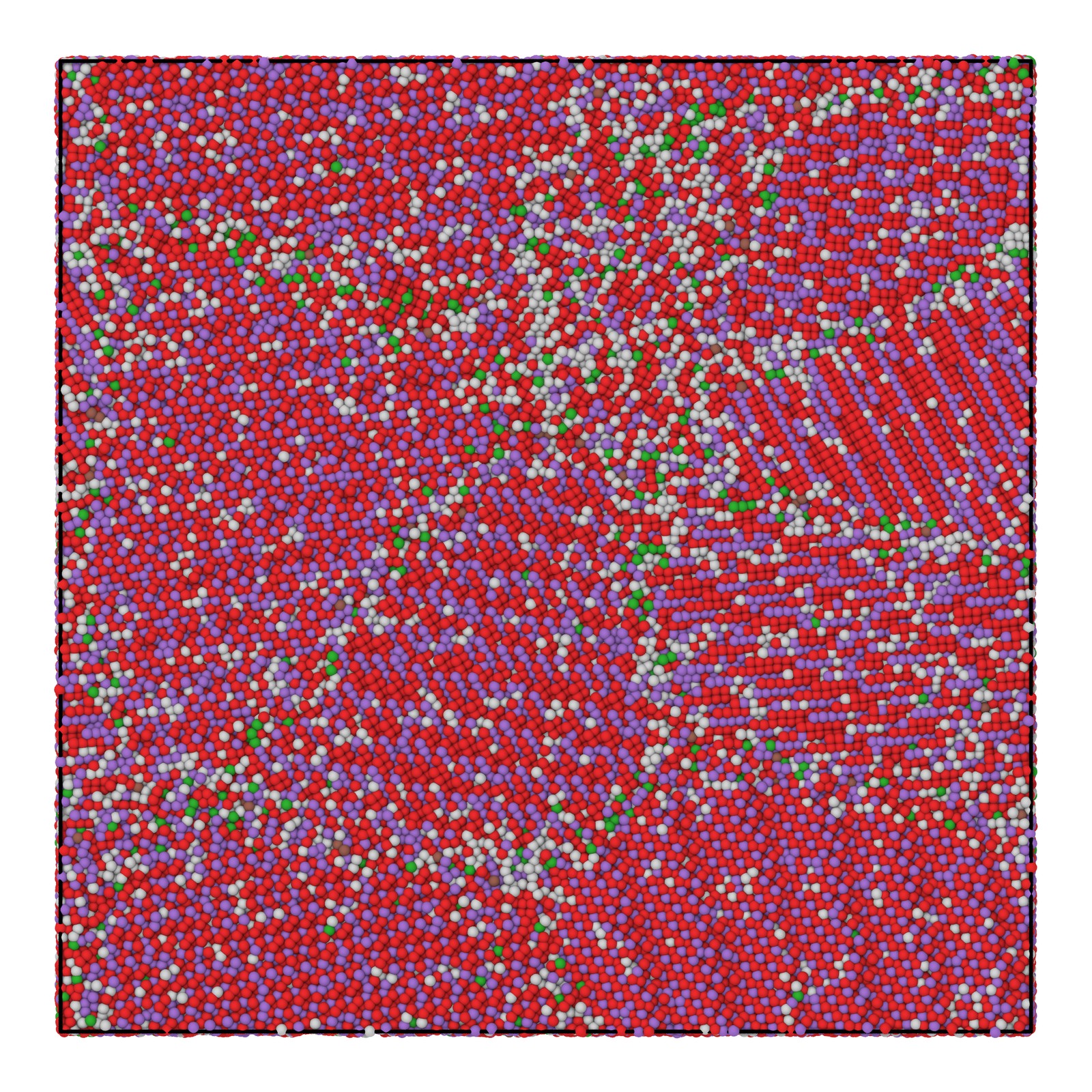}{(c)}
\caption{The polycrystalline $\omega$-Ti sample of
  Figure~\ref{fig:omega_poly}(a), classified by three conventional
  methods: (a)~common neighbor analysis (CNA), (b)~polyhedral template
  matching (PTM; RMSD cutoff $0.15$, default templates), (c)~Ackland--Jones.  Atoms are
  colored by each method's structure-type assignment---FCC (green), HCP (red), BCC (purple), icosahedral
  (brown), and other
  (gray).  CNA leaves nearly all atoms as other; PTM labels
  roughly $30\%$ as BCC or HCP; Ackland--Jones labels
  the majority as HCP or BCC.  None has a reference for the
  $\omega$ structure, in contrast to \VoroTop with the analytical
  $\omega$ filter (Figure~\ref{fig:omega_poly}(a)).}
\label{fig:omega_methods}
\end{figure}

Figure~\ref{fig:omega_methods} renders the $0.5\,T_{\omega\beta} =
585$\,K snapshot under each of the three conventional methods, with consistent
coloring across panels: other atoms (not matching any predefined
structure type) in light gray, FCC in green, HCP in red, BCC in purple,
and icosahedral in brown.  The same snapshot classified by \VoroTop
is shown in Figure~\ref{fig:omega_poly}(a).

Adaptive common neighbor analysis (CNA), which has no tunable threshold, classifies $97.85\%$ of atoms
as other, with only a small scattered remainder ($2.15\%$) labeled
HCP, BCC, or FCC: an honest abstention.  Polyhedral template matching
(PTM), run with the FCC, HCP, and BCC templates that are enabled by
default, is parametrized by an RMSD cutoff that controls
how poor a fit between an atom's local environment and a template
is acceptable, and behaves very differently depending on
this cutoff (Table~\ref{tab:omega_ptm_sweep}).  At the default cutoff
(RMSD $= 0.10$) PTM classifies $98.80\%$ of atoms as
other---like CNA, a clean abstention.  At a more permissive
RMSD $= 0.15$ (the value we use in
Figure~\ref{fig:omega_methods}(b)), $69.09\%$ of atoms are
classified as other; the remaining roughly $30\%$ are labeled BCC or
HCP.  The BCC labels follow the 1a sublattice: $70\%$ of the atoms
with unshared 1a cell types are labeled BCC, while $86\%$ of those with
unshared 2d cell types remain other.  At larger cutoffs the other fraction
continues to fall, and with no cutoff at all PTM labels roughly
$50\%$ of atoms HCP and $40\%$ BCC.  At no cutoff does PTM identify
$\omega$-Ti, since neither site is in its template set, although a
custom $\omega$ template could in principle be supplied.  With the
icosahedral template also enabled, $1.7\%$ of atoms are labeled
icosahedral at RMSD $0.15$.

\begin{table}[h]
\centering
\caption{PTM per-class atom counts for the $585$\,K snapshot under
  several RMSD cutoffs, with the FCC, HCP, and BCC templates that are
  enabled by default; the icosahedral template is off, hence the
  dashes.  As the cutoff is loosened, the method passes smoothly from
  abstention to labeling most atoms HCP or BCC.  The final row
  ($\infty$) removes the cutoff entirely; the atoms still labeled
  other are those for which PTM finds no valid template
  correspondence at any RMSD.}
\label{tab:omega_ptm_sweep}
\begin{tabular}{lrrrrr|r}
\toprule
RMSD cutoff & other & FCC & HCP & BCC & ICO & non-other (\%) \\
\midrule
0.10 & $202{,}070$ &     68 &  $1{,}454$ &     931 & --- &  1.20 \\
0.12 & $189{,}035$ &    254 &  $3{,}472$ & $11{,}762$ & --- &  7.57 \\
0.15 & $141{,}295$ & $1{,}053$ &  $8{,}793$ & $53{,}382$ & --- & 30.91 \\
0.18 &  $78{,}057$ & $2{,}076$ & $55{,}720$ & $68{,}670$ & --- & 61.83 \\
0.20 &  $56{,}380$ & $3{,}141$ & $75{,}203$ & $69{,}799$ & --- & 72.43 \\
$\infty$ & $12{,}836$ & $6{,}562$ & $101{,}421$ & $83{,}704$ & --- & 93.72 \\
\bottomrule
\end{tabular}
\end{table}

The Ackland--Jones method~\cite{ackland2006jones}, also known as
bond-angle analysis~\cite{stukowski2012structure}, has no
equivalent tunable threshold and produces a
single result: $49.07\%$ HCP, $29.27\%$ BCC, $17.56\%$ other,
$3.52\%$ FCC, and $0.58\%$ icosahedral.  Its angular-distribution
heuristic finds enough similarity between the $\omega$ Wyckoff sites
and its HCP and BCC references to assign the
majority of atoms to those classes.  The assignment follows the
sublattices: $74\%$ of the atoms with unshared 1a cell types are
labeled BCC, and $78\%$ of those with unshared 2d cell types are
labeled HCP.  It does not, however, separate the grain boundaries from
the bulk, since $72\%$ of the atoms that \VoroTop classifies as other
also receive a crystalline label.  Per-class atom counts for the three
methods and for \VoroTop are given in
Table~\ref{tab:omega_methods_breakdown}.

\begin{table}
\centering
\caption{Per-class atom counts at $0.5\,T_{\omega\beta} = 585$\,K
  for CNA, PTM (RMSD cutoff $0.15$, default templates), Ackland--Jones, and \VoroTop;
  $204{,}523$ atoms in total.  The icosahedral template is not among PTM's defaults, hence the dash.}
\label{tab:omega_methods_breakdown}
\small
\setlength{\tabcolsep}{4pt}
\begin{tabular}{lrrrrrr}
\toprule
Method & other & FCC & HCP & BCC & ICO & \\
\midrule
CNA              & $200{,}128$ &       78 &     $1{,}781$ &   $2{,}513$ &        23 & \\
PTM              & $141{,}295$ & $1{,}053$ &     $8{,}793$ &  $53{,}382$ &         --- & \\
Ackland--Jones   &  $35{,}920$ & $7{,}201$ &   $100{,}351$ &  $59{,}857$ & $1{,}194$ & \\
\midrule
\multicolumn{6}{l}{\VoroTop $\omega$-1a + $\omega$-2d (bulk):}
                            & \multicolumn{1}{r}{$161{,}832$ ($79.13\%$)} \\
\multicolumn{6}{l}{\VoroTop other (mainly grain boundaries):}
                            & \multicolumn{1}{r}{ $42{,}691$ ($20.87\%$)} \\
\bottomrule
\end{tabular}
\end{table}

Across the three methods the outcomes range from abstention to
confident labeling under the names of other structures.  The labels
are not arbitrary, as the Ackland--Jones result shows, but without an
$\omega$ reference these methods cannot tell $\omega$ from genuine BCC
or HCP material, and they do not separate the grain boundaries from
the bulk.  \VoroTop's reference is the filter, generated automatically
from the crystal structure.

\end{document}